\documentclass[a4paper,11pt]{article}
\usepackage[utf8]{inputenc}
\usepackage{amsmath,amsfonts,amssymb,array,graphicx,color,comment}
\usepackage{graphicx}
\usepackage[nosort]{cite}
\usepackage{amssymb}
\usepackage{amsthm}
\usepackage{caption}
\usepackage{mathtools}
\usepackage{subcaption}
\usepackage[linktocpage]{hyperref}

\newcommand{\Nf}{N}

\usepackage[margin=1in]{geometry}

\graphicspath{{Figures/}}

\numberwithin{equation}{section}

\newcommand{\rmmode}[2]{L^{(#1)}_{\,#2}}

\newcommand{\ket}[1]{|#1 \rangle}
\newcommand{\bra}[1]{\langle #1 |}
\newcommand{\kket}[1]{|#1 \rangle\!\rangle}
\newcommand{\aver}[1]{\langle #1 \rangle}
\newcommand{\bigaver}[1]{\Big\langle #1 \Big\rangle}
\newcommand{\skel}[1]{\Bigg\rangle\overset{\displaystyle #1}{\!\!\!-\!\!\!-\!\!\!-\!\!\!-\!\!\!-\!\!\!-\!\!\!}\Bigg\langle}
\newcommand{\cblock}[5]{\sideset{_{\displaystyle #2}^{\displaystyle #1}}{_{\displaystyle #4}^{\displaystyle #3}}{\mathop{\skel{#5}}}}
\newcommand{\cblockv}[5]{\sideset{_{\displaystyle #2\ }^{\displaystyle #1\ }}{_{\ \displaystyle #4}^{\ \displaystyle #3}}{\mathop{\rotatebox[origin=c]{90}{$\skel{\rotatebox{-90}{\scalebox{0.9}{$#5$}}}$}}}}

\newcommand{\Tb}{\overline{T}}

\newcommand{\hb}{\bar{h}}

\newcommand{\wb}{\bar{w}}

\newcommand{\zb}{\bar{z}}

\newcommand{\M}{\mathcal{M}}
\newcommand{\Cbb}{\mathbb{C}}
\newcommand{\Dbb}{\mathbb{D}}
\newcommand{\Hbb}{\mathbb{H}}
\newcommand{\Rbb}{\mathbb{R}}
\newcommand{\Sbb}{\mathbb{S}}
\newcommand{\Zbb}{\mathbb{Z}}
\newcommand{\id}{\mathbf{1}}
\newcommand{\OVir}{\mathrm{OVir}}
\newcommand{\FGauss}{{}_2 \mathrm{F}_1}
\newcommand{\wt}{\widetilde}
\newcommand{\hh}{\widehat{h}}

\newcommand{\acon}[3]{\mathcal{A}^{(#1)}_{#2,#3}}
\newcommand{\bcon}[4]{\mathcal{B}^{(#1)}_{#2,#3,#4}}

\begin{document}
\title{R\'enyi entropies for one-dimensional quantum systems with mixed boundary conditions}
\author{Benoit Estienne, Yacine Ikhlef, Andrei Rotaru}

\date{\today}
\maketitle

\begin{abstract}
We present a general method for calculating R\'enyi entropies in the ground state of a one-dimensional critical system with mixed open boundaries, for an interval starting at one of its ends. 

Technically, we consider the case when the boundary operator implementing the change of boundary conditions is degenerate under the Virasoro algebra. By exploiting the null-vectors conditions,

we derive the ordinary differential equations that govern the scaling functions associated to the entropies. In particular, we provide an explicit formula for the second R\'enyi entropy. Additionally, we identify and compute the leading finite-size corrections to compare our theoretical results with numerical data for the Ising and three-state Potts critical chains.

\end{abstract}

\clearpage
\tableofcontents
\clearpage

\section{Introduction}
\label{sec:intro}

The understanding of quantum entanglement has proved to be a research topic of continued and central interest for physicists working in domains as diverse as high energy physics, condensed matter theory and quantum information. Entanglement measures have turned out to be useful diagnostic tools for tensor network algorithms, quantities of interest for the AdS/CFT correspondence, and, most relevantly for the present work, a powerful tool for probing the physics of quantum many-body systems. With respect to the latter, the study of entanglement has proved crucial to the study of phase transitions in one-dimensional quantum systems, by allowing their detection and the characterization of their critical exponents and corresponding central charge \cite{vidal_entanglement_2003,calabrese_entanglement_2004,2008JHEP...11..076C,furukawa_mutual_2009}. Important applications of entanglement are found in higher dimensions as well. We mention, for two-dimensional systems,  the establishment of intrinsic topological order and various anyonic quantum dimensions \cite{Kitaev_2006,Levin_2006} and the detection and counting of critical Dirac fermions \cite{2009PhRvB..80k5122M,2017JSMTE..04.3104C,2018SciA....4.5535Z,2021PhRvB.103w5108C}. Finally, entanglement can also be used in two \cite{PhysRevB.88.155314,Crepel:2018ycz,2019NatCo..10.1860C,PhysRevLett.123.126804,PhysRevB.101.115136} or higher dimensions \cite{2021PhRvB.103p1110H,2021ScPP...11...16E}  to reveal gapless interface modes. While the focus on entanglement entropies has been mostly theoretical, in recent years, experimental proposals as well as actual experiments have been designed to measure them \cite{brydges_probing_2019,niknam_experimental_2021,doi:10.1126/science.aau0818,2021arXiv210107814V,Neven_2021,PhysRevLett.125.120502}.

The basic setup is as follows: one considers a quantum system in a pure state $\ket{\Psi}$, and a spatial bipartition of said system into two complementary subregions $A$ and $B$. The entanglement between them is then encoded in the reduced density matrix $\rho_A = \mathrm{Tr}_B \ket{\Psi} \langle\Psi|$ and it can be quantified through entanglement measures, such as the \textit{R\'enyi entanglement entropies} \cite{2008RvMP...80..517A,Calab_Cardy_Doyon2009,RevModPhys.82.277,LAFLORENCIE20161,headrick2019lectures}
\begin{equation}
  S_N(A)=\frac{1}{1-N} \log \mathrm{Tr}_A\left(\rho_A^N\right) \,,
\end{equation}
and in particular the $N\to 1$ limit, corresponding to the well-known \textit{von Neumann entropy}:
\begin{equation}
  S(A) = -\mathrm{Tr}_A\left(\rho_A \log \rho_A\right) \,.
\end{equation}
For strongly correlated quantum systems, the theoretical computation of entanglement entropies is a technically challenging endeavor. However, if these systems are one-dimensional and critical, the formidable toolbox of two-dimensional Conformal Field Theory (CFT) is available to tackle such computations. Arguably the most renowned result obtained in this framework is the \textit{universal} asymptotic growth for the ground state entanglement entropy of an interval $A$ of length $\ell$ in an infinite system: \cite{vidal_entanglement_2003,calabrese_entanglement_2004,CALLAN199455,holzhey_geometric_1994,latorre_ground_2004,calabrese_cardy_2009}
\begin{equation} \label{eq:EE_1}
  S_N(\ell) \sim
  \frac{c}{6} \, \frac{N+1}{N} \, \log\ell \qquad (\ell \to \infty) \,,
\end{equation}
where $c$ is the central charge of the one-dimensional critical system under consideration. 
The calculation of entanglement entropies through such methods rests on two crucial insights. The first one is that, for integer values of $N$, and a subsystem $A=\cup_i [u_i,v_i]$ built as the union of some disjoint intervals, the moments of the reduced density matrix $\mathrm{Tr}_A\left(\rho_A^N\right)$ can be expressed as the partition function of an $N$-sheeted Riemann surface with conical singularities corresponding to the endpoints of the intervals $[u_i,v_i]$ \cite{calabrese_entanglement_2004,CALLAN199455}. Such partition functions can be evaluated, with significant toil, for free theories and some special cases of interacting models \cite{furukawa_mutual_2009,calabrese_entanglement_2009,2012JSMTE..02..016R,coser_renyi_2014,calabrese_entanglement_2011,alba_entanglement_2010,alba_entanglement_2011,datta_renyi_2014,2016JSMTE..05.3109C,2018JSMTE..11.3101R,2024JHEP...05..236E}. 
In general, however, a second insight is needed to make progress: the replication of the \textit{spacetime} of the theory can be ``exchanged''  for the replication of the \textit{target space} of the CFT \cite{dixon_conformal_1987,klemm_orbifolds_1990,borisov_systematic_1998}. Such a construction, known in the literature as the \textit{cyclic orbifold CFT} \cite{borisov_systematic_1998}, is built from the permutation symmetric product of $N$ copies of the original CFT (referred to as \textit{the mother CFT}), by modding out the discrete subgroup $\Zbb_N$ of cyclic permutations. In this framework,  the conical singularities of the mother CFT defined on the replicated surface are accounted for by insertions of \textit{twist operators} \cite{dixon_conformal_1987} in cyclic orbifold correlators. Thus, by computing correlators of twist operators, one can evaluate $\mathrm{Tr}_A\left(\rho_A^N\right)$ for a variety of setups. Furthermore, one can easily adapt this twist operator formalism to encode modified boundary conditions around the branch points \cite{dupic_entanglement_2018}, which is fitting for computations of more refined entanglement measures such as the  symmetry-resolved entanglement entropy \cite{Laflor_2014,PhysRevLett.120.200602,PhysRevB.98.041106,Capizzi_2020,10.21468/SciPostPhys.10.3.054,bonsignori_boundary_2021} or for explorations of entanglement in non-unitary systems \cite{dupic_entanglement_2018,Bianch_2014}.

\begin{figure}[h!]
  \centering
  \includegraphics[scale=0.8]{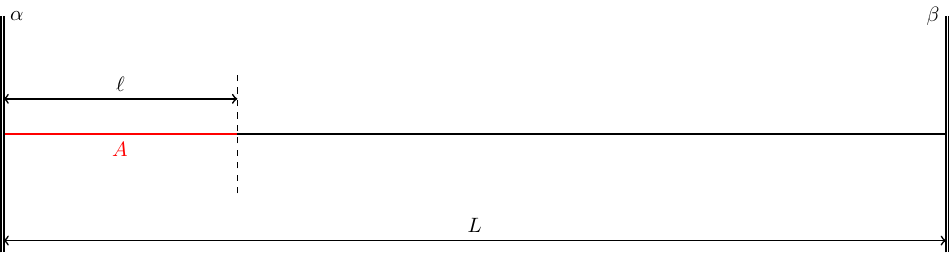}
  \caption{An interval of length $\ell$ in a 1d critical chain with mixed BC $(\alpha\beta)$ and length $L$.}
  \label{fig:intro1dchain}
\end{figure}

In this article, we consider the R\'enyi entanglement entropy in a one-dimensional critical system, with open boundary condition (BC) $\alpha$ at one end of the chain and $\beta$ at the other end, when the subregion $A$ is a single interval \emph{adjacent to one boundary} (see Figure~\ref{fig:intro1dchain}). In the scaling limit, such a critical system is described by a Boundary Conformal Field Theory (BCFT), with a well-understood \cite{cardy_boundary_1989,cardy_bulk_1991,cardy_finite-size_1988,behrend_classification_1998} correspondence between the Virasoro representations and the \textit{conformal boundary conditions}, and an algebra of boundary operators that interpolate between them, namely the boundary condition changing operators (BCCOs) $\psi^{(\alpha\beta)}$. 

The case of identical BCs, namely $\alpha=\beta$, has been thoroughly analysed using either conformal field theory methods \cite{calabrese_entanglement_2004,calabrese_cardy_2009,taddia_entanglement_2013,taddia2013entanglement,Taddia_2016,Estienne:2021xzo} or exact free fermion techniques \cite{fagotti_universal_2011,xavier_entanglement_2020,jafarizadeh_entanglement_2021}. These analytical studies have been complemented with numerical implementations based on density-matrix renormalization group (DMRG) techniques \cite{laflorencie_boundary_2006,Ren_2008,PhysRevB.99.195445,zhou_entanglement_2006} and other methods \cite{Orus_2014}.
In that setup, 
the determination of the R\'enyi entanglement entropy reduces to computing a one-point twist function on the upper half-plane or on the disk, and yields the following universal behaviour,
as the lattice spacing $a \to 0$ (keeping $\ell$ and $L$ finite) 
\cite{calabrese_entanglement_2004}: 
\begin{equation} \label{eq:EE_2}
  S_\Nf^{(\alpha\alpha)}(\ell) \sim \frac{c}{12} \frac{\Nf+1}{\Nf} \log \left[
    \frac{2 L}{\pi a}\sin\left(\frac{\pi\ell}{L}\right) 
  \right] + \log g_\alpha + C_N  \,,
\end{equation}
where $g_\alpha$ is the \textit{universal boundary entropy} \cite{affleck_universal_1991} associated to the boundary condition $\alpha$, and $C_N$ is a non-universal constant that does not depend on $\alpha$. 

In the case of mixed boundary conditions $\alpha \neq \beta$, conformal invariance yields the general form of the R\'enyi entropy
\begin{equation} \label{eq:EE_3}
  S_N^{(\alpha\beta)}(\ell) \sim \frac{c}{12} \frac{N+1}{N} \log \left[
    \frac{2 L}{\pi a}\sin\left(\frac{\pi\ell}{L}\right)
  \right] + \frac{1}{1-N} \log \mathcal{F}_N^{(\alpha\beta)}(\ell/L)  + C_N \,,
\end{equation}
where $\mathcal{F}_N^{(\alpha\beta)}$ is a universal scaling function, which interpolates smoothly between the values $g_\alpha^{1-N}$ and $g_\beta^{1-N}$. The aim of the present work is to find an exact determination of this scaling function for $\ell \in [0,L]$. We focus on the situation when the BCCO $\psi^{(\alpha\beta)}$ is degenerate at level two under the Virasoro algebra, \textit{i.e.} when $\psi^{(\alpha\beta)} \equiv \psi_{12}$ in the usual Kac notation.
Through Cardy's doubling trick, the problem then amounts to computing a (chiral) four-point function of the form $\aver{\Psi_{12}\sigma^\dag\sigma\Psi_{12}}$ in a cyclic orbifold CFT, where $\sigma$ is the twist operator, and $\Psi_{12}=(\psi_{12})^{\otimes N}$ is the replicated degenerate operator. We perform this calculation in various examples, by deriving a linear Ordinary Differential Equation (ODE) for this four-point function, through the use of null-vector conditions under the orbifold Virasoro algebra. We also compute the leading finite-size corrections to \eqref{eq:EE_3} by applying a similar approach to the correlators of the form $\aver{\Psi_{12}\sigma^\dag_\phi\sigma_\phi^{\phantom\dag}\Psi_{12}}$, where $\sigma_\phi$ is a \textit{composite} twist operator, associated to a primary operator $\phi$ in the mother CFT. Typically, $\phi$ is an ``energy-like'' operator, \textit{i.e.} a relevant primary operator in the invariant sector under the discrete symmetries of the model.

The article is organised as follows. In Section~\ref{sec:setup_summary}, we recall the general setup leading to the generic form \eqref{eq:EE_3}, and we provide a summary of our analytical and numerical results for a generic BCFT, and for the Ising and three-state Potts critical chains. In Section~\ref{sec:orbifold} we quickly review the cyclic orbifold construction, emphasising its implementation on the upper-half plane, including the calculation of some OPE structure constants, which are useful for our particular purpose.
In Section~\ref{sec:ODEs} we expose the derivation of ODEs for the scaling functions $\mathcal{F}_N^{(\alpha\beta)}$ in various cases. The more technical details are included in the Appendix.

\section{Summary of results}
\label{sec:setup_summary}

\subsection{Scaling functions for R\'enyi entropies}
\label{sec:scaling-functions}

We consider a quantum system at criticality on an open chain of size $L$, with lattice spacing $a$.
The left and right ends of the system have distinct boundary conditions, denoted as $\alpha$ and $\beta$, respectively. As the system size becomes sufficiently large, the universal properties of the bulk are described by a two-dimensional CFT, whereas the behaviour of the boundaries can be effectively characterised using \emph{conformal boundary conditions}. Notably, we are excluding any consideration of subleading corrections resulting from irrelevant boundary perturbations \cite{Eriksson_2011,2016JHEP...10..140C,PhysRevB.96.075153}.

\begin{figure}[h!]
    \centering
    \includegraphics{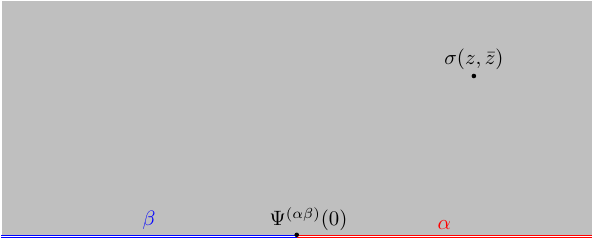}
    \caption{
    Operator insertions for the $\Zbb_N$ orbifold BCFT correlators of interest on the upper-half plane with mixed BC ($\alpha$ for $x>0$ and $\beta$ for $x<0$ ). There is another BCCO at $x=\infty$, which is not depicted here.}
    \label{fig:UHP_BCCO}
\end{figure}

Our objective is to evaluate the entanglement entropy between the interval $A=[0,\ell]$ and its complement $B=[\ell,L]$, in the ground state, as depicted in Figure \ref{fig:intro1dchain}.
The scaling limit is defined by letting $a \to 0$, while keeping the lengths $L$ and $\ell$ finite. The $N{}^{\text{th}}$ R\'enyi entropy $S_N^{(\alpha\beta)}(\ell)$ is given (up to a non-universal additive constant) by the one-point function of the twist operator $\sigma$\cite{Calabrese:2004eu}:
\begin{align}
  S_N^{(\alpha\beta)}(\ell) = \frac{1}{1-N}  \log \langle \sigma(\ell,\bar\ell) \rangle^{(\alpha \beta)}_{\mathbb{S}_L },
\end{align}
where $\mathbb{S}_L$ denotes the infinite strip (with imaginary time running along the imaginary axis) 
\begin{align}
  \mathbb{S}_L = \{w \in \mathbb{C}, \, 0 < \textrm{Re}(w) < L\} \,,
\end{align}
with conformal BC $\alpha$ (resp. $\beta$) on the left (resp. right) boundary.
The dimension of the twist operator $\sigma$ is given by \cite{calabrese_entanglement_2004,klemm_orbifolds_1990,borisov_systematic_1998}
\begin{equation}
  h_\sigma = \frac{c}{24} \left(N-\frac1N \right) \,,
\end{equation}
where $c$ is the central charge.
Upon mapping to the upper half plane $\Hbb$ via the map
\begin{align}\label{eq:maptoUHP}
    \begin{cases}
        \Sbb_L \to \Hbb \\
        w \mapsto z(w) =  \exp(i\pi w/L) \,,
    \end{cases}
\end{align}
one gets
\begin{align}
  \aver{\sigma(w,\wb)}^{(\alpha\beta)}_{\mathbb{S}_L }
  =  \left|\frac{\pi z}{L}\right|^{2h_{\sigma}}
  \aver{\sigma(z,\zb)}_{\Hbb}^{(\alpha\beta)} \,.
\end{align}
On the right-hand side, the upper half-plane $\Hbb$ has boundary conditions $\alpha$ for $x>0$ and $\beta$ for $x<0$, as in Fig.~\ref{fig:UHP_BCCO}. 
The corresponding R\'enyi entropy reads
\begin{equation} \label{eq:S_N.vs.F_N}
  \boxed{S_N^{(\alpha\beta)}(\ell) = \frac{c}{12}\frac{N+1}{N}
  \log\left[ \frac{2L}{\pi a} \sin\left(\frac{\pi\ell}{L}\right)\right]
  + \frac{1}{1-N} \log \mathcal{F}_N^{(\alpha\beta)}(e^{i\pi\ell/L},e^{-i\pi\ell/L}) \,,}
\end{equation}
where we reintroduced the lattice spacing $a$, and the universal function $\mathcal{F}_N^{(\alpha\beta)}$ is given by
\begin{equation} \label{eq:F_N}
  \boxed{\mathcal{F}_N^{(\alpha\beta)}(z,\zb) = |z-\zb|^{2h_\sigma} \,
  \aver{\sigma(z,\zb)}^{(\alpha\beta)}_{\Hbb} \,,
  \qquad z\in\Hbb \,.}
\end{equation}
For $\alpha=\beta$, the function $\mathcal{F}_N^{(\alpha\alpha)}$ reduces to a one-point function on $\Hbb$, or equivalently on the unit disc $\Dbb$:
\begin{equation}
  \mathcal{F}_N^{(\alpha\alpha)}(z,\zb) = \aver{\sigma(0)}^{(\alpha)}_{\Dbb} = g_\alpha^{1-N} \,,
\end{equation}
[see \eqref{eq:A_sigma}], and one recovers the well-known expression~\eqref{eq:EE_2} of \cite{calabrese_entanglement_2004}.
In the generic situation $\alpha \neq \beta$, the function $\mathcal{F}_N^{(\alpha\beta)}$ is non-trivial. It tends to a finite constant on each part of the boundary. Indeed, the leading behaviour is given by the bulk-boundary OPE $\sigma \to \id$, which yields
\begin{equation} \label{eq:BC.F_N}
  \lim_{\mathrm{Arg}(z) \to 0} \mathcal{F}_N^{(\alpha\beta)}(z,\zb) = g_\alpha^{1-N} \,,
  \qquad \lim_{\mathrm{Arg}(z) \to \pi} \mathcal{F}_N^{(\alpha\beta)}(z,\zb) = g_\beta^{1-N} \,.
\end{equation}

The leading asymptotic behaviour of the entanglement entropy as the lattice spacing $a$ approaches zero is captured by (\ref{eq:S_N.vs.F_N}--\ref{eq:F_N}), up to an additive constant that is independent of the boundary condition and arises from the normalisation of the twist operator.
When comparing with a finite-size system, it is necessary to account for various sources of corrections originating from irrelevant deformations of the Hamiltonian in both the bulk and the boundary \cite{Cardy:2010zs,eriksson_corrections_2011} or parity effects \cite{calabrese_parity_2010}.
Furthermore, it is well established \cite{taddia_entanglement_2013,rajabpour_entanglement_2012} that for open systems, these finite-size corrections to the entanglement entropy are more pronounced than for a periodic chain. 
The most significant corrections arise from the subleading scaling of the lattice twist operators \cite{dupic_entanglement_2018}. Like any other lattice operator, the lattice twist operator $\sigma_{\text{lattice}}$ can be expressed, in the continuum limit, as a local combination of scaling operators
\begin{align}
  \sigma_{\text{lattice}} \sim \sum_{\phi \text{ primary}}  C_\phi \, a^{2 h_{\sigma_{\phi}}} \sigma_{\phi} + 
  \dots
\end{align}
where $\dots$ denotes the contribution of descendants under the orbifold Virasoro algebra.
Here, $a$ is
the lattice spacing, the sum is over primary operators of the mother theory, and $\sigma_\phi$ denotes the composite twist operator \cite{Bianch_2014,Horvath:2021rjd,dupic_entanglement_2018} associated to the primary operator $\phi$.
The conformal dimension of the composite twist operator is given by
\begin{equation}
  h_{\sigma_\phi} = h_\sigma + \frac{h_\phi}{N} \,,
\end{equation}
where $h_\phi$ is the conformal dimension of $\phi$.
The constants $C_\phi$ are non-universal, dimensionful amplitudes.
Thus, the expectation value of the twist operator reads, in the scaling limit,
\begin{equation} \label{eq_generic_subleading_corrections}
  \aver{\sigma_{\text{lattice}}(\ell)}_{\Sbb_L}^{(\alpha\beta)} \sim \sum_{\phi \text{ primary}}
    C_\phi \, \left(\frac{\pi a}{2L\sin\frac{\pi\ell}{L}}\right)^{2 h_{\sigma_{\phi}}}
    \mathcal{F}_{N,\phi}^{(\alpha\beta)}(z,\zb) + \dots
\end{equation}
where $z=\exp(i\pi\ell/L)$, and the typical scaling function is given by
\begin{align} \label{eq:F_{N,phi}} 
  \boxed{\mathcal{F}_{N,\phi}^{(\alpha\beta)}(z,\zb)
  =|z-\zb|^{2h_{\sigma_\phi}} \aver{\sigma_\phi(z,\zb)}^{(\alpha\beta)}_{\Hbb} \,, \qquad z\in\Hbb \,.}
\end{align}
It obeys the boundary conditions
\begin{equation} \label{eq:BC.F_{N,phi}}
  \lim_{\mathrm{Arg}(z) \to 0} \mathcal{F}_{N,\phi}^{(\alpha\beta)}(z,\zb) = A_{\phi,\id}^{(\alpha)} \, g_\alpha^{1-N} \,,
  \qquad \lim_{\mathrm{Arg}(z) \to \pi} \mathcal{F}_{N,\phi}^{(\alpha\beta)}(z,\zb) = A_{\phi,\id}^{(\beta)} \, g_\beta^{1-N} \,,
\end{equation}
where $A_{\phi,\id}^{(\alpha)}$ and $A_{\phi,\id}^{(\beta)}$ are bulk-boundary OPE coefficients in the mother CFT [see \eqref{eq:bulk_boundary_OPE}].

One can implement the change of boundary conditions by the insertion of the BCCO $\Psi^{(\alpha \beta)} = (\psi^{(\alpha \beta)})^{\otimes N}$ associated to the change of BC $\alpha \to \beta$ on all replicas in the orbifold BCFT:
\begin{align}
  \aver{X}^{(\alpha \beta)}_{\Hbb}
  = \frac{\aver{\Psi^{(\beta\alpha)}(\infty) X \Psi^{(\alpha\beta)}(0)}_{\Hbb}}{\aver{\Psi^{(\beta\alpha)}(\infty) \Psi^{(\alpha\beta)}(0)}_{\Hbb}} \,,
\end{align}
where $X$ denotes any product of bulk operators.
An alternative expression for the scaling functions~\eqref{eq:F_N} and \eqref{eq:F_{N,phi}} can then be obtained through a conformal map from the upper half-plane $\Hbb$ to the unit disk $\Dbb$, namely $u \mapsto (u-z)/(u-\zb)$. This yields
\begin{equation} \label{eq:F_{N,phi}bis} 
    \boxed{\begin{aligned}
    & \mathcal{F}_{N}^{(\alpha\beta)}(z,\zb)
    =  \frac{\aver{\Psi^{(\beta\alpha)}(1) \sigma(0) \Psi^{(\alpha\beta)}(z/\zb)}_{\Dbb}}{\aver{\Psi^{(\beta\alpha)}(1)  \Psi^{(\alpha\beta)}(z/\zb)}_{\Dbb}} \,, \\
    & \mathcal{F}_{N,\phi}^{(\alpha\beta)}(z,\zb)
    =  \frac{\aver{\Psi^{(\beta\alpha)}(1) \sigma_{\phi}(0) \Psi^{(\alpha\beta)}(z/\zb)}_{\Dbb}}{\aver{\Psi^{(\beta\alpha)}(1)  \Psi^{(\alpha\beta)}(z/\zb)}_{\Dbb}} \,.
\end{aligned}
}
\end{equation}

As is customary with the $\mathbb{Z}_{\Nf}$ cyclic orbifold, one possible approach to studying the scaling function~\eqref{eq:F_{N,phi}bis} is to work on the $N$-sheeted branched covering of the base manifold. In this approach, a three-point function as in the RHS of \eqref{eq:F_{N,phi}bis} becomes a disk partition function of the mother CFT, with the boundary partitioned into $2\Nf$ connected components, with alternating boundary conditions $\alpha$ and $\beta$, as described in \cite{taddia_entanglement_2013}. This can be generalized to the case of an excited twist field, yielding the expression
\begin{align} \label{eq:F_{N,phi}unfolded} 
    \mathcal{F}_{N,\phi}^{(\alpha\beta)}(z,\zb)
    =  \frac{1}{N^{2N h_{\psi}}}
    \frac{\bigaver{\prod_{a=0}^{N-1}\psi^{(\beta\alpha)}(\omega^a)\psi^{(\alpha\beta)}\left(\omega^a (z/\zb)^{1/N}\right) \, \phi(0)}_{\Dbb}}
    {\bigaver{\psi^{(\beta\alpha)}(1)\psi^{(\alpha\beta)}(z/\zb)}^N_{\Dbb}} \,,
\end{align}
where $\omega=\exp(2i\pi/N)$. One of the main limitations of this unfolding method is the dramatic increase of moduli space. What starts as a simple one-variable problem, involving only the variable $\eta=z/\zb$, becomes a more complicated problem with $2N$ variables. Although, in the specific case under study, these $2N$ positions of the operators are not entirely independent, calculating the $2N$-point correlation functions typically requires treating them as if they were. As a result, this method is only practical for certain CFTs, where the computation of such $2N$-point correlation functions can be carried out explicitly, such as in the Ising model or the free scalar field \cite{taddia_entanglement_2013}. We thus adopt an alternative approach to calculate the scaling function.

Rather than unfolding the correlation function into the cumbersome expression \eqref{eq:F_{N,phi}unfolded}, one can instead operate within the framework of the cyclic orbifold.  Cardy's doubling trick guarantees that the correlation function $\aver{\Psi^{(\beta\alpha)}(\infty) \sigma_{\phi}(z,\zb) \Psi^{(\alpha\beta)}(0)}_{\Hbb}$ is a linear combination of conformal blocks of the form 
\begin{equation} \label{eq:conformal_block(z,zb)}
    \cblock{\Psi^{(\beta\alpha)}(\infty)}{\ \Psi^{(\alpha\beta)}(0)}{\sigma_{\phi}^\dag(\zb)}{\sigma_{\phi}(z)}{\Psi_k} \,. 
\end{equation}
After rescaling and observing that $|z-\zb|=(z-\zb)/i$ for any $z \in \Hbb$, we thus obtain the scaling function as a linear combination of rescaled conformal blocks:
\begin{equation} \label{eq_generic_block_expansion}
    \mathcal{F}_{N,\phi}^{(\alpha\beta)}(z,\zb)
    = (\eta-1)^{2h_{\sigma_\phi}} \sum_k \, X_k   \cblock{\Psi^{(\beta\alpha)}(\infty)}{\ \Psi^{(\alpha\beta)}(0)}{\sigma_{\phi}^\dag(1)}{\sigma_{\phi}(\eta)}{\Psi_k} \,,
    \qquad \eta = z/\zb \,.
\end{equation}
The allowed intermediate states $\Psi_k$ are in correspondence with untwisted boundary operators \cite{2023JPhA...56T5403E}, determined by the bulk-boundary OPE $\sigma_{\phi} \to \Psi_k$ as the operator $\sigma_{\phi}$ tends to the boundary and the boundary OPE $\Psi^{(\alpha\beta)}\cdot\Psi^{(\beta\alpha)} \to \Psi_k$, and the coefficients $X_k$ can be related to the OPE structure constants.
Naturally, the complexity is conserved to some extent, and the main challenge here lies in dealing with orbifold conformal blocks, which are significantly more intricate than standard Virasoro ones \cite{2018JSMTE..11.3101R,ares_crossing-symmetric_2021}. However, it has been shown in \cite{dupic_entanglement_2018} that provided 
the operators $\psi^{(\alpha\beta)}$ and/or $\phi$ are degenerate under the Virasoro algebra in the mother CFT,
it is possible in some cases to derive a linear ODE for these conformal blocks. This is the approach we follow in this paper, under the assumption that the operator $\psi^{(\alpha\beta)}$ is degenerate at level two.

\subsection{Generic BCFT}

We consider a
mother CFT with central charge $c <1$.
We shall use the Kac parameterisation for the central charge and the conformal dimensions
\begin{equation} \label{eq:Kac}
  c = 1-\frac{6(1-g)^2}{g} \,, \qquad h_{rs} = \frac{(r-sg)^2-(1-g)^2}{4g} \,,\qquad 0<g<1 \,,
\end{equation}
and we denote by $\phi_{rs}$ the primary operator of dimension $h_{rs}$. As mentioned above, we restrict our analysis to the particular case when the operator $\psi^{(\alpha\beta)}$
has conformal dimension $h_\psi=h_{12} = (3g -2 )/4$, in which case we have 
\begin{equation} \label{eq:null_vec}
    L_{-1}^2  \psi^{(\alpha\beta)} = g \, L_{-2} \psi^{(\alpha\beta)} \,.
\end{equation}
Under this assumption, we derive a linear ODE for the leading and subleading scaling functions at $N=2$, and for the leading scaling function at $N=3$. Using the Frobenius method, the solutions of these ODEs can always be represented as explicit, exact power series, which in some cases simplify to well-known special functions, such as hypergeometric functions. 
The results presented below are expressed in terms of the parameterisation (see Appendix~\ref{app:change_of_var})
\begin{equation} \label{eq:def.zeta}
   \mathcal{F}(z,\zb) = F[\zeta(z/\zb)] \,,
   \qquad \zeta(\eta) = \frac{(1+\sqrt\eta)^2}{4\sqrt\eta} \,.
\end{equation}
In polar coordinates, if $z=r \exp(i\theta)$, then $\eta=\exp(2i\theta)$ and $\zeta(z/\zb)=\cos^2(\theta/2)$.
The contribution to $\mathcal{F}(z,\zb)$ from the conformal block with intermediate state $\Psi_k$ in the channel $\mathrm{Arg}(z)\to 0$  is typically a power-series solution to the ODE, of the form
\begin{equation}
    F_k(\zeta) = (1-\zeta)^{h_{\Psi_k}/2} \sum_{n=0}^\infty a_{n,k} \, (1-\zeta)^n \,.
\end{equation}
Alternatively, one can work with the basis of solutions $\{F_k(1-\zeta)\}$, corresponding to the channel $\mathrm{Arg}(z)\to \pi$.

\paragraph{The $N=2$ R\'enyi entropy.} For $N=2$, we find that, upon the change of variables \eqref{eq:def.zeta}, the scaling function $\mathcal{F}_2^{(\alpha\beta)}$ in \eqref{eq:S_N.vs.F_N} obeys the hypergeometric ODE (see Appendix~\ref{app:hypergeom}).
Together with the boundary conditions \eqref{eq:BC.F_N}, this fixes the function $\mathcal{F}_2^{(\alpha\beta)}$ completely:
\begin{equation} \label{eq:F2}
  \boxed{\mathcal{F}_2^{(\alpha\beta)}(z,\zb)
  = g_\alpha^{-1}\, G\left(\cos^2\frac{\theta}{2}\right) + g_\beta^{-1}\, G\left(\sin^2\frac{\theta}{2}\right) \,,
  \qquad \theta=\mathrm{Arg}(z) \,,}
\end{equation}
where the function $G$ is the unique solution of \eqref{eq:hypergeom.ode.zeta} satisfying $G(0) = 0$ and $G(1) =1$. This function can be expressed in terms of Gauss's hypergeometric function $\FGauss$ as 
\begin{equation} \label{eq:G}
  \boxed{G(\zeta) = \zeta^{h_{13}} \times
  \frac{\FGauss\left(\left. g, 1-g;  2g \right| \zeta \right)}
  {\FGauss(g, 1-g; 2g| 1)} \,,}
\end{equation}
where $g$ is related to the central charge through \eqref{eq:Kac}, and the exponent in the prefactor reads $h_{13}=2g-1$. This simple form of the solution is valid only if $h_{13}>0$.

Any scaling function $\mathcal{F}_{2,\phi}^{(\alpha\beta)}$ \eqref{eq:F_{N,phi}} contributing to the finite-size corrections of $S^{(\alpha\beta)}_2$ involves a composite twist operator $\sigma_\phi$.
For generic $\phi$, we derive the \textit{fourth-order} linear ODE for $\mathcal{F}_{2,\phi}^{(\alpha\beta)}$, upon the change of variable \eqref{eq:def.zeta}
\begin{equation} \label{eq:ODE-fourth-order}
  \boxed{\begin{aligned}
    & (g-1) g \left[16 (1-2 g) h_{\sigma_\phi}-3 (3 g-2) (16 h_{\sigma_\phi}-1) \zeta (\zeta-1)\right] F(\zeta)\\
    & +2(g-1) (1-2 \zeta) \left[(4 g^2-8 g+3)+3 \left(6 g^2 -15g +10-8g h_{\sigma_\phi}\right) \zeta(\zeta-1) \right] F'(\zeta) \\
    & +\zeta(\zeta-1)  \left[8 (2g^2-5g+3)+(66 g^2-177g+120-16g h_{\sigma_\phi})\zeta (\zeta-1) \right] F''(\zeta) \\
    & +5 (2 g-3) \zeta^2(\zeta-1)^2  (1-2 \zeta) F'''(\zeta)
    + 2 \zeta^3 (\zeta-1)^3 F^{(4)}(\zeta)=0 \,.
  \end{aligned}}
\end{equation}
The local exponents of the ODE at $\zeta \to 0$ or $\zeta \to 1$
yield the dimensions of intermediate states $\Psi_k$ in the conformal blocks \eqref{eq_generic_block_expansion}:
\begin{equation}
  \id \,, \quad (\id\otimes\phi_{13}+\phi_{13}\otimes\id) \,, \quad \phi_{13}\otimes\phi_{13} \,, \quad \phi_{13}\otimes\phi_{13} \,.
\end{equation}
Let us denote by $F_1(\zeta),F_2(\zeta),F_3(\zeta),F_4(\zeta)$  the solutions of the ODE corresponding to these conformal blocks, and normalised so that $F_k(\zeta) \sim (\zeta-1)^{h_{\Psi_k}/2}$ as $\zeta \to 1$. The functions $F_k(\zeta)$ can be described as explicit power series and evaluated numerically to arbitrary precision for $\zeta \in [0,1]$, through the Frobenius method.
The physical solution then reads, up to an overall phase factor,
\begin{equation}
  \mathcal{F}_{2,\phi}^{(\alpha\beta)}(z,\zb) = \lambda_1 \, F_1(\zeta) + \lambda_2 \, F_2(\zeta) + \lambda_3 \, F_3(\zeta) + \lambda_4 \, F_4(\zeta) \,,
  \quad \zeta=\cos^2\frac{\mathrm{Arg}\,z}{2} \,,
\end{equation}
where the constants $\lambda_1,\dots,\lambda_4$ are related to the orbifold OPE structure constants in a simple manner.

\paragraph{The $N=3$ R\'enyi entropy.}
For $N=3$, the scaling function $\mathcal{F}_3^{(\alpha\beta)}$ obeys the third-order ODE upon the change of variable \eqref{eq:def.zeta}
\begin{equation}  \label{eq_ODE_3_leading}
  \boxed{\begin{aligned}
    & 4 AB(6A+B+3) (1-2\zeta) F(\zeta) + 2[9AB - C\zeta(1-\zeta)]F'(\zeta) \\
    & - 9(4A+B-2)(1-2\zeta)\zeta(1-\zeta) F''(\zeta)
    + 18\zeta^2(1-\zeta)^2 F'''(\zeta)=0 \,,
  \end{aligned}}
\end{equation}
where
\begin{equation}
  A=g-1 \,, \qquad B=6g-5 \,, \qquad C = 36A^2+2B^2+30AB-18A-3B \,.
\end{equation}
The local exponents at $\zeta \to 0$ or $\zeta \to 1$ correspond to the intermediate states 
\begin{equation}
  \id \,, \quad (\id\otimes\psi_{13}\otimes\psi_{13}+\psi_{13}\otimes\id\otimes\psi_{13}+\psi_{13}\otimes\psi_{13}\otimes\id) \,, \quad \psi_{13}\otimes\psi_{13}\otimes\psi_{13} \,,
\end{equation}
and $\mathcal{F}_3^{(\alpha\beta)}$ is given by a linear combination of the power series solutions $F_1,F_2,F_3$.

\subsection{The critical Ising model}

We consider the critical Ising quantum chain, with free ($f$) or fixed ($+$ or $-$) open BCs. There are two inequivalent choices with different left and right BCs, namely $(\alpha,\beta)=(+,-)$, and $(\alpha,\beta)=(+,f)$.

The leading behaviour of R\'enyi entropies for generic $N$ in the critical Ising model was previously obtained in \cite{taddia_entanglement_2013} through a different approach, and compared to numerical data of the actual Ising chain obtained using DMRG.  It is worth noting that even for relatively large systems of $\sim 10^2$ sites, large deviations were observed between the lattice data and the leading CFT prediction. 

Our method allows to recover the leading CFT contribution to the R\'enyi entropy (for $N=2$ and $N=3$), but also to compute the first subleading correction.  We observe that the agreement with the lattice data becomes excellent once this correction is taken into account,  even for small systems of $\sim 26 $ sites accessible to exact diagonalisation.

\begin{figure}[h]
  \centering
  \begin{subfigure}[b]{0.49\textwidth}
    \centering
    \includegraphics[width=\textwidth]{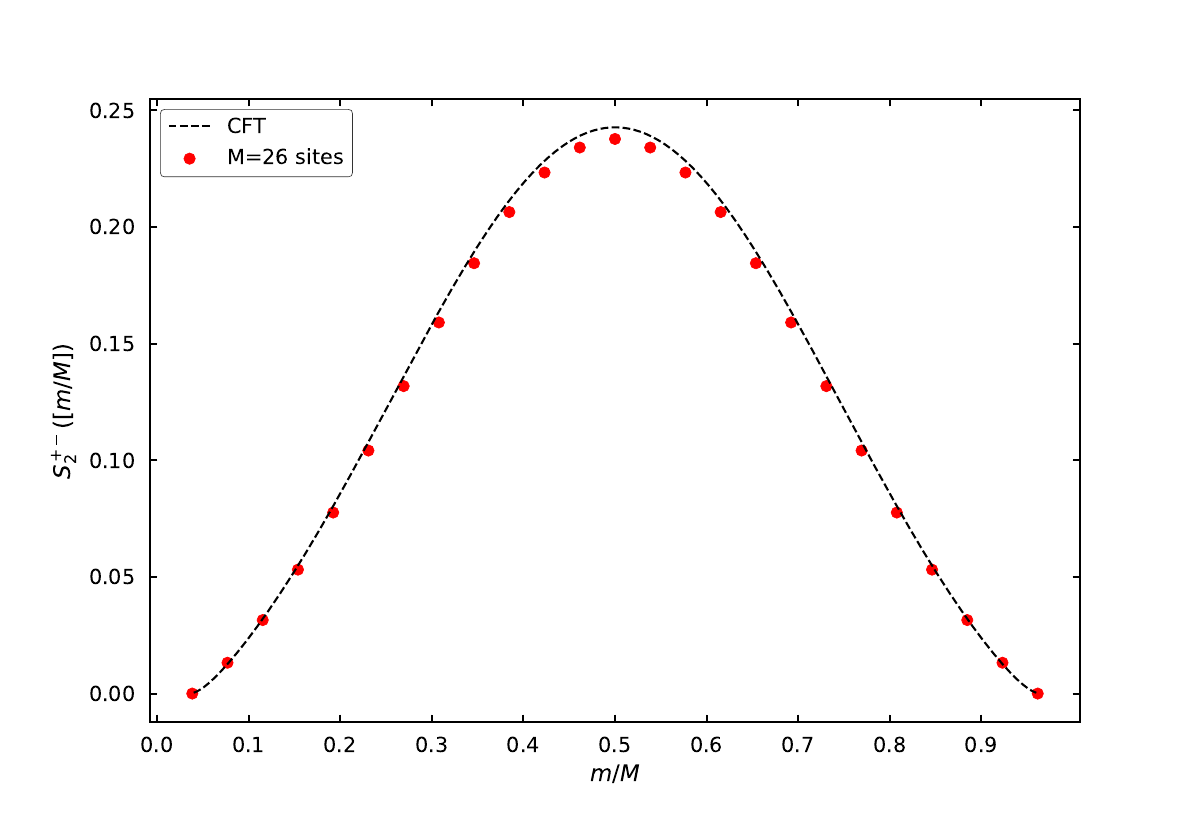}
    \caption{Fixed mixed BC ($+-$)}
    \label{renyi2fixedmixedBC}
  \end{subfigure}
  \begin{subfigure}[b]{0.49\textwidth}
    \centering
    \includegraphics[width=\textwidth]{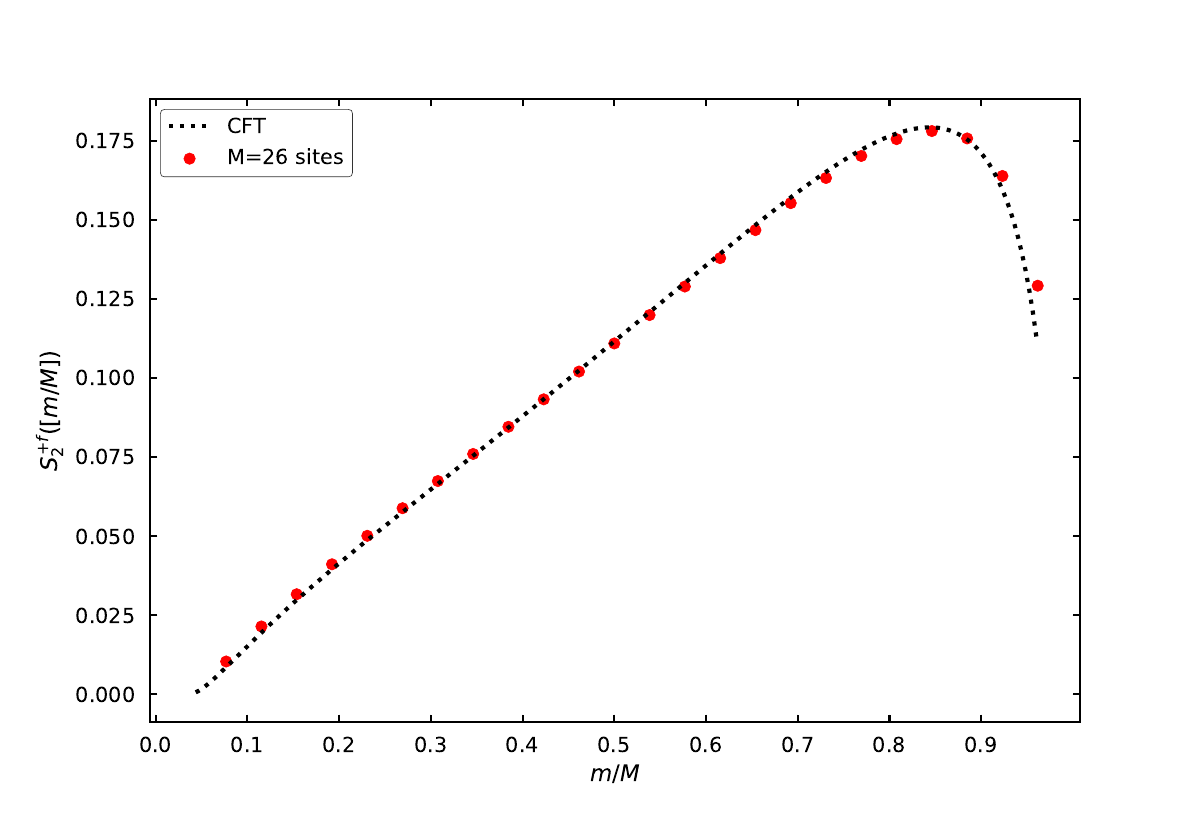}
    \caption{Fixed-free mixed BC ($+f$)}
    \label{renyi2fixedfreeBC}
  \end{subfigure}
  \caption{The second R\'enyi entropy of the interval $[0,m]$ in the critical Ising chain with two types of mixed BC, for a chain of size $M=26$. In both cases, the interval is adjacent to the $\alpha=+$ boundary.}
  \label{fig:renyi2graphs}
\end{figure}

For $N=2$,  the scaling functions appearing in the expansion \eqref{eq_generic_subleading_corrections} read
\begin{equation}
  \begin{aligned}
    &\mathcal{F}_2^{(+-)}(re^{i\theta},re^{-i\theta}) = \sqrt{2} \left( 1-\tfrac{1}{4} \sin^2\theta \right) \,,
    &&\qquad \mathcal{F}_{2,\varepsilon}^{(+-)}(re^{i\theta},re^{-i\theta}) = \sqrt{2} \left( 1-\tfrac{9}{4} \sin^2\theta \right) \,, \\
    &\mathcal{F}_2^{(+f)}(re^{i\theta},re^{-i\theta}) = \sqrt{2} \cos \frac{\theta}{4} \,,
    &&\qquad \mathcal{F}_{2,\varepsilon}^{(+f)}(re^{i\theta},re^{-i\theta}) = \sqrt{2} \cos \frac{3\theta}{4} \,.
  \end{aligned}
\end{equation}
The expressions for the leading contribution $\mathcal{F}_2^{(\alpha\beta)}$ agree with the results reported in \cite{taddia_entanglement_2013}. As far as we know, the subleading corrections $\mathcal{F}_{2,\varepsilon}^{(\alpha\beta)}$ had not been computed previously. We show in Figure \ref{fig:renyi2graphs} the remarkable agreement between our CFT calculations (which include the leading and subleading contributions) and the numerical results from the exact diagonalisation of the Ising Hamiltonian, for the second R\'enyi entropy of the interval $[0,m]$ in a system of $M$ sites.

\begin{figure}[h!]
    \centering
    \includegraphics[scale=0.6]{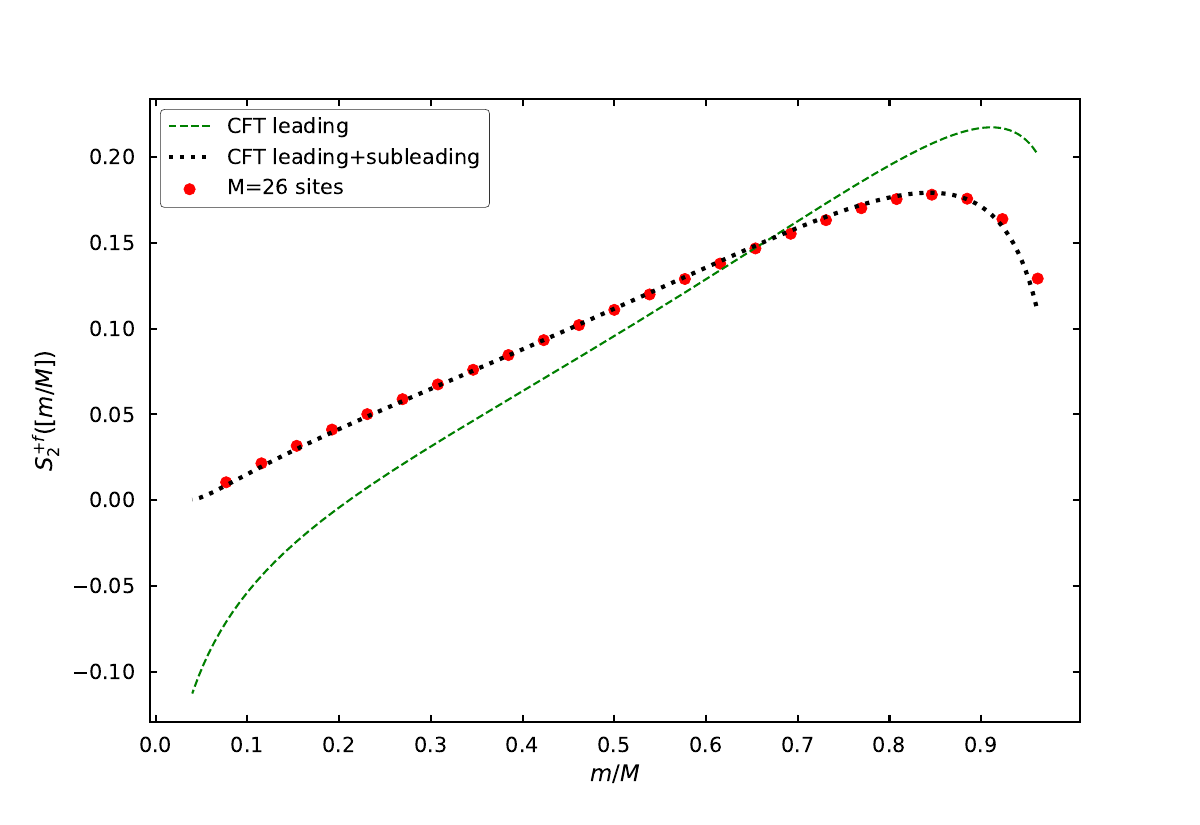}
    \caption{The second R\'enyi entropy of the interval $[0,m]$ in the critical Ising chain of size $M=26$ with mixed free fixed BC. Inclusion of the subleading term in the expansion \eqref{eq_generic_subleading_corrections} is crucial for obtaining a satisfying agreement between CFT predictions and lattice results.}
    \label{fig:FSScomparison}
\end{figure}

To illustrate the large amplitude of finite-size effects, we show in Figure \ref{fig:FSScomparison} how the CFT prediction fares against the lattice results with and without the incorporation of the subleading term. Even for the curve including both subleading and leading terms in \eqref{eq_generic_subleading_corrections}, the agreement with lattice data becomes worse close to the boundary. This can be traced to the presence of corrections from \textit{descendants} of twist operators, which introduce terms of $\mathcal{O}(M^{-h_{\varepsilon}-1})$ relative to the bare twist contribution.

\begin{figure}[h!]
  \centering
  \begin{subfigure}[b]{0.49\textwidth}
    \centering
    \includegraphics[width=\textwidth]{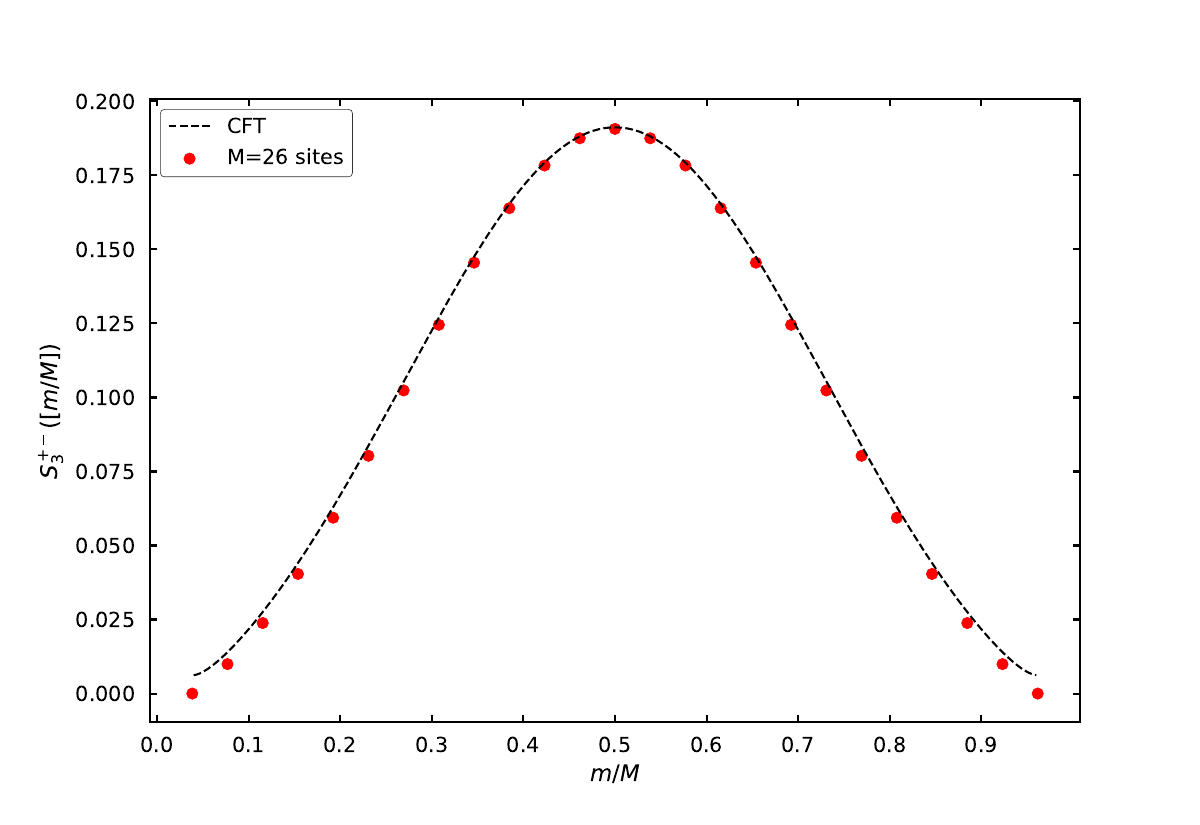}
    \caption{Fixed mixed BC}
    \label{fig:renyi3fixedmixed}
  \end{subfigure}
  \begin{subfigure}[b]{0.49\textwidth}
    \centering
    \includegraphics[width=\textwidth]{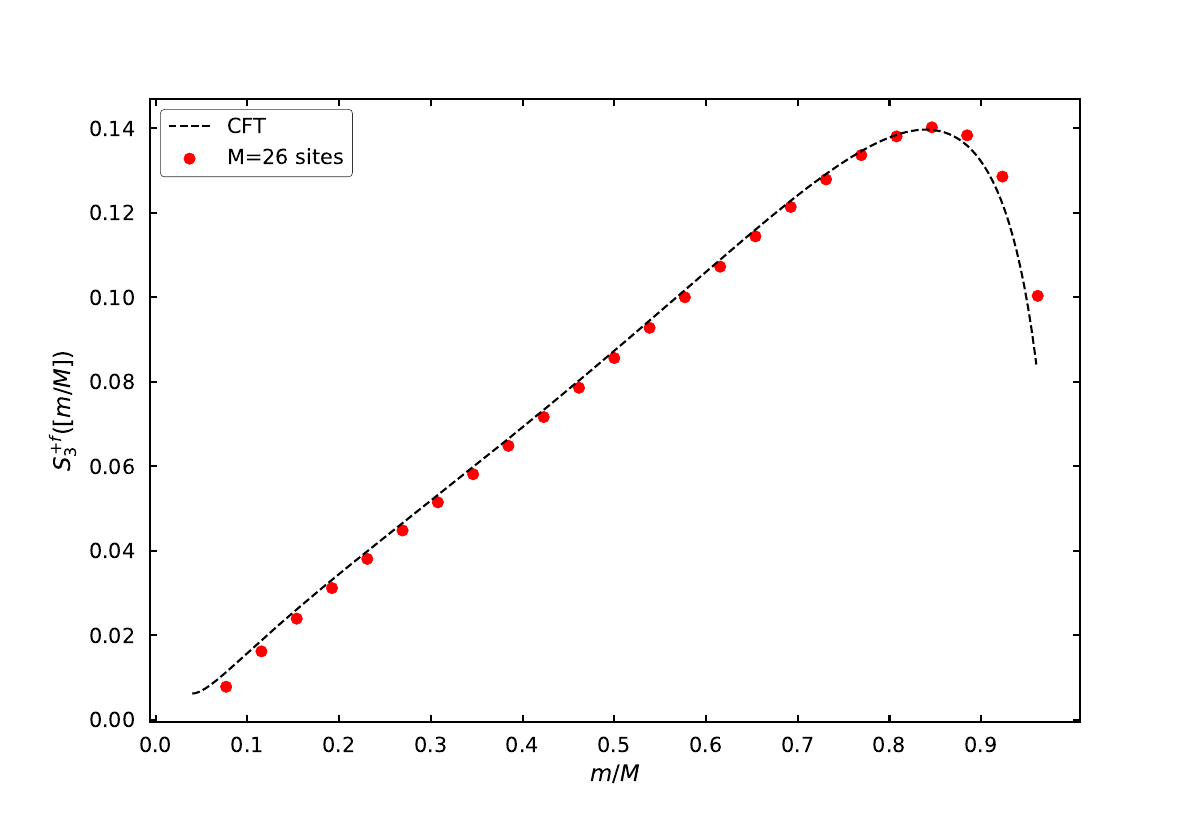}
    \caption{Fixed-free mixed BC}
    \label{fig:renyi3fixedfree}
  \end{subfigure}
  \caption{The third R\'enyi entropy of the interval $[0,m]$ in the critical Ising chain with two types of mixed BC, for a chain of size $M=26$. In both cases, the interval is adjacent to the $\alpha=+$ boundary.}
  \label{fig:renyi3graphs}
\end{figure}

We can perform the same kind of analysis for the third R\'enyi entropies $S_3^{(\alpha\beta)}$. In this case, we obtain the scaling functions
\begin{equation}
  \begin{aligned}
    & \mathcal{F}_3^{(+-)}(re^{i\theta},re^{-i\theta}) 
    = 2 \left(1- \tfrac{4}{9} \sin^2\theta \right) \,,
    && \qquad \mathcal{F}_{3,\varepsilon}^{(+-)}(re^{i\theta},re^{-i\theta})
    = 2 \left(1- \tfrac{16}{9} \sin^2\theta \right) \,, \\
    & \mathcal{F}_3^{(+f)}(re^{i\theta},re^{-i\theta})
    = 2 \cos \frac{\theta}{3} \,,
    && \qquad \mathcal{F}_{3,\varepsilon}^{(+f)}(re^{i\theta},re^{-i\theta})
    = 2 \cos \frac{2\theta}{3} \,.
  \end{aligned}
\end{equation}
In Figure \ref{fig:renyi3graphs}, we once again compare our CFT calculations (including both the leading and subleading term) with the lattice results for the third R\'enyi entropy $S_3^{(\alpha\beta)}$, to good agreement for mixed fixed BC (Fig.~\ref{fig:renyi3fixedmixed}) and mixed free fixed BC (Fig.~\ref{fig:renyi3fixedfree}). As for the $N=2$ results, including the CFT subleading contribution to $S_3^{(\alpha\beta)}$ is necessary to obtain a satisfying match with the lattice results. Further finite-size corrections in this case decay as $\mathcal{O}(M^{-2 h_{\varepsilon}/3-1})$. 

As announced in the beginning of the section, our results for the bare twist correlators (for all configurations of mixed BC) are compatible with the ones of \cite{taddia_entanglement_2013}.
The subleading contribution to the R\'enyi entropies from the excited twist correlator is largely responsible for the mismatch between the lattice and CFT data in the aforementioned article. Finite-size corrections of this magnitude can be suppressed only with much larger system sizes $M\sim 10^3$ sites \cite{Estienne:2021xzo} .

\subsection{The critical three-state Potts model}

In the three-state Potts quantum chain, we denote by $R,G,B$ the three possible values of each spin. We consider two types of boundary conditions: fixed ($R$, $G$ or $B$), and restricted ($GB$, $RB$ or $RG$). For example, the condition $GB$ means that the boundary spin can take the values $G$ or $B$, with equal energy.

The lattice expectation value $\aver{\sigma_{\text{lattice}}(\ell)}^{(GB,R)}_{\Sbb_L}$ follows an expansion of the form \eqref{eq_generic_subleading_corrections}. Here, the conformal dimensions of twist operators are $h_\sigma=1/20$ and $h_{\sigma_\varepsilon}=1/4$. We estimate numerically the non-universal amplitudes $C_{\id},C_{\varepsilon}$ by a simple analysis on the critical three-state Potts chain with \textit{periodic boundary conditions}.

The scaling function $\mathcal{F}_2^{(GB,R)}$ is obtained by setting $g=6/5$ in (\ref{eq:F2}--\ref{eq:G}). 
The leading finite-size correction $\mathcal{F}_{2,\varepsilon}^{(GB,R)}$ is determined by applying the Frobenius method to the fourth-order ODE \eqref{eq:ODE-fourth-order} -- see Section~\ref{sec:Potts}. Remarkably, in this case, only a single intermediate state $\Psi=\id$ contributes in the channel associated to $\mathrm{Arg}(z) \to \pi$.

\begin{figure}[h!]
    \centering
    \includegraphics[scale=0.7]{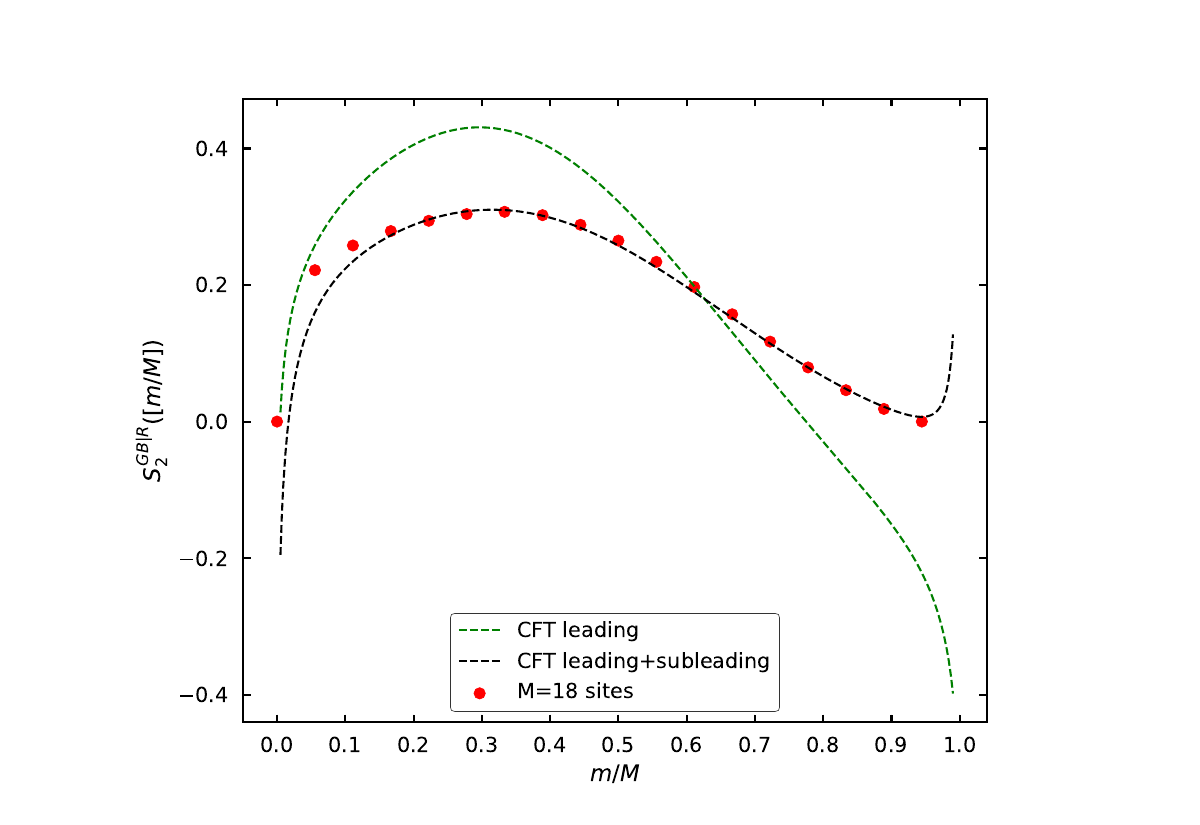}
    \caption{The second R\'enyi entropy of the interval $[0,m]$ in the critical three-state Potts chain of $M=18$ sites, with mixed $(GB,R)$ BC.}
    \label{fig:pottsy_potts}
\end{figure}

Putting everything together, we can compare the lattice computation for the second R\'enyi entropy $S_2^{(GB,R)}$ with our analytic results in Figure \ref{fig:pottsy_potts}. We observe that the inclusion of the subleading term gives an analytic curve that is closer to the lattice data, and in fact matches it quite well for most of the data points. We have also checked that the discrepancy between the CFT and the lattice data is supressed for a system of $M=18$ sites, compared to smaller system sizes.

Firstly, due to the operator content of the D-series $\mathcal{M}_{6,5}$ CFT, we expect the higher order corrections to $S_2^{(GB,R)}$ to have a slower power law decay than in the case of the Ising CFT. We conjecture that the next-to-subleading contribution will decay as $\sim M^{-2h_{\sigma_{\varepsilon}}+1} $. These corrections arise from the operators $\sigma_X$ and $\sigma_\varepsilon^{(1)}$. While $\mathcal{F}_{2,X}^{(R,GB)}$ can be calculated by a repeat of the method employed for the subleading term, the correlator involving a descendant twist operator would require the derivation of a new differential equation, which is beyond the scope of this work.

Furthermore, the quantum chain sizes we can reach are diminished in the case of the three-state Potts model, since the size of the space of states grows as $\sim 3^M$. This memory constraint prevents us from reaching sizes at which higher order corrections are suppressed, using our computational methods. This limitation can be, perhaps, bypassed through the usage of more sophisticated numerical tools, such as DMRG or tensor network methods, to access system sizes $M$ for which the unknown higher-order correction terms are further suppressed.

\begin{figure}
    \centering
    \includegraphics[scale=0.6]{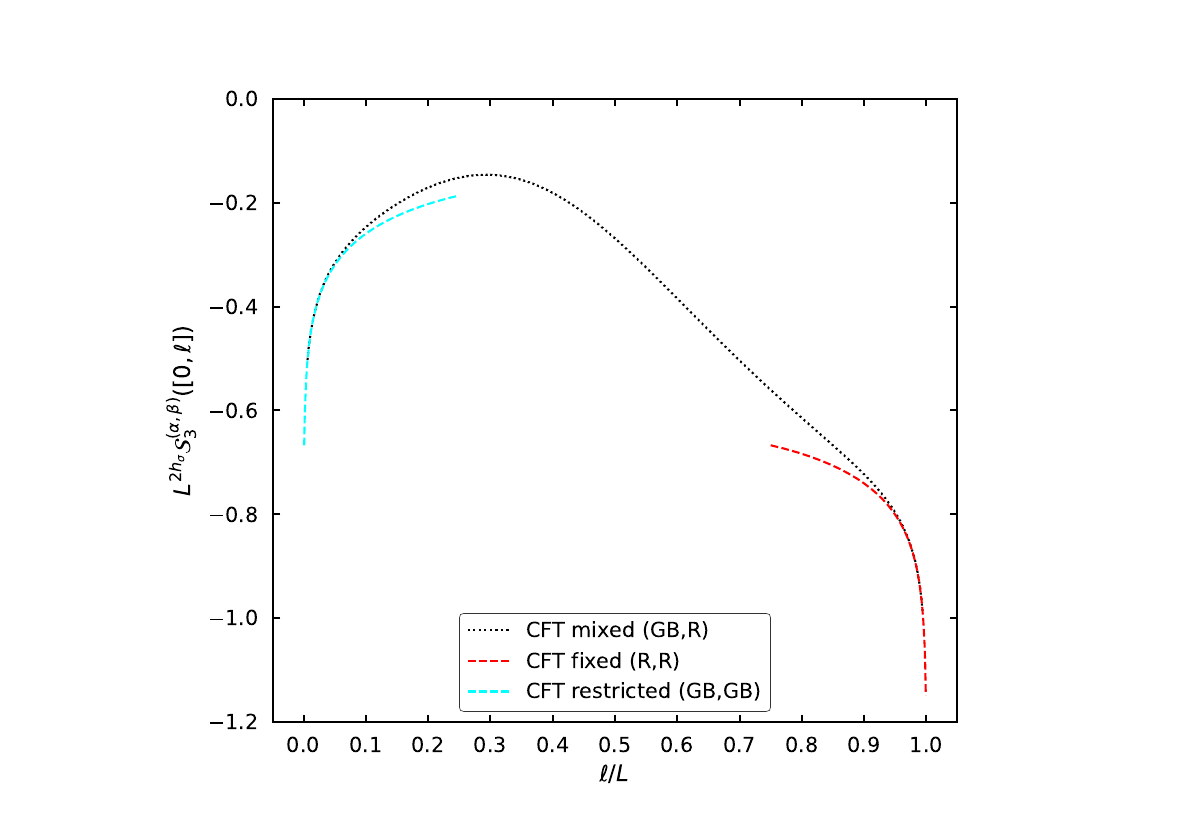}
    \caption{The third R\'enyi entropy of the interval $[0,\ell]$ in the critical three-state Potts chain of size $L$ with boundary conditions $(GB,R)$, $(R,R)$ and $(GB,GB)$, in the scaling limit. The mixed BC curve can be seen to interpolate between the identical BC results.}
    \label{fig:CFTinterpolate_N3}
\end{figure}

We obtain the scaling limit of $S_3^{(GB,R)}$, by solving the third-order ODE \eqref{eq_ODE_3_leading} for $\mathcal{F}_3^{(GB,R)}$. 
Like for $\mathcal{F}_{2,\varepsilon}^{(GB,R)}$, only one intermediate state $\Psi=\id$ contributes, and we compute the corresponding power series by solving the ODE numerically with the Frobenius method. However, in this case, we do not have access to the leading finite-size correction $\mathcal{F}_{3,\varepsilon}^{(GB,R)}$, with relative amplitude $M^{-4/15}$. Thus, our CFT results are limited to the dominant term of $S_3^{(GB,R)}$, and the comparison with numerical results is expected to be poor. The CFT predictions are shown in Fig.~\ref{fig:CFTinterpolate_N3}.

\clearpage

\section{The cyclic orbifold in the presence of a boundary}
\label{sec:orbifold}

Here we review briefly the cyclic orbifold construction \cite{klemm_orbifolds_1990,borisov_systematic_1998,2023JPhA...56T5403E}, and we examine specifically the features of the corresponding orbifold boundary CFT, in the upper-half plane geometry. We follow the notations of \cite{2023JPhA...56T5403E}. 

\subsection{The cyclic orbifold construction}

Consider the original CFT model $\M$, which we shall call the mother CFT. The cyclic orbifold CFT $\M_N$ is obtained by taking $N$ identical decoupled replicas of $\M$, and allowing for defect lines along which the copies are connected cyclically.

The underlying algebra for $\M_N$ consists of currents $T^{(r)}(z)$, with $r \in \Zbb_N$ playing the role of the discrete Fourier index with respect to the cyclic permutation of replicas. The total energy-momentum tensor of $\M_N$ is given by $T^{(0)}$, and the other currents $T^{(r)}$ can be viewed as additional conserved currents in an extended conformal algebra. The associated Laurent modes are denoted $L_m^{(r)}$, 
where the index $m$ belongs to $\Zbb/N$. They generate the orbifold Virasoro algebra $\mathrm{OVir}_N$
\begin{equation}
  \left[L_m^{(r)},L_n^{(s)} \right] = (m-n)L_{m+n}^{(r+s)} + \frac{Nc}{12}m(m^2-1) \, \delta_{m+n,0} \, \delta_{r+s,0} \,,
\end{equation}
where $c$ is the central charge of the mother CFT.
The invariant subalgebra (of the universal enveloping algebra) with respect to the $\Zbb_N$ action is called the \emph{neutral subalgebra} $\mathcal{A}_N$. It is generated by monomials with a vanishing total Fourier index : 
\begin{equation}
  \mathcal{A}_N = \bigaver{L_{m_1}^{(r_1)} \dots L_{m_p}^{(r_p)} \,, \quad r_1+\dots+r_p  \equiv 0 \mod N} \,.
\end{equation}
The orbifold Virasoro algebra and the neutral subalgebra have antiholomorphic counterparts $\overline{\rm OVir}_N$ and $\overline{\cal A}_N$, respectively.

In the holomorphic part of the CFT $\M_N$ for prime $N$, the operator content is classified in terms of the primary operators under the neutral subalgebra $\mathcal{A}_N$. It is shown in \cite{2023JPhA...56T5403E} that in the bulk CFT $\M_N$, the corresponding scalar operators form the full set of $\mathrm{OVir}_N \otimes \overline{\rm OVir}_N$ primary operators which are invariant under cyclic permutations of the replicas.
There are three kinds of $\mathcal{A}_N$ primary operators.
The \textit{untwisted non-diagonal operators} are symmetrised products of non-identical primary operators $\phi_1,\dots,\phi_N$ on the decoupled copies
\begin{equation} \label{eq:invariant.primary.1}
  [\phi_1,\dots,\phi_N] = \# \sum_{a=0}^{N-1} \phi_{a+1}\otimes \dots \otimes \phi_{a+N} \,,
\end{equation}
The \textit{untwisted diagonal operators} associated to a primary operator $\phi$ of $\M$ are of the form
\begin{equation} \label{eq:invariant.primary.2}
  \Phi^{(0)}=\Phi = \phi\otimes \dots \otimes \phi \,,
  \qquad \Phi^{(r)} = \begin{cases}
    \# L_{-1}^{(r)} \cdot (\phi\otimes \dots \otimes \phi) & \text{if } \phi \neq \id \\[5pt]
    \# L_{-2}^{(r)} \cdot (\id\otimes \dots \otimes \id) & \text{if } \phi = \id
  \end{cases}
\end{equation}
where $r=1,\dots,N-1$. 
The \textit{twist operators} are the non-local operators inserting the endpoints of defect lines connecting the replicas cyclically. The bare twist operator $\sigma$ corresponds to a cyclic permutation $a \mapsto a+1$ across the defect, with boundary conditions around the branch point given by the vacuum state. The composite twist operator associated to the primary operator $\phi$ of $\M$ is the form \cite{Castro_Alvaredo_2011,Bianch_2014}
\begin{equation} \label{eq:invariant.primary.3}
  \sigma_\phi(z) = \# \lim_{\epsilon \to 0} \left[\epsilon^{(N-1)h/N} \, [\phi,\id,\dots,\id](z+\epsilon) \, \sigma(z) \right] \,,
\end{equation}
where $h$ is the conformal dimension of $\phi$. 
For $\phi=\id$, one simply sets $\sigma_\id=\sigma$.
The twisted sector of the operator content also includes the conjugate twist operators $\sigma_\phi^\dag$ associated to $a \mapsto a-1$, and more generally the operators $\sigma_\phi^{[k]}$ associated to $a \mapsto a+k$ for $k \in \Zbb_N^\times$.
The orbifold Virasoro modes $L_m^{(r)}$ acting on an untwisted operator must have an integer index $m$, otherwise they vanish. In contrast, due to the non-trivial monodromy of $T^{(r)}$ around a branch point, the modes $L_m^{(r)}$ mode acting on $\sigma$ (resp. $\sigma^\dag$) must have an index $m \in \Zbb-r/N$ (resp. $m \in \Zbb+r/N$), and similarly for $\sigma_\phi$ and $\sigma^\dag_\phi$.
The following operators are also primary under $\mathcal{A}_N$:
\begin{equation} \label{eq:invariant.primary.4}
    \sigma_\phi^{(r)} = \# \, L_{-r/N}^{(r)} \cdot \sigma_\phi \,,
    \qquad \sigma_\phi^{\dag(r)} = \# \, L_{-(N-r)/N}^{(r)} \cdot \sigma_\phi^\dag \,,
\end{equation}
for $r=1,\dots,N$.

The conformal dimensions of the above operators read
\begin{equation}
  \begin{aligned}
    & h_{[\phi_1,\dots,\phi_N]} = h_{\phi_1}+ \dots+h_{\phi_N} \,,
    \qquad h_\Phi = Nh_\phi \,,
    \qquad h_{\Phi^{(r)}} = \begin{cases}
      Nh_\phi+1 &\text{if } \phi \neq \id \\
      2 & \text{if } \phi = \id
    \end{cases} \\
    & h_{\sigma_\phi} = h_{\sigma_\phi^\dag} = \frac{c}{24}\left(N-\frac{1}{N}\right) +\frac{h_\phi}{N} \,,
    \qquad h_{\sigma_\phi^{(r)}} = h_{\sigma_\phi^{\dag(N-r)}} = h_{\sigma_\phi} + \frac{r}{N} \,,
  \end{aligned}
\end{equation}
where $h_\phi$ denotes the conformal dimension of $\phi$.
The normalisation factors $\#$ in (\ref{eq:invariant.primary.1}--\ref{eq:invariant.primary.4}) are chosen so that the two-point function of any $\mathcal{A}_N$ primary operator $\mathcal{O}$ on the full plane satisfies $\aver{\mathcal{O}(0)\mathcal{O}^\dag(1)}_{\Cbb}=1$.

\subsection{The cyclic orbifold on the upper half-plane}

To construct the cyclic orbifold BCFT, we will work on the upper half-plane $\Hbb$, with the boundary along the real axis. We parameterise $\Hbb$ by $z=x+iy$ with $x\in \Rbb$ and $y>0$, and we impose the gluing condition on the boundary for the stress-energy tensor components:
\begin{equation}
  T^{(0)}(x)=\Tb^{(0)}(x) \quad \text{for} \quad x \in \mathbb{R} \,,
\end{equation}
which ensures that the boundary is conformal, \textit{i.e.} it preserves the total stress-energy tensor \cite{cardy_conformal_1985}. The $\Zbb_N$ orbifold, however, has an extended symmetry, and we must choose if and how the components of the additional currents $T^{(r\neq 0)}$ are glued at the boundary. Our usage of the replica trick provides a clear indication for these choices: since we are considering $N$ copies of the \textit{same} mother BCFT, we must impose  the gluing condition $T_a(x)=\Tb_a(x)$  on each of them. By taking the Fourier transform of this relation, we find that in the orbifold CFT we are effectively imposing:
\begin{equation}\label{eq:gluingcondition}
  \forall r \in \Zbb_N \,, \quad
  \forall x \in \Rbb \,, \qquad
  T^{(r)}(x)=\Tb^{(r)}(x) \,.
\end{equation}
This implies that the boundary  preserves a full copy of the $\OVir_N$ algebra. By the same reasoning on CFT replicas, the physically relevant orbifold boundary states correspond to having the same conformal BC on the $N$ copies of the mother CFT.

On the upper half-plane, we will set the conformal BC $\alpha$ on the positive real axis $x>0$ and the conformal BC $\beta$ on $x<0$. To implement such mixed conformal BC in a BCFT, we shall work with the formalism of \textit{boundary condition changing operators} (BCCOs) \cite{cardy_boundary_1989}. These operators, restricted to live on the boundary, are placed at the points of suture of regions of different BC. The full operator algebra of a BCFT is then formed by considering the OPEs between both BCCOs and bulk operators.
For a given pair of conformal BCs $(\alpha,\beta)$, there can be several primary BCCOs implementing the change $\alpha \to \beta$: we denote such an operator $\psi_h^{(\alpha\beta)}$, where $h$ specifies its conformal dimension. The most relevant BCCO implementing $\alpha \to \beta$ is simply referred to as $\psi^{(\alpha\beta)}$.

In the $\Zbb_N$ orbifold CFT, we only consider \textit{untwisted} BCCOs. We define the diagonal BCCOs as 
\begin{equation}\label{eq:diagBCCO}
  \Psi_h^{(\alpha\beta)}=\underbrace{\psi_h^{(\alpha\beta)}\otimes\dots \otimes \psi_h^{(\alpha\beta)}}_{N \text{ times}} \,,
\end{equation}
and likewise for $ \Psi^{(\alpha\beta)}$, while the non-diagonal BCCOs are 
\begin{equation} \label{eq:Psi-nondiag}
  [\psi_{h_1},\dots,\psi_{h_N}]^{(\alpha\beta)} := \frac{1}{\sqrt N} \sum_{a=0}^{N-1} \psi_{h_{a+1}}^{(\alpha\beta)}\otimes \dots \otimes \psi_{h_{a+N}}^{(\alpha\beta)} \,,
\end{equation}
where at least two dimensions $h_i,h_j$ are distinct.
Orbifold correlators with mixed BC are obtained by inserting the most relevant diagonal BCCOs:
\begin{equation}\label{eq:correlationfunction}
  \aver{\mathcal{O}_1(z_1,\zb_1)\dots\mathcal{O}_n(z_n,\zb_n)}^{\alpha\beta}_{\Hbb}
  =\frac{\aver{\Psi^{(\beta\alpha)}(\infty)\, \mathcal{O}_1(z_1,\zb_1) \dots\mathcal{O}_n(z_n,\zb_n)  \Psi^{(\alpha \beta)}(0)}_{\Hbb}}{\aver{\Psi^{(\beta\alpha)}(\infty) \Psi^{(\alpha \beta)}(0)}_{\Hbb}} \,.
\end{equation}
By Cardy's doubling trick \cite{cardy_conformal_1985,cardy_bulk_1991}, such $(n+2)$-point correlators are given by linear combinations of the $(2n+2)$-point conformal blocks on the Riemann sphere $\Cbb$ with external operators
\begin{equation}
  \Psi(\infty),  \mathcal{O}_1(z_1), \overline{\mathcal{O}}_1(\zb_1),
  \dots, \mathcal{O}_n(z_n), \overline{\cal O}_n(\zb_n), \Psi(0) \,,
\end{equation}
where $\overline{\mathcal{O}_i}(\zb)$ is the antiholomorphic counterpart of $\mathcal{O}_i(z)$. In more precise terms, $\overline{\mathcal{O}_i}$ is the operator conjugate to $\mathcal{O}_i$ with respect to the symmetry algebra preserved by the boundary \cite{recknagel_relevant_2000}. For $\Zbb_N$ twist operators, conjugation acts as $\overline{\sigma}_\phi=\sigma^\dag_\phi$ \cite{dixon_conformal_1987}. For instance, the one-twist function
\begin{equation}
  \aver{\sigma(z,\zb)}_{\Hbb}^{(\alpha\beta)}
  = \frac{\aver{\Psi^{(\beta\alpha)}(\infty)\sigma(z,\zb)\Psi^{(\alpha\beta)}(0)}_{\Hbb}}{\aver{\Psi^{(\beta\alpha)}(\infty)\Psi^{(\alpha\beta)}(0)}_{\Hbb}}
\end{equation}
is given by a linear combination of conformal blocks of the form
\begin{equation}
 \cblock{\Psi^{(\beta\alpha)}(\infty)}{\ \Psi^{(\alpha\beta)}(0)}{\sigma^\dag(\zb)}{\sigma(z)}{\Psi_k} \quad.
\end{equation}
The intermediate state $\Psi_k$ is a diagonal or non-diagonal untwisted boundary operator. It must be compatible both with the boundary condition $\alpha$, and the boundary condition $\beta$. For instance, if $\Psi_k$ is diagonal, it must be of the form $\Psi_k=\psi_h\otimes\dots\otimes\psi_h$, where $h$ is such that both BCCOs $\psi_h^{(\alpha\alpha)}$ and $\psi_h^{(\beta\beta)}$ are allowed in the mother CFT.

In the mother BCFT, bulk-boundary and boundary-boundary OPEs read
\begin{align}
    & \phi(x+iy,x-iy) \underset{y \to 0}{\sim}
    \sum_h A_{\phi,\psi_h}^{(\alpha)} \, (2y)^{h-2h_\phi} \, \psi_h(x) 
    + \dots  \label{eq:bulk_boundary_OPE} \\
  & \psi_{h_1}^{(\alpha\beta)}(x_1) \psi_{h_2}^{(\beta\gamma)}(x_2)
  \underset{x_1 \to x_2}{\sim}
  \sum_{h_3} B^{(\alpha\beta\gamma)}_{h_1,h_2,h_3}
  (x_1-x_2)^{-h_1-h_2+h_3} \psi_{h_3}^{(\alpha\gamma)}(x_2)
  + \dots  \label{eq:boundary_boundary_OPE}
\end{align}
In \eqref{eq:bulk_boundary_OPE}, $\phi$ denotes a scalar bulk operator and $\alpha$ is the BC on the real axis. In \eqref{eq:boundary_boundary_OPE},
$\psi_{h_1}^{(\alpha\beta)}(x_1)$ and $\psi_{h_2}^{(\beta\gamma)}(x_2)$ are two BCCOs, with $x_1 > x_2$. The $\dots$ denote the contributions from the Virasoro descendants.
We write these OPEs in shorthand notation as
\begin{align}
    \phi \big|_\alpha \to \sum_h \psi_h \,, \qquad
    \psi_{h_1}^{(\alpha\beta)} \cdot \psi_{h_2}^{(\beta\gamma)} \to 
    \sum_{h_3} \psi_{h_3}^{(\alpha\gamma)} \,.
\end{align}
The OPE structure constants can be expressed as correlation functions:
\begin{align}
    & A_{\phi,\psi_h}^{(\alpha)} = 
    \aver{\phi(i/2,-i/2)\psi_h(0)}_{\Hbb}^{(\alpha)}
    = \aver{\phi(0)\psi_h(1)}_{\Dbb}^{(\alpha)} \,, \\
    & B^{(\alpha\beta\gamma)}_{h_1,h_2,h_3}
    = \aver{\psi_3^{(\gamma\alpha)}(\infty) \psi_1^{(\alpha\beta)}(1) \psi_2^{(\beta\gamma)}(0)}_{\Hbb} \,.
\end{align}

Correspondingly, in the orbifold BCFT, we have the OPEs
\begin{align}
  & \sigma_\phi(x+iy,x-iy) \underset{y \to 0}{\sim}
  \sum_{\Psi^{(\alpha\alpha)}} \acon{\alpha}{\sigma_\phi}{\Psi} \, (2y)^{h_\Psi-2 h_{\sigma_\phi}} \, \Psi^{(\alpha\alpha)}(x) +\dots  \label{eq:twist_boundary_OPE} \\
  & \Psi_1^{(\alpha\beta)}(x_1) \Psi_2^{(\beta\gamma)}(x_2)
  \underset{x_1 \to x_2}{\sim}
  \sum_{\Psi_3^{(\alpha\gamma)}} \bcon{\alpha\beta\gamma}{\Psi_1}{\Psi_2}{\Psi_3}
  (x_1-x_2)^{-h_{\Psi_1}-h_{\Psi_2}+h_{\Psi_3}} \Psi_3^{(\alpha\gamma)}(x_2)
  + \dots \label{eq:orb_boundary_OPE}
\end{align}
where the sums in the RHS run over BCCOs of the form $\Psi_h$ or $[\psi_{h_1},\dots,\psi_{h_N}]$, and the $\dots$ denote $\mathcal{A}_N$ descendants.
We write, in short-hand notation
\begin{align}
     \sigma_\phi \big|_\alpha \to \sum_\Psi \Psi \,, \qquad
     \Psi_1^{(\alpha\beta)} \cdot \Psi_2^{(\beta\gamma)} \to 
    \sum_{\Psi_3} \Psi_3^{(\alpha\gamma)} \,.
\end{align}
For  $\Psi^{(\alpha\alpha)}=\id$, the structure constants in \eqref{eq:twist_boundary_OPE} read
\begin{equation} \label{eq:A_sigma}
    \acon{\alpha}{\sigma}{\id} = \aver{\sigma(0)}_{\Dbb}^{(\alpha)}
    = g_\alpha^{1-N} \,,
    \qquad \acon{\alpha}{\sigma_\phi}{\id} = g_\alpha^{1-N} \, A_{\phi,\id}^{(\alpha)} \,,
\end{equation}
where $g_\alpha$ is the partition function on the unit disc with BC $\alpha$.
This quantity is also known as the \textit{universal boundary entropy}, since it can be written as $g_\alpha=\aver{0|\alpha}$, where  $\ket\alpha$ is the boundary state associated to the conformal BC $\alpha$, and $\ket 0$ is the vacuum state. More generally, the bulk-boundary structure constant for a twist operator reads
\begin{align}
    \acon{\alpha}{\sigma_{\phi}}{\Psi_h} = 
    g_{\alpha}^{1-N} \, N^{-Nh} \aver{\phi(0) \psi_h^{(\alpha\alpha)}(1) \psi_h^{(\alpha\alpha)}(\omega) \cdots \psi_h^{(\alpha\alpha)}(\omega^{N-1})}^{(\alpha)}_{\mathbb{D}} \,,
\end{align}
in the case when $\Psi$ is a diagonal BCCO, $\Psi=\Psi_h$ (this can be easily extended to non-diagonal BCCOs).

The structure constants in \eqref{eq:orb_boundary_OPE} read
\begin{align}\label{eq:bconDDD}
  & \bcon{\alpha\beta\alpha}{\Psi_{h_i}}{\Psi_{h_j}}{\Psi_{h_k}}
  = \left( B^{(\alpha\beta\alpha)}_{h_i,h_j,h_k} \right)^N \,, \\
  & \bcon{\alpha\beta\alpha}{\Psi_{h_i}}{\Psi_{h_j}}{[\psi_{h_{k_1}},\dots,\psi_{h_{k_N}}]}
  =\sqrt{N}\prod_{a=1}^N B^{(\alpha\beta\alpha)}_{h_i,h_j,h_{k_a}} \,.
\end{align}

\section{Differential equations for the scaling functions}
\label{sec:ODEs}

In this section, we outline the derivation of the ODEs announced in Sec.~\ref{sec:setup_summary}, following the approach of \cite{dupic_entanglement_2018}. The derivation relies on the null-vector equations satisfied by the orbifold primary operators. 

\subsection{The function $\aver{\Psi_{12}\cdot \sigma \cdot \Psi_{12}}$ in a generic $\Zbb_2$ orbifold}
\label{sec:baretwistN2}

The entropy scaling function $\mathcal{F}_2^{(\alpha\beta)}$, given by equation \eqref{eq:F_N}, can be expressed in terms of the $\mathbb{Z}_2$ orbifold conformal blocks
\begin{equation}
  J_k(\eta) = \cblock{\Psi_{12}(\infty)}{\Psi_{12}(0)}{\sigma(1)}{\sigma(\eta)}{\Psi_k}
 \,,
\end{equation}
as in \eqref{eq_generic_block_expansion}. In~\cite{dupic_entanglement_2018}, an ODE was obtained for a related object, namely the bulk correlation function
$$\aver{\Psi_{12}(\infty)\sigma^\dag(1)\sigma(\eta,\bar\eta)\Psi_{12}(0)}_{\Cbb} \,.$$
Let us recall this derivation, as it will serve as a prototype to derive more involved ODEs in the present context. In the mother CFT, we have the null-vector conditions
\begin{equation}
  (gL_{-2}-L_{-1}^2) \cdot \psi_{12} \equiv 0 \,,
  \qquad L_{-1}\cdot \id \equiv 0 \,,
\end{equation}
where $g$ is related to the central charge through \eqref{eq:Kac}.
The above relations translate respectively into null-vector conditions for the replicated operator
\begin{equation} \label{eq:N2untwistednulllevel2}
    \begin{aligned}
    & \left[2g\, L_{-2}^{(0)}-\left(L_{-1}^{(0)} \right)^{2} \left(L_{-1}^{(1)}\right)^{2}\right]
    \cdot \Psi_{12} \equiv 0,   \\
    & \left[gL_{-2}^{(1)}- L_{-1}^{(0)} L_{-1}^{(1)} \right] \cdot \Psi_{12} \equiv 0 \,,
    \end{aligned}
\end{equation}
and for the twist operator
\begin{equation} \label{eq:N2twistnullvector}
   L_{-1/2}^{(1)} \cdot \sigma \equiv 0 \,,
\end{equation}
the latter equation deriving from the induction procedure \cite{borisov_systematic_1998}.
Additionally, since $\Psi_{12}$ is invariant by cyclic permutations of copies, we have
\begin{equation} \label{eq:L_0^1.Psi=0}
  L_0^{(1)}\cdot \Psi_{12} = 0 \,.
\end{equation}
The first step in the derivation of the ODE starts with the following contour integral along the close contour $C_\infty$ enclosing the singular points $0$, $1$ and $\eta$: 
\begin{align} \label{eq:closed-contour}
  & \oint_{C_\infty} dw \, (w-1)^{1/2} (w-\eta)^{1/2} \, \wt{J}_k(\eta,w) 
\end{align}
where $ \wt{J}_k$ is the modified conformal block
\begin{align}
  \wt{J}_k(\eta,w)=\cblock{\phantom{T^{(1)}(w)L_{-1}^{(1)}\cdot}\Psi_{12}(\infty)}{T^{(1)}(w)L_{-1}^{(1)}\cdot\Psi_{12}(0)}{\sigma(1)}{\sigma(\eta)}{\Psi_k} \,.
\end{align}
This contour integral must vanish, given that the relation \eqref{eq:L_0^1.Psi=0} yields  
\begin{align}
\wt{J}_k(\eta,w) = O(w^{-3}) \qquad \textrm{as} \qquad w \to \infty \,. 
\end{align}
Moreover, the null-vector relation \eqref{eq:N2twistnullvector} ensures that the integrand has no pole at $w = 1$ and $w =\eta$. The only poles occur at $w=0$, as can be read from the OPE
\begin{align}
  & T^{(1)}(w) L_{-1}^{(-1)}\ket{\Psi_{12}}  = \left( w^{-3}  L_1^{(1)}L_{-1}^{(-1)} + w^{-2}  L_0^{(1)}L_{-1}^{(-1)}+ w^{-1}  L_{-1}^{(1)}L_{-1}^{(-1)} + O(1) \right)  \ket{\Psi_{12}} \,.
\end{align}
Since $\Psi_{12}$ and $\sigma$ are primary, we have
\begin{align}
  & T^{(1)}(w) L_{-1}^{(-1)}\cdot\ket{\Psi_{12}}
  = \sum_{n \leq 1} w^{-n-2} L_n^{(1)}L_{-1}^{(-1)}\cdot\ket{\Psi_{12}} \,, \\
  & \bra{\Psi_{12}}\cdot T^{(1)}(w) = \sum_{n \geq 0} w^{-n-2} \bra{\Psi_{12}}\cdot L_n^{(1)} \,,
\end{align}
and
\begin{align}
  & T^{(1)}(w) \sigma(1) = \sum_{n\leq 0} (w-1)^{-n-3/2}L_{n-1/2}^{(1)}\cdot \sigma(1) \,, \\
  & T^{(1)}(w) \sigma(\eta) = \sum_{n\leq 0} (w-\eta)^{-n-3/2}L_{n-1/2}^{(1)}\cdot \sigma(\eta) \,.
\end{align}
As a consequence, the integration contour for \eqref{eq:closed-contour} can be replaced by a contour $C_0$ enclosing only the singular point $0$. The vanishing of this integral yields the orbifold Ward identity
\begin{equation}
  \cblock{\phantom{\mu\cdot}\Psi_{12}(\infty)}{\mu\cdot\Psi_{12}(0)}{\sigma(1)}{\sigma(\eta)}{\Psi_k} =0 \,,
\end{equation}
where
\begin{equation}
  \mu = \left(\alpha(0) L_{-1}^{(1)} + \alpha'(0) L_{0}^{(1)} + \frac{\alpha''(0)}{2} L_1^{(1)}\right)\cdot L_{-1}^{(1)}  \,,
  \qquad \alpha(w)=(1-w)^{1/2}(\eta-w)^{1/2} \,.
\end{equation}
Using the orbifold commutation relations, we can write
\begin{equation}
  \mu\cdot \ket{\Psi_{12}}
  = \left[\alpha(0) \left(L_{-1}^{(1)}\right)^2 + \alpha'(0) L_{-1}^{(0)} + 2 h_{12}\alpha''(0) \right]
  \cdot \ket{\Psi_{12}} \,.
\end{equation}
Inserting the first null-vector condition \eqref{eq:N2untwistednulllevel2}, we obtain 
\begin{equation}
  \mu\cdot \ket{\Psi_{12}}
  = \left[2g \alpha(0) L_{-2}^{(0)} - \alpha(0) \left(L_{-1}^{(0)}\right)^2 + \alpha'(0) L_{-1}^{(0)} + 2 h_{12}\alpha''(0) \right]
  \cdot \ket{\Psi_{12}} \,.
\end{equation}
Finally, since the modes $L_m^{(0)}$ are simply the Virasoro modes of the orbifold CFT, we can use the standard procedure to express their action on a correlation function (or conformal block) as a differential operator.
We find that the conformal blocks $J_k$ satisfy the ODE
\begin{equation} \label{eq:N2baretwistdiffeq}
  \begin{aligned}
    & 64 g^2 \eta^2 (\eta-1)^2 \, J''(\eta)
    + 16g \eta (\eta-1) \, \left[ (-14g^2+23g-6)\eta +2g (1-4g) \right] \, J'(\eta) \\
    & +(3g-2) \, \left[ +3(5g-6)(1-2g)^2\eta^2 + 12g(1-2g)\eta + 16g^2(g-1)
    \right] \, J(\eta) = 0 \,.
  \end{aligned}
\end{equation}
The local exponents for $J(\eta)$ are given by
\begin{center}
  \begin{tabular}{ccc}
    0 & 1 &  $\infty$ \\
    \hline $-2 h_{12} $ & $-2 h_{\sigma}$  &  $2h_{\sigma}-2 h_{12}$ \\
    $-2 h_{12}+h_{13}/2$ &  $-2 h_{\sigma}+ 2 h_{13}$  &$2h_{\sigma}-2 h_{12}+h_{13}/2$ 
  \end{tabular}
\end{center}
This corresponds to the intermediate states $\{ \id, \Phi_{13} \}$ in the channel $\eta \to 1$, and $\{\sigma,\sigma_{13}\}$ in the channels $\eta \to 0$ and $\eta \to \infty$.
Note that, when the mother CFT is a minimal model $\mathcal{M}_{p,p'}$, one can check for various values of $(p,p')$ that these are exactly the intermediate states allowed by the orbifold fusion rules found in \cite{2023JPhA...56T5403E}, and that they all have multiplicity one.

Let us now return to the entropy scaling function $\mathcal{F}_2^{(\alpha\beta)}$ \eqref{eq_generic_block_expansion}. The typical power-series solutions of the ODE around $\eta=0$ or $\eta=\infty$ diverge close to the physical domain for $\eta$, namely the unit circle, and the analogous solutions around $\eta=1$ only converge on a portion of the unit circle. A more convenient choice is the variable $\zeta$:
 \begin{equation} \label{eq:zeta(eta)}
  \zeta(\eta) = \frac{(1+\sqrt\eta)^2}{4\sqrt\eta} \,.
\end{equation}
This change of variable offers several advantages. First the ODE remains of Fuchsian type (see Appendices~\ref{app:hypergeom} and \ref{app:change_of_var}). Additionally,  the physical domain corresponds to $0 < \zeta < 1$, since   
\begin{equation}
    \zeta=\cos^2 \frac{\pi \ell}{2L} \,,
    \qquad 0<\ell <L \,.
\end{equation}
This means that $\zeta$ is not only real but also remains in the domain where the Frobenius method yields convergent power series expansions (both around $\zeta =0$ and $\zeta =1$). Finally, imposing the appropriate boundary conditions as $\ell \to 0$ and $\ell \to L$ becomes simpler, as these correspond to $\zeta \to 1$ and $\zeta \to 0$ respectively, whereas both limits collapse to $\eta \to 1$ in the variable $\eta$. \\

After some elementary algebra (see Appendix~\ref{app:hypergeom}), the scaling function is found to satisfy the following ODE:
\begin{equation} \label{eq:ODE-order2}
  \zeta(\zeta-1) F''(\zeta) + 2(1-g)(2\zeta-1) F'(\zeta) + (2-3g)(1-g) F(\zeta) = 0 \,.
\end{equation}
where we have introduced the change of variable $\mathcal{F}_2^{(\alpha\beta)}(z,\zb)=F[\zeta(z/\zb)]$.
We recognise the hypergeometric equation \eqref{eq:hypergeom.ode.zeta} with parameters
\begin{equation}
  A=2-3g \,, \qquad B=\frac{3}{2}-2g \,, \qquad C=\frac{3}{2}-g=A-B+1 \,.
\end{equation}
Provided that $h_{13}  >0$ (that is $g>1/2$), any holomorphic solution on $0<\zeta<1$ can be written as
\begin{equation}
  F(\zeta) = \lambda \, G(\zeta) + \mu \, G(1-\zeta) \,,
\end{equation}
where $\lambda$ and $\mu$ are constants, and 
\begin{equation}
  G(\zeta)= \zeta^{2g-1} \times
  \frac{\FGauss\left(\left. g, 1-g;  2g \right| \zeta \right)}
  {\FGauss(g, 1-g; 2g| 1)} \,.
\end{equation}
Indeed, $G$ is given by the function $\wt{I}_2$ of Appendix~\ref{app:hypergeom}, normalised so that $G(0)=0$ and $G(1)=1$.
As argued in Section~\ref{sec:scaling-functions}, the limiting values at $\ell=0$ and $\ell=L$ are given by the boundary entropies $g_\alpha^{-1}$ and $g_\beta^{-1}$ respectively. This fixes the values of $\lambda$ and $\mu$, and we obtain
\begin{equation}
  \mathcal{F}_2^{(\alpha\beta)}(z,\zb)
  = g_\alpha^{-1} G[\zeta(z/\zb)] + g_\beta^{-1} G[1-\zeta(z/\zb)] \,.
\end{equation}

\subsection{The function $\aver{\Psi_{12} \cdot \sigma_{\phi} \cdot \Psi_{12}}$ in a generic $\Zbb_2$ orbifold}
\label{sec:compositetwistN2}

We now turn to the universal scaling functions $\mathcal{F}_{2,\phi}^{(\alpha\beta)}(z,\zb)$ encoding the subleading corrections in \eqref{eq_generic_subleading_corrections}. Such a function corresponds to the conformal blocks
\begin{equation}
  J_k(\eta) = \cblock{\Psi_{12}(\infty)}{\ \Psi_{12}(0)}{\sigma_{\phi}(1)}{\sigma_{\phi}(\eta)}{\Psi_k}
  \,.
\end{equation}
Here we consider the case of a generic composite twist operator $\sigma_{\phi}$, with conformal dimension
\begin{equation}
  \hh = h_\sigma + \frac{h}{2} \,,
\end{equation}
where $h$ is the conformal dimension of the operator $\phi$ in the mother CFT. This means that we do not assume any null-vector condition on $\sigma_\phi$, and in particular \eqref{eq:N2twistnullvector} no longer holds.
For $\Psi_{12}$, by combining the null-vector conditions \eqref{eq:N2untwistednulllevel2} with the commutation relation $[L_{-1}^{(1)},L_{-2}^{(1)}]=L_{-3}^{(0)}$, we obtain the relation
\begin{equation}
    \rmmode{1}{-2}\rmmode{1}{-1}\ket{\Psi_{12}}
    =\left[2 \rmmode{0}{-1}\rmmode{0}{-2}-\rmmode{0}{-3} -\frac{1}{g}\left(\rmmode{0}{-1}\right)^3\right]\ket{\Psi_{12}} \,.
\end{equation}
We obtain two Ward identities by computing the following closed-contour integrals
\begin{equation}
  \oint dw (w-1)^{3/2}(w-\eta)^{3/2} \wt{J}_k(\eta,w) \qquad \text{and} \qquad
  \oint dw (w-1)^{3/2}(w-\eta)^{3/2} w^{-1} \wt{J}_k(\eta,w) \,,
\end{equation}
where
\begin{equation}
  \wt{J}_k(\eta,w):=\cblock{\phantom{T^{(1)}(w)L_{-1}^{(1)}\cdot}\Psi_{12}(\infty)}{T^{(1)}(w)L_{-1}^{(1)}\cdot\Psi_{12}(0)}{\sigma_{\phi}(1)}{\sigma_{\phi}(\eta)}{\Psi_k} \,.
\end{equation}
Combining these relations, we obtain an equation of the form
\begin{equation}
  \sum_j \cblock{\lambda_j\cdot\Psi_{12}(\infty)}{\mu_j\cdot\Psi_{12}(0)}{\sigma_{\phi}(1)}{\sigma_{\phi}(\eta)}{\Psi_k} =0 \,,
\end{equation}
where $\mu_j$ and $\lambda_j$ are elements of the Virasoro subalgebra generated by the modes $L_m^{(0)}$. This yields a linear ODE of order four on the conformal blocks $J_k$. In terms of the scaling function \eqref{eq:F_{N,phi}} $\mathcal{F}_{2,\phi}^{(\alpha\beta)}(z,\zb)=F[\zeta(z/\zb)]$, the ODE reads
\begin{equation} \label{eq:ODE-order4}
  \begin{aligned}
    & (g-1) g \left[16 (1-2 g) \hh-3 (3 g-2) (16 \hh -1) \zeta (\zeta-1)\right] F(\zeta)\\
    & +2(g-1) (1-2 \zeta) \left[(4 g^2-8 g+3)+3 (6 g^2 -15g +10-8g\hh) \zeta(\zeta-1) \right] F'(\zeta) \\
    & +\zeta(\zeta-1)  \left[8 (2g^2-5g+3)+(66 g^2-177g+120-16g\hh)\zeta (\zeta-1) \right] F''(\zeta) \\
    & +5 (2 g-3) \zeta^2(\zeta-1)^2  (1-2 \zeta) \mathcal{F}'''(\zeta)
    + 2 \zeta^3 (\zeta-1)^3 F^{(4)}(\zeta)=0 \,.
  \end{aligned}
\end{equation}
This ODE is manifestly invariant under $\zeta \mapsto 1-\zeta$, which reflects the invariance of the rescaled conformal blocks $(\eta-1)^{2\hh}J_k(\eta)$ under $\eta \mapsto 1/\eta$.
If we parametrize the conformal dimension of $\phi$ as $h=h_{\lambda,0}$ with $\lambda \in \Rbb$, the ODE takes the form
\begin{equation}
  \begin{aligned}
    & [\zeta(\zeta-1)]^3 F^{(4)}(\zeta)
    -\tfrac{5}{2} (2g-3) [\zeta(\zeta-1)]^2 (2\zeta-1) F'''(\zeta) \\
    & +z(z-1) \left[4 (g-1)(2g-3)+ (64-97g+37g^2-\lambda^2) \zeta(\zeta-1) \right] F''(\zeta) \\
    & +(g-1) (1-2\zeta) \left[(2g-3) (2 g-1)+\tfrac{3}{2}(28-47 g+20 g^2-2 \lambda^2) \zeta(\zeta-1) \right] F'(\zeta) \\
    & +(g-1) \left[\tfrac{1}{2}(2g-1) (8-17 g+8 g^2-2 \lambda^2)+3 (3g-2) (2g-2-\lambda) (2g-2+\lambda)\zeta(\zeta-1)\right] F(\zeta) \,.
  \end{aligned}
\end{equation}
The local exponents for $F(\zeta)$ are given by 
\begin{equation} \label{eq:exponents}
  \begin{tabular}{ccc}
    $0$ & $1$ & $\infty$ \\
    \hline
    $0$ & $0$ & $-4h_{12}$ \\
    $\frac{h_{13}}{2}$ & $\frac{h_{13}}{2}$     & $-4h_{12}+1$ \\
    $h_{13}$   & $h_{13}$   & $-4h_{12}+(1+\lambda)/g$ \\
    $h_{13}+1$ & $h_{13}+1$ & $-4h_{12}+(1-\lambda)/g$ \\
  \end{tabular}
\end{equation}
The correspondence between singular points in the variables $\eta$ and $\zeta$ is given by
\begin{align} \label{eq:eta->1}
    \zeta \underset{\eta \to 0}{\sim} \frac{1}{4\sqrt\eta} \,,
    \qquad \begin{cases}
        (1-\zeta) \underset{\eta \to 1}{\sim} \left(\frac{\eta-1}{4i}\right)^2 & \text{if } \mathrm{Im}\,\eta>0 \,,\\
        \zeta \underset{\eta \to 1}{\sim} \left(\frac{1-\eta}{4i}\right)^2
         & \text{if } \mathrm{Im}\,\eta<0 \,,
    \end{cases}
    \qquad \zeta \underset{\eta \to \infty}{\sim} \frac{\sqrt\eta}{4} \,.
\end{align}
In particular, the limits $\mathrm{Arg}(z)\to 0$ and $\mathrm{Arg}(z)\to \pi$ are mapped to $\zeta \to 1$ and $\zeta \to 0$, respectively. In these limits, we have $\eta \to 1$, and the contribution of $J_k$ to $\mathcal{F}_{2,\phi}^{(\alpha\beta)}$ \eqref{eq_generic_block_expansion} scales as $(\eta-1)^{h_{\Psi_k}}$, which is proportional to $(1-\zeta)^{h_{\Psi_k}/2}$  and $\zeta^{h_{\Psi_k}/2}$, respectively. Similar relations can be established for $\zeta \to \infty$, in correspondence with $\eta\to 0$ and $\eta \to \infty$.
Hence, the local exponents \eqref{eq:exponents} correspond to the intermediate states (counted with their multiplicities):
\begin{equation}
  \begin{aligned}
    & \{ \id, [\id,\psi_{13}], \Psi_{13}, \Psi_{13}\}
    && \text{in the channels } \zeta \to 0 \text{ and } \zeta \to 1 \,, \\
    & \{ \sigma_{\lambda,0}, \sigma_{\lambda,0}^{(1)}, \sigma_{\lambda,2}, \sigma_{\lambda,-2}\}
    && \text{in the channel }  \zeta \to \infty \,.
  \end{aligned}
\end{equation} 
Recall that the conformal blocks are labelled by primary operators under the neutral subalgebra $\mathcal{A}_N$.

When the mother CFT is a minimal model, one can check on various examples that the orbifold fusion rules derived in \cite{borisov_systematic_1998} from the Verlinde formula are consistent with these intermediate states. 
\medskip

The physical solution $\mathcal{F}_{2,\phi}^{(\alpha\beta)}$ can be written as a linear combination
\begin{equation} \label{eq_F_2_phi_0}
  \mathcal{F}_{2,\phi}^{(\alpha\beta)}(z,\zb) = \lambda_1 \, F_1(\zeta) + \lambda_2 \, F_2(\zeta) + \lambda_3 \, F_3(\zeta) + \lambda_4 \, F_4(\zeta) \,,
\end{equation}
where $\zeta=\zeta(z/\zb)$, and $F_1,F_2,F_3,F_4$ are the holomorphic solutions of \eqref{eq:ODE-order4} of the form
\begin{equation}
    F_k(\zeta) = (1-\zeta)^{h_{\Psi_k}/2} \sum_{n=0}^{\infty} a_{k,n} \, (1-\zeta)^n \,,
\end{equation}
normalised so that $a_{k,0}=1$.
The differential equation \eqref{eq:ODE-order4} can be solved via the Frobenius method, yielding infinite series which are convergent for $|1-\zeta| <1$. By symmetry, a basis of solutions around $\zeta =0$ is simply given by the functions $F_k(1-\zeta)$.
In practice, using truncated series is extremely efficient to evaluate these functions numerically to arbitrary precision.

The last step is to describe a general algorithm for the determination of the coefficients $\lambda_k$. Using the OPEs \eqref{eq:twist_boundary_OPE} and \eqref{eq:orb_boundary_OPE}, one obtains
\begin{equation}
    \mathcal{F}_{2,\phi}^{(\alpha\beta)}(z,\zb) \underset{\mathrm{Arg}(z)\to 0}{=} \sum_{k=1}^4 \left( \frac{z-\zb}{i} \right)^{h_{\Psi_k}} 
    \left( \mathcal{A}_{\sigma_\phi,\Psi_k}^{(\alpha)} \, \mathcal{B}_{\Psi_{12}\Psi_{12}\Psi_k}^{(\alpha\beta\alpha)} + \dots \right) \,,
\end{equation}
where $\dots$ denotes the subdominant contributions from the descendant operators. Using \eqref{eq:eta->1}, the dominant term of $F_k(\zeta)$ in this limit is $[(\eta-1)/4i]^{h_{\Psi_k}} \sim [(z-\zb)/4i]^{h_{\Psi_k}}$, and thus we obtain for $k=1,2$:
\begin{equation}
    \lambda_k = 4^{h_{\Psi_k}} \, \mathcal{A}_{\sigma_\phi,\Psi_k}^{(\alpha)} \, \mathcal{B}_{\Psi_{12}\Psi_{12}\Psi_k}^{(\alpha\beta\alpha)} \,.
\end{equation}
The determination of $\lambda_3,\lambda_4$ is less direct, due to the two-fold degeneracy of the intermediate state $\Psi_{13}$. We use the alternative expansion
\begin{equation}
  \mathcal{F}_{2,\phi}^{(\alpha\beta)}(z,\zb) = \mu_1 \, F_1(1-\zeta) + \mu_2 \, F_2(1-\zeta) + \mu_3 \, F_3(1-\zeta) + \mu_4 \, F_4(1-\zeta) \,.
\end{equation}
Like in the expansion $\zeta\to 1$, here we have
\begin{equation}
    \mu_k = 4^{h_{\Psi_k}} \, \mathcal{A}_{\sigma_\phi,\Psi_k}^{(\beta)} \, \mathcal{B}_{\Psi_{12}\Psi_{12}\Psi_k}^{(\beta\alpha\beta)} \,,
    \qquad \text{for } k=1,2 \,.
\end{equation}
Using the change of basis
\begin{equation}
    F_k(\zeta) = \sum_{\ell=1}^4 p_{k\ell} F_\ell(1-\zeta) \,,
\end{equation}
and comparing the two expansions of $\mathcal{F}_{2,\phi}^{(\alpha\beta)}(z,\zb)$, we obtain the relations
\begin{equation}
    \sum_{k=1}^4 \lambda_k p_{k\ell} =\mu_\ell \,.
\end{equation}
Setting $\ell=1,2$ in these equations then yields the values of $\lambda_3,\lambda_4$ in terms of $\lambda_1,\lambda_2,\mu_1,\mu_2,$ and the coefficients $p_{k\ell}$. The latter can be obtained numerically, using the numerical power series expansions of the functions $F_k(\zeta)$ and $F_\ell(1-\zeta)$ in the intersection of their convergence domains.

\subsection{The function $\aver{\Psi_{12} \cdot \sigma \cdot \Psi_{12}}$ in a generic $\Zbb_3$ orbifold}
\label{sec:baretwistN3}

Here, the relevant conformal blocks are of the form
\begin{equation}
  J_k(\eta) = \cblock{\Psi_{12}(\infty)}{\ \Psi_{12}(0)}{\sigma^\dag(1)}{\sigma(\eta)}{\Psi_k} \,.
\end{equation}
We give the null-vectors of $\Psi_{12}$ at levels two and three:
\begin{align}
  & \rmmode{r}{-2} \cdot \Psi_{12} =
  \frac{1}{3g}\sum_{s=0}^{2}\rmmode{r-s}{-1}\rmmode{s}{-1}\cdot \Psi_{12} \,, \\
  & \rmmode{3-r}{-1}\rmmode{r}{-2}\cdot \Psi_{12} =
  \frac{1}{3g}\left[2 \rmmode{0}{-1}\rmmode{1}{-1}\rmmode{2}{-1}+\left(\rmmode{3-r}{-1}\right)^3\right]\cdot \Psi_{12} \,,
\end{align}
for $r\in\{0,1,2\}$.

To derive an ODE for the conformal blocks, we employ seven orbifold Ward identities, together with the six  null-vector conditions above. To not overload the presentation of this section with technical details, we relegate the specifics of the derivation to Appendix \ref{app:baretwistthirdrenyi}.

We obtain the ODE for the scaling function $\mathcal{F}_3^{(\alpha\beta)}$ in the variable $\zeta$ \eqref{eq:zeta(eta)}
\begin{equation} \label{eq:ODE-order3}
  \begin{aligned}
    & 4 AB(6A+B+3) (1-2\zeta) F(\zeta) + 2[9AB - C\zeta(1-\zeta)]F'(\zeta) \\
    & - 9(4A+B-2)(1-2\zeta)\zeta(1-\zeta) F''(\zeta)
    + 18\zeta^2(1-\zeta)^2 F'''(\zeta)=0 \,,
  \end{aligned}
\end{equation}
where
\begin{equation}
  A=g-1 \,, \qquad B=6g-5 \,,
\end{equation}
and
\begin{equation}
  C = 36A^2+2B^2+30AB-18A-3B \,.
\end{equation}
For generic $(A,B)$, if $C$ has the above value, then the local exponents for $\mathcal{F}(\zeta)$ are
\begin{center}
  \begin{tabular}{ccc}
    $0$ & $1$ & $\infty$ \\
    \hline
    $0$ & $0$ & $-2A-\frac{B}{3}-1$  \\
    $2A+1$ & $2A+1$ & $-\frac{2B}{3}$ \\
    $\frac{B}{2}+1$ & $\frac{B}{2}+1$ & $-2A$
  \end{tabular}
\end{center}
For the above values of $(A,B)$, this gives
\begin{center}
  \begin{tabular}{ccc}
    $0$ & $1$ & $\infty$ \\
    \hline
    $0$ & $0$ & $-6h_{12}+\frac{2}{3} h_{12}$ \\
    $h_{13}$ & $h_{13}$ & $-6h_{12}+\frac{2}{3} h_{12}+ \frac{2}{3}$ \\
    $\frac{3h_{13}}{2}$ & $\frac{3h_{13}}{2}$ & $-6h_{12}+\frac{2}{3}h_{14}$ 
  \end{tabular}
\end{center}
These local exponents correspond to the intermediate states:
\begin{equation}
  \begin{aligned}
    & \{ \id, [\id,\psi_{13},\psi_{13}], \Psi_{13}\}
    && \text{in the channels } \zeta \to 0 \text{ and } \zeta \to 1\,, \\
    & \{ \sigma_{12}, \sigma_{12}^{(1)} , \sigma_{14} \}
    && \text{in the channel }  \zeta \to \infty \,.
  \end{aligned}
\end{equation}
where we use the shorthand notation $\sigma_{rs}=\sigma_{\phi_{rs}}$.

From this point on, one can proceed as in the case of $\mathcal{F}_{2,\phi}^{(\alpha\beta)}$, except that the complication coming from non-trivial multiplicities of the local exponents is absent.

It is worth pointing out that the existence of this third order ODE was anticipated in \cite{dupic_imaginary_2019}, where only a weaker, fourth order ODE had been obtained. This third order ODE is optimal for generic CFTs, as confirmed by conformal block counting using the orbifold Verlinde formula \cite{2023JPhA...56T5403E}.

\section{The critical Ising model}
\label{sec:Ising}

\subsection{The quantum chain and the BCFT}

The Hamiltonian of the Ising quantum chain with open BC, describing $M$ spins with generic BC at the boundary, is given by:
\begin{equation}
  H_{\alpha\beta} = -\sum_{j=1}^{M-1} s_j^z s_{j+1}^z - h\,\sum_{j=1}^M s_j^x -h_\alpha s^z_1-h_\beta s^z_M\,,
\end{equation}
where $s^{x,y,z}_j$ denote Pauli spin operators acting on site $j$. We denote the lattice spacing by $a$, so that the length of the chain is $L=Ma$. The parameters $h_\alpha,h_\beta$ denote external magnetic fields (in the $z$ direction) acting at the boundary sites $j=1$ and $j=M$. The ground state of this Hamiltonian is then found by \textit{exact diagonalization} (ED) for system sizes $M\leq 26$ sites, and from it, the R\'enyi entropies are evaluated. 

To take the \textit{scaling limit} of the critical chain, we send $M\rightarrow \infty, a\rightarrow 0$ while keeping $L=Ma$ fixed. In this limit, bulk criticality is achieved for $h=1$, while each boundary admits three \textit{critical points} $h_\alpha,h_\beta\in\{0,\pm \infty\}$, corresponding to free, $+$ and $-$ BC. 

From a CFT perspective, the scaling limit of the critical Ising chain with open boundaries is very well understood. It is described by the BCFT with central charge $c=1/2$ and a bulk operator spectrum consisting of three primary operators -- the identity $\id$ $(h_{\id}=0)$, energy $\varepsilon$ $(h_\varepsilon=0)$ and spin  $s$ $(h_{s}=1/16)$-- and their descendants \cite{di_francesco_conformal_1997}. The three boundary critical points correspond to the three conformal boundary conditions for the Ising BCFT, which, in the framework of radial quantization on the annulus, allow the construction of the following physical boundary states \cite{cardy_boundary_1989}:
\begin{align}
  \ket{f} &= \kket{\id} - \kket{\varepsilon}
  && \textit{(free BC)} \,, \\
  \ket{\pm} &= \frac{1}{\sqrt{2}} \kket{\id}
  + \frac{1}{\sqrt{2}} \kket{\varepsilon} \pm \frac{1}{2^{1/4}} \kket{s}
  && \textit{(fixed BC)} \,,
\end{align}
where $\kket{\phi}$ denotes the Ishibashi state \cite{ishibashi_boundary_1989} corresponding to the primary operator $\phi$. The physical boundary states $\ket\alpha$ are in one-to-one correspondence with the primary operators of the bulk CFT\footnote{This statement is strictly true if the bulk CFT is diagonal, see \cite{runkel_structure_2000} for a detailed discussion.}: $\ket{f} \leftrightarrow s$ and  $\ket\pm \leftrightarrow \id/\varepsilon$. The boundary operators that interpolate between
two conformal BCs can be inferred from this correspondence \cite{cardy_boundary_1989,lewellen_sewing_1992}. The spectrum of primary boundary operators $\psi_{i}^{(\alpha\beta)}$ of the Ising BCFT is summarized in  Table \ref{tab:IsingBCCOs}.

\begin{table}[h]
  \begin{tabular}{|c|c|c|c|}
    \hline
    $(\alpha\beta)$ & $+$ & $-$ & $f$ \\ \hline
    $+$    &  $\psi_\id$ & $\psi_\varepsilon$  & $\psi_s$   \\ \hline
    $-$   &  $\psi_\varepsilon$ & $\psi_\id$  & $\psi_s$  \\ \hline
    $f$    &  $\psi_s$ & $\psi_s$  & $\psi_\id,\psi_\varepsilon$  \\ \hline
  \end{tabular}
  \centering
  \caption{Boundary operator spectrum of the Ising BCFT}
  \label{tab:IsingBCCOs}
\end{table}

The ground state entropies read
\begin{equation}
    g_+ = g_- = \frac{1}{\sqrt 2} \,,
    \qquad g_f = 1 \,,
\end{equation}
and the bulk-boundary OPE coefficients for the energy operator are given by
\begin{equation}
   A_{\varepsilon,\id}^{(+)} = A_{\varepsilon,\id}^{(-)} = 1 \,,
   \qquad A_{\varepsilon,\id}^{(f)} = -1 \,.
\end{equation}

\subsection{R\'enyi entropies}
According to the general discussion \eqref{eq_generic_subleading_corrections}, in the scaling limit with $a\to 0$, the lattice twist operator admits a \textit{local} expansion into scaling operators of the corresponding orbifold CFT, leading to 
\begin{equation}\label{eq:onepointlatticetwistUHP}
  \aver{\sigma_{\text{lattice}}(\ell)}_{\Sbb_L}^{(\alpha\beta)}
  = C_{\id} \left(\frac{\pi a}{2L\sin\frac{\pi\ell}{L}}\right)^{2h_\sigma} \mathcal{F}_N^{(\alpha\beta)}(z,\zb)
  + C_{\varepsilon} \left(\frac{\pi a}{2L\sin\frac{\pi\ell}{L}}\right)^{2h_{\sigma_\varepsilon}}
  \mathcal{F}_{N,\varepsilon}^{(\alpha\beta)}(z,\zb)
  +\dots 
\end{equation} 
where
\begin{equation}
  \mathcal{F}_N^{(\alpha\beta)}(z,\zb) = |z-\zb|^{2h_\sigma} \aver{\sigma(z,\zb)}_{\Hbb}^{(\alpha\beta)} \,,
  \qquad \mathcal{F}_{N,\varepsilon}^{(\alpha\beta)}(z,\zb) =  |z-\zb|^{2h_{\sigma_\varepsilon}} \aver{\sigma_\varepsilon(z,\zb)}_{\Hbb}^{(\alpha\beta)} \,,
\end{equation}
and $z=\exp(i\pi\ell/L)$. The amplitudes $C_{\id}$ and $C_{\varepsilon}$ in \eqref{eq:onepointlatticetwistUHP} are not universal quantities, but they can be obtained with very high accuracy from a numerical analysis of the infinite Ising chain. Here one can employ the free fermion techniques of \cite{vidal_entanglement_2003} and the well-known analytical results for the R\'enyi entropy of an interval in an infinite system \cite{holzhey_geometric_1994,calabrese_entanglement_2004} to obtain numerical fits for $C_{\id}$ and $C_{\varepsilon}$, with great accuracy.

In our computational setup, the system sizes accessible through exact diagonalization are limited to $M\leq 26$ and for technical reasons
we only consider even system sizes.
With system sizes of this order of magnitude, finite-size corrections are quite significant. In fact the finite-size effects are still important even at the much larger system sizes ($M\sim 100$ sites) accessible through DMRG methods \cite{taddia_entanglement_2013}. The most relevant corrections we have found arise from the subleading scaling of the lattice twist operator, given in equation (\ref{eq:onepointlatticetwistUHP}). The relative amplitude of the subleading term with respect to the leading one scales like $\mathcal{O}\left({M}^{-2h_{\varepsilon}/\Nf}\right)$. Since we do not have access, numerically, to system sizes large enough to suppress these corrections, it is necessary to take into account the first two terms in the expansion of (\ref{eq:onepointlatticetwistUHP}) to find a good agreement with the lattice data. Such subleading contributions to the lattice twist operator have previously been understood, through the path-integral formalism on the corresponding replicated surface, under the name of ``unusual corrections''  \cite{eriksson_corrections_2011,cardy_unusual_2010}. 

\subsubsection{Scaling functions for the entropy $S_N^{(+-)}$}

In the critical Ising model, the only primary BCCO between the boundary conditions $\alpha=+$ and $\beta=-$ has dimension $h_\varepsilon=1/2$, and thus we denote $\Psi^{(+-)}=\Psi_\varepsilon$. Here, we set $g=4/3$ in \eqref{eq:Kac}, so that the central charge is $c=1/2$, and $h_{12}=1/2$. In the limits $\mathrm{Arg}(z)\to 0$ and $\mathrm{Arg}(z)\to \pi$, for any twisted operator $\sigma_\phi$, the boundary OPEs reduce to $\sigma_\phi\big|_+ \to \id$ and $\sigma_\phi\big|_- \to \id$, respectively. As a consequence, for both $\mathcal{F}_N^{(+-)}$ and $\mathcal{F}_{N,\varepsilon}^{(+-)}$, only the conformal block with intermediate state $\Psi_k=\id$ contributes in the expansion \eqref{eq_generic_block_expansion}, and thus, in each case, the scaling function is given by the solution of the ODE associated to the local exponent $0$ at $\zeta\to 1$ or $\zeta\to 0$.

\paragraph{The case $N=2$.} The scaling function $\mathcal{F}_2^{(+-)}$ obeys the hypergeometric equation~\eqref{eq:ODE-order2}
\begin{equation}
    \zeta(\zeta-1) F''(\zeta) -\tfrac{2}{3}(2\zeta-1)F'(\zeta)+\tfrac{2}{3} F(\zeta) = 0 \,.
\end{equation}
The solution with local exponent $0$ at $\zeta\to 1$ can be determined by the Frobenius method. It turns out that the linear recursion for the coefficients yields a sequence $(a_n)$ which vanishes for $n>2$. We obtain 
\begin{equation}
    F_1(\zeta) = 1-(1-\zeta)+(1-\zeta)^2 = 1-\zeta(1-\zeta) \,.
\end{equation}
In terms of the variable $\theta=\mathrm{Arg}(z)$, we have $\zeta=\cos^2(\theta/2)$, which yields
\begin{equation}
   F_1 \left(\cos^2\frac{\theta}{2} \right) = 1- \tfrac{1}{4} \sin^2\theta \,.
\end{equation}
The physical solution is then given by $F_1$, up to a multiplicative factor, which is determined by the BCs \eqref{eq:BC.F_N}. We obtain:
\begin{equation}
    \mathcal{F}_2^{(+-)}\left(re^{i\theta},re^{-i\theta} \right) =
    \sqrt{2} \left( 1 - \tfrac{1}{4} \sin^2\theta \right) \,.
\end{equation}

Similarly, for $\mathcal{F}_{2,\varepsilon}^{(+-)}$, after setting $g=4/3$ and $\hh=h_{\sigma_\varepsilon}=9/32$ in the fourth-order equation \eqref{eq:ODE-order4}, we consider the solution with local exponent $0$, and apply the Frobenius method. This yields the function
\begin{equation}
     F_1(\zeta) = 1-9\zeta(1-\zeta) \,.
\end{equation}
Using the BCs \eqref{eq:BC.F_{N,phi}}, we obtain
\begin{equation}
    \mathcal{F}_{2,\varepsilon}^{(+-)}(re^{i\theta},re^{-i\theta}) =
    \sqrt{2} \left( 1 - \tfrac{9}{4} \sin^2\theta \right) \,.
\end{equation}

\paragraph{The case $N=3$.} With a similar approach, we solve the third-order ODE \eqref{eq:ODE-order3} and the BCs \eqref{eq:BC.F_N} for $\mathcal{F}_{3}^{(+-)}$, and we obtain
\begin{equation}
    \mathcal{F}_3^{(+-)}(re^{i\theta},re^{-i\theta}) =
    2 \left( 1 - \tfrac{4}{9} \sin^2\theta \right) \,.
\end{equation}

In contrast, we do not have an ODE for $\mathcal{F}_{3,\varepsilon}^{(+-)}$, but we can obtain its expression as follows. Using the multiplicities found in \cite{2023JPhA...56T5403E}, we have the fusion rules in the $\Zbb_3$ orbifold of the Ising CFT:
\begin{equation}
    \sigma_\varepsilon \times \sigma_\varepsilon \to \id \,,
    \qquad \Psi_\varepsilon \times \Psi_\varepsilon \to \id \,,
    \qquad \Psi_\varepsilon \times \sigma_\varepsilon \to \sigma^{(2)} \,.
\end{equation}
Hence, there is only one conformal block of the form \eqref{eq:conformal_block(z,zb)}. This means that the conformal blocks in the three possible channels are all proportional.
For the rescaled conformal blocks, we have
\begin{equation}
    J(\eta) = \cblock{\Psi_\varepsilon(\infty)}{\ \Psi_\varepsilon(0)}{\sigma_\varepsilon^\dag(1)}{\sigma_\varepsilon(\eta)}{\id} 
    \ \propto\ \cblockv{\Psi_\varepsilon(\infty)}{\ \Psi_\varepsilon(0)}{\sigma_\varepsilon^\dag(1)}{\sigma_\varepsilon(\eta)}{\sigma^{(2)}}
    \ \propto\ \cblockv{\ \Psi_\varepsilon(0)}{\Psi_\varepsilon(\infty)}{\sigma_\varepsilon^\dag(1)}{\sigma_\varepsilon(\eta)}{\sigma^{(2)}}
\end{equation}
The local exponents at $\eta \to 0$ and $\eta \to 1$ are $\mu=-h_{\Psi_\varepsilon}-h_{\sigma_\varepsilon}+h_{\sigma^{(2)}}=-1$ and $-2h_{\sigma_\varepsilon}$, respectively.
Hence, the function defined as
\begin{equation}
    K(\eta) := \eta \, (\eta-1)^{2h_{\sigma_\varepsilon}} \, J(\eta)
\end{equation}
is an entire function on $\Cbb$. In the limit $\eta \to \infty$, one has
\begin{equation}
    J(\eta) \underset{\eta \to \infty}{\propto} \eta^{h_{\Psi_\varepsilon}-h_{\sigma_\varepsilon}-h_{\sigma^{(2)}}}
    \qquad \Rightarrow \qquad
    K(\eta) \underset{\eta \to \infty}{\propto} \eta^{-2\mu}=\eta^2 \,.
\end{equation}
Thus, $K(\eta)$ is a polynomial of degree two.
Due to the symmetry  $J(1/\eta)=\eta^{2h_{\sigma_\varepsilon}}J(\eta)$, the polynomial $K(\eta)$ is of the form
\begin{equation}
    K(\eta) = a (\eta^2+1) + b \eta \,.
\end{equation}
Using the OPE at $\eta \to 1$, for generic $N$
\begin{equation}
    \sigma_\varepsilon(\eta) \sigma_\varepsilon^\dag(1)
    = (\eta-1)^{-2h_{\sigma_\varepsilon}} \id
    + \frac{2h_{\sigma_\varepsilon}}{Nc} (\eta-1)^{-2h_{\sigma_\varepsilon}+2} \, T^{(0)}(1) + \dots
\end{equation}
we obtain the expansion
\begin{equation}
    J(\eta)
    = (\eta-1)^{-2h_{\sigma_\varepsilon}} 
    + \frac{2h_{\sigma_\varepsilon}}{Nc}  \, \aver{\Psi_\varepsilon|T^{(0)}(1)|\Psi_\varepsilon} \, (\eta-1)^{-2h_{\sigma_\varepsilon}+2} + \dots
\end{equation}
The expectation value is easily computed:
\begin{equation}
    \aver{\Psi_\varepsilon|T^{(0)}(1)|\Psi_\varepsilon}= Nh_\varepsilon = \frac{3}{2} \,.
\end{equation}
This yields
\begin{equation}
    K(\eta) = \eta + \frac{4}{9} (\eta-1)^2 =\eta \left(1-\tfrac{16}{9}\sin^2\theta \right) \,.
\end{equation}
Finally, by imposing the BC \eqref{eq:BC.F_N}, we obtain
\begin{equation}
    \mathcal{F}_{3,\varepsilon}^{(+-)}(re^{i\theta},re^{-i\theta}) =
    2 \left( 1 - \tfrac{16}{9} \sin^2\theta \right) \,.
\end{equation}

\subsubsection{Scaling functions for the entropy $S_N^{(+f)}$}

Here also, there is only one primary BCCO between the boundary conditions $\alpha=+$ and $\beta=f$, and it has dimension $h_s=1/16$. Hence, we denote the replicated BCCO as $\Psi^{(+f)}=\Psi_s$. We set $g=3/4$ in \eqref{eq:Kac}, so that $c=1/2$ and $h_{12}=1/16$. Since, in the limit $\mathrm{Arg}(z)\to 0$, we have the OPE $\sigma_\phi \big|_+ \to \id$, only the conformal block with intermediate state $\Psi_k=\id$ contributes to the scaling function.

\paragraph{The case $N=2$.} The scaling function $\mathcal{F}_2^{(+f)}$ obeys the hypergeometric equation~\eqref{eq:ODE-order2}
\begin{equation}
    \zeta(\zeta-1) F''(\zeta) +\tfrac{1}{2}(2\zeta-1)F'(\zeta)-\tfrac{1}{16} F(\zeta) = 0 \,.
\end{equation}
We perform the change of variable $f(\theta)=F[\cos^2(\theta/2)]$, which yields the regular ODE
\begin{equation}
    f''(\theta) + \tfrac{1}{16} f(\theta) = 0 \,.
\end{equation}
The unique solution which obeys the BCs \eqref{eq:BC.F_N} is given by
\begin{equation}
    \mathcal{F}_2^{(+f)}\left(re^{i\theta},re^{-i\theta} \right) =
    \sqrt{2} \cos\frac{\theta}{4} \,.
\end{equation}

For $\mathcal{F}_{2,\varepsilon}^{(+f)}$, we set $g=3/4$ and $\hh=h_{\sigma_\varepsilon}=9/32$ in the fourth-order equation \eqref{eq:ODE-order4}. The scaling function is then given by
\begin{equation}
    \mathcal{F}_{2,\varepsilon}^{(+f)}\left(re^{i\theta},re^{-i\theta} \right) =
    \sqrt{2} \cos\frac{3\theta}{4} \,.
\end{equation}
Indeed, one can easily verify that it solves the ODE, and obeys the BCs \eqref{eq:BC.F_{N,phi}}.

\paragraph{The case $N=3$.} For the scaling function $\mathcal{F}_3^{(+f)}$, the third-order ODE \eqref{eq:ODE-order3} and the BC \eqref{eq:BC.F_N} yield
\begin{equation}
    \mathcal{F}_3^{(+f)}\left(re^{i\theta},re^{-i\theta} \right) =
    2 \cos\frac{\theta}{3} \,.
\end{equation}

For the scaling function $\mathcal{F}_{3,\varepsilon}^{(+f)}$, we have not derived an ODE. In this case, the multiplicities computed in \cite{2023JPhA...56T5403E} yield the fusion rules
\begin{equation}
    \sigma_\varepsilon \times \sigma_\varepsilon
    \to \id + [\id,\varepsilon,\varepsilon] \,,
    \qquad \Psi_s \times \Psi_s
    \to \id + [\id,\varepsilon,\varepsilon] \,,
    \qquad \Psi_s \times \sigma_\varepsilon
    \to \sigma_s + \sigma_s^{(2)} \,.
\end{equation}
Hence, the space of conformal blocks is two-dimensional. We \emph{assume} that these conformal blocks obey a second-order Fuchsian ODE. Then, the above fusion rules yield the local exponents of the ODE for $\mathcal{F}_{3,\varepsilon}^{(+f)}$, in terms of the variable $\zeta$:
\begin{equation}
    \begin{tabular}{ccc}
    0 & 1 & $\infty$ \\
    \hline $0$ & $0$ & $\tfrac{2}{3}$ \\
    $\tfrac{1}{2}$ & $\tfrac{1}{2}$ & $-\tfrac{2}{3}$
  \end{tabular}
\end{equation}
These are exactly the local exponents of the hypergeometric ODE \eqref{eq:hypergeom.ode.zeta}, with parameters $a=\tfrac23$ and $b=0$.
Changing variables to $\zeta=\cos^2(\theta/2)$, we find that the function $\cos(2\theta/3)$ corresponds to the solution with local exponent $0$ at $\zeta \to 1$. By imposing the BC \eqref{eq:BC.F_{N,phi}}, we obtain
\begin{equation}
    \mathcal{F}_{3,\varepsilon}^{(+f)}\left(re^{i\theta},re^{-i\theta} \right) =
    2 \cos\frac{2\theta}{3} \,.
\end{equation}

\section{The critical three-state Potts model}
\label{sec:Potts}

\subsection{The lattice quantum chain}
A natural extension of the Ising chain, the three-state Potts model allows the spins at each site to take one of three possible values $\{R,G,B\}$, which we can also conveniently parametrize by third roots of unity $\{1,\omega,\bar{\omega}\}$, with $\omega=\exp(2\pi i/3)$. The  Hamiltonian of the three-state Potts model, tuned to its bulk critical point \cite{zou_conformal_2018,affleck_boundary_1998,henkel2013conformal} is given by:
\begin{equation} \label{eq:Potts_hamiltonian}
  \begin{aligned}
    H_{\alpha\beta} = -v\left[\sum_{j=1}^{M-1} \left(Z_j Z_{j+1}^{\dagger}+Z_j^{\dagger}Z_{j+1}\right)+\sum_{j=2}^{M-1}\left(X_j+X_j^\dag\right)
      +H_1^{(\alpha)}+ H_M^{(\beta)} \right] \,,
  \end{aligned}
\end{equation}
where $v=\sqrt{3}/(2\pi^{3/2})$ is the conformal normalization factor \cite{henkel2013conformal} and the operators $Z_j$ and $X_j$ act at site $j$ as:
\begin{equation}
  Z=\left(
    \begin{array}{ccc}
      1 & 0 & 0 \\
      0 & \omega & 0 \\
      0 & 0 & \omega^2
    \end{array}\right) \,,
  \qquad X=\left(
    \begin{array}{ccc}
      0 & 1 & 0 \\
      0 & 0 & 1 \\
      1 & 0 & 0
    \end{array}\right) \,.
\end{equation}
The terms $H_1^{(\alpha)}$ and  $H_M^{(\beta)}$ set the BCs at the ends of the chain. For the purpose of this analysis, we will set \textit{fixed} BC of type $R$ at site $M$ and  \textit{restricted} boundary conditions of type $\{G,B\}$ at site $1$ -- the spin at site $1$ is forbidden from taking the value $R$. This is implemented through the boundary terms:
\begin{equation}
  H^{(R)}=h\left(
    \begin{array}{ccc}
      1 & 0 & 0 \\
      0 & 0 & 0 \\
      0 & 0 & 0
    \end{array}\right) \,,
  \qquad H^{(GB)}=h\left(
    \begin{array}{ccc}
      -1 & 0 & 0 \\
      0 & 1 & 0 \\
      0 & 0 & 1
    \end{array}\right) \,,
\end{equation}

The critical points of interest for the boundaries correspond to $h=+\infty$. However, for any $h>0$, the boundaries will flow towards the same critical points, up to irrelevant boundary perturbations\cite{graham_renormalisation_2000} . These are typically inconsequential when $h$ takes a large positive value. Alternatively, one can implement $|h| = \infty$ by simply restricting the local Hilbert spaces of the boundary sites to exclude the $\{G,B\}$ and $\{R\}$ configurations on the left and right boundary, respectively.

\subsection{The boundary CFT}
The scaling limit of this critical chain is captured by the D-series BCFT  $\mathcal{M}(6,5)$  with central charge $c=4/5$ and a bulk primary operator spectrum that contains the scalar operators given in Table \ref{tab:Pottsoperators} as well as the non-scalar operators with conformal dimensions $(h,\hb)=(2/5,7/5),(7/5,2/5),(3,0),(0,3)$.
One can, as shown in Table \ref{tab:Pottsoperators}, assign a $\mathbb{Z}_3$ charge to the scalar operators, and their respective conformal families, that is consistent with the fusion rules between them. The conjugation in Table \ref{tab:Pottsoperators} is thus used to distinguish the operators with the same conformal dimension, but opposite $\mathbb{Z}_3$ charge.

\renewcommand{\arraystretch}{1.1}
\begin{table}[h!]
  \centering
  \begin{tabular}{ccc}
    Operator & Conformal dimension  & $\Zbb_3$ charge \\
    \hline $\id$   & $0$  & $0$ \\
    $\varepsilon\equiv\phi_{12}$  & $\frac{2}{5}$ & $0$ \\
    $X \equiv\phi_{13}$  &  $\frac{7}{5}$ & $0$ \\
    $\phi_{14}$  & $3$ & $0$ \\
    $s,s^\dag \equiv \phi_{33}$ & $\frac{1}{15}$ & $\pm 1$ \\
    $\psi,\psi^\dag \equiv \phi_{34}$   & $\frac{2}{3}$ & $\pm 1$ \\
    \hline \hline
  \end{tabular}
  \caption{Spectrum of scalar primary operators in the three-state Potts CFT.}
  \label{tab:Pottsoperators}
\end{table}
In the scaling limit, the fixed and restricted boundary critical points will correspond, naturally, to the fixed and restricted\footnote{In \cite{cardy_boundary_1989} they are referred to as "mixed" BC. } conformal boundary states \cite{cardy_boundary_1989,caldeira_free_2003}.
The conformal boundary states are labelled by the primary operators of Table \ref{tab:Pottsoperators}, and they can be expressed in terms of the Ishibashi states $\kket{\phi}$ as
\begin{equation} \label{eq:boundarystatesPotts}
  \begin{aligned}
    & \ket\id = \mathcal{N} [ (\kket\id + \kket\psi + \kket{\psi^\dag})
    +\lambda(\kket\varepsilon + \kket{s} + \kket{s^\dag}) ]
    && \text{(fixed $R$)} \\
    & \ket\psi =\mathcal{N} [ (\kket\id + \omega\kket\psi + \bar\omega\kket{\psi^\dag})
    +\lambda(\kket\varepsilon + \omega\kket{s} + \bar\omega\kket{s^\dag}) ]
    && \text{(fixed $G$)} \\
    &\ket{\psi^\dag} =\mathcal{N} [ (\kket\id + \bar\omega\kket\psi + \omega\kket{\psi^\dag})
    +\lambda(\kket\varepsilon + \bar\omega\kket{s} + \omega\kket{s^\dag}) ]
    && \text{(fixed $B$)} \\
    & \ket\varepsilon = \mathcal{N} [ \lambda^2(\kket\id + \kket\psi + \kket{\psi^\dag})
    - \lambda^{-1}(\kket\varepsilon + \kket{s} + \kket{s^\dag}) ]
    && \text{(restricted $GB$)} \\
    &\ket{s} = \mathcal{N} [ \lambda^2(\kket\id + \omega\kket\psi + \bar\omega\kket{\psi^\dag})
    - \lambda^{-1}(\kket\varepsilon + \omega\kket{s} + \bar\omega\kket{s^\dag}) ]
    && \text{(restricted $RB$)} \\
    & \ket{s^\dag} = \mathcal{N} [ \lambda^2(\kket\id + \bar\omega\kket\psi + \omega\kket{\psi^\dag})
    - \lambda^{-1}(\kket\varepsilon + \bar\omega\kket{s} + \omega\kket{s^\dag}) ]
    && \text{(restricted $RG$)} \,,
  \end{aligned}
\end{equation}
where 
\begin{equation}
  \mathcal{N} =\sqrt{\frac{2}{\sqrt{15}} \sin \frac{\pi}{5}} \,,
  \qquad \lambda=\sqrt{\frac{\sin (2 \pi / 5)}{\sin (\pi / 5)}} \,.
\end{equation}

Due to the $\Zbb_3$ symmetry of the Hamiltonian, we have some freedom to set which conformal boundary state corresponds to the \textit{fixed} boundary condition $R$ in the chain. However, this uniquely determines the CFT boundary state that corresponds to the  \textit{restricted} boundary conditions $GB$. This can be understood by considering the spectrum of boundary operators that interpolate between these conformal BC \cite{behrend_classification_1998}, and ensuring the results are consistent with the underlying $\mathbb{Z}_3$ symmetry. In our case, choosing \textit{fixed} $\,R\leftrightarrow |\id\rangle$ forces us to assign \textit{restricted} $ \, GB\leftrightarrow |\varepsilon \rangle$. 
The most relevant boundary operator interpolating between these BCs is $\psi^{(GB,R)}_{12}$ \cite{behrend_classification_1998}, with conformal dimension $h_{\varepsilon}=2/5$.

The ground state degeneracies read \cite{affleck_boundary_1998}
\begin{equation} \label{eq:groundstate degeneracies}
  g_R=\left(\frac{5-\sqrt{5}}{30}\right)^{\frac{1}{4}} \,,
  \qquad g_{GB}=2g_R \cos(\pi/5) \,.
\end{equation}

\subsection{Determination of an entropy scaling function}
In this section, we illustrate the calculation of the scaling function $\mathcal{F}_{N,\varepsilon}^{(\alpha\beta)}$ for $N=2$, in the critical three-state Potts chain with boundary conditions $\alpha=GB$ on the left boundary and $\beta=R$ on the right. This function determines the subleading contributions to the R\'enyi entropy $S_2^{(GB,R)}$, arising from the composite twist operator $\sigma_\varepsilon$, where $\varepsilon$ is the energy operator of the Potts model. Hence, the mother BCFT is the Virasoro minimal model $\mathcal{M}(6,5)$, and in Kac notation, we have $\varepsilon=\phi_{12}$.

Setting $g=6/5$ and $\hh=h_\sigma+h_\varepsilon/2=1/4$ in \eqref{eq:ODE-order4}, we obtain the ODE
\begin{equation}
\begin{aligned}
    & -24 [18\zeta(\zeta-1)+7] F(\zeta) 
    +6 (2\zeta-1) [44\zeta(\zeta-1)+7] F'(\zeta)  \\
    & -30 \zeta(\zeta-1) [9\zeta(\zeta-1)+4]  F''(\zeta)
    +375 [\zeta(\zeta-1)]^2 (2\zeta-1) F^{(3)}(\zeta)  \\
    & +250 [\zeta(\zeta-1)]^3 F^{(4)}(\zeta) = 0\,.
\end{aligned}
\end{equation}
The Riemann scheme \eqref{eq:exponents} yields
\begin{equation} \label{eq:exponents-Potts}
  \begin{tabular}{ccc}
    $0$ & $1$ & $\infty$ \\
    \hline $0$ & $0$ & $-\frac{8}{5}$ \\
    $\frac{7}{10}$ & $\frac{7}{10}$ & $-\frac{3}{5}$ \\
    $\frac{7}{5}$ & $\frac{7}{5}$ & $1$ \\
    $\frac{12}{5}$ & $\frac{12}{5}$ & $-\frac{9}{5}$
  \end{tabular}
\end{equation}

The solutions around $\zeta=0$ are obtained in terms of a power series, using the Frobenius method. They take the form 
\begin{equation}\label{eq:frobeniusansatz}
    F_k(1-\zeta) = \zeta^{\gamma_k} \sum_{n=0}^{\infty} a_{k,n} \zeta^n \,, \qquad \text{with} \quad a_{k,0}=1 \,.
\end{equation}
To determine the coefficients $a_{k,n}$, it is convenient to rewrite the ODE using the differential operator $\theta = \zeta \partial_{\zeta}$. It takes the form 
\begin{equation}
 \left[ P_0(\theta) + \zeta  P_1(\theta) 
 +  \zeta^2 P_2(\theta) + \zeta^3  P_3(\theta)  \right] 
 \cdot \mathcal{F} =0 \,,
\end{equation}
with 
\begin{equation}
\begin{aligned}
P_{0}(\theta) & = -\theta  (5 \theta -12) (5 \theta -7) (10 \theta -7) \\
P_1(\theta) & = 6 (5 \theta -7) (4 + 21 \theta -65 \theta^2  + 25 \theta^3) \\
P_2(\theta)  & =-3 (5 \theta -8) \left( 18 - 13 \theta - 95 \theta^2 + 50 \theta^3 \right) \\
P_3(\theta) & = 2 (\theta +1) (5 \theta -9) (5 \theta -8) (5 \theta -3) \,.
\end{aligned}
\end{equation}
The local exponents $\gamma_k$ are the roots of the \textit{characteristic polynomial} $P_0$:
\begin{equation}
\gamma_1 =0, \quad \gamma_2 = \frac{7}{10}, \quad \gamma_3 = \frac{7}{5}, \quad \gamma_4 = \frac{12}{5} \,,
\end{equation}
and they correspond to the first and second column of \eqref{eq:exponents-Potts}. The coefficients $a_{k,n}$ are given by the linear recursion relation (where, by convention, $a_{k,n}=0$ if $n<0$):
\begin{align}
a_{k,n} = - \frac{P_1(\gamma_k + n-1) a_{k,n-1} + P_2(\gamma_k + n-2) a_{k,n-2} + P_3(\gamma_k + n-3) a_{k,n-3}  }{P_0(\gamma_k + n)}
\end{align}
provided $P_0(\gamma_k + n)$ does not vanish for $n \geq 1$. This condition holds except for the case $\gamma_4 = \gamma_3+1$, which can be traced back to the degenerate fusion $\sigma_{\phi} \times \sigma_{\phi} \to   2 \Psi_{13}$.  In particular for $F_3$ we have $P_0(\gamma_3+1) = P_1(\gamma_3)=0$, and  $a_{3,1}$ becomes a free parameter (changing the value of $a_{3,1}$ amounts to adding a multiple of $F_4$ to $F_3$). The actual physical solution corresponds to a certain choice of $a_{3,1}$, while $F_4$ is unphysical. 

From the above recursion relation one can obtain the series expansion of the all solutions \eqref{eq:frobeniusansatz} to a high order with minimal computational cost. For instance, for $k=1$, the first coefficients $a_{1,n}$ of the series are
\begin{equation}
\begin{aligned}
    & 1, -4, \frac{24}{13}, \frac{98}{299}, \frac{553}{3887}, \frac{13566}{167141}, \frac{10826144 }{203744879}, \\
    &\frac{7703652}{203744879}, \frac{423213759}{14873376167}, 
    \frac{27519615504 }{1234490221861}, \frac{5654297376}{314101062367},
    \dots 
\end{aligned}
\end{equation}
The truncated power series are accurate enough to evaluate the functions $F_k(\zeta)$ to arbitrary precision, provided that $|\zeta|$ does not approach $1$ too closely. For values of $\zeta$ near $1$, the other basis of solutions $F_k(\zeta)$ can be used instead.

In the determination of the physical scaling function, a huge simplification occurs, since the boundary condition $\alpha=R$ is incompatible with the boundary operator $\psi_{13}$, and hence the OPE as $\mathrm{Arg}(z)\to \pi$ simplifies to
\begin{equation}
    \sigma_{\varepsilon}(z,\zb) \Big|_R \to \id^{(R,R)} \,.
\end{equation}
This implies that, out of the four independent solutions to the ODE in the domain $|\zeta|<1$, only $F_1(1-\zeta)$ contributes,
and we get 
\begin{equation}\label{eq:excited_onepoint_potts}
  \mathcal{F}_{2,\phi}^{(GB,R)}(z,\zb) 
  = \mu_1 \, F_1(1-\zeta) \,.
\end{equation}
The constant $\mu_1$ is given by
\begin{equation}
    \mu_1=\acon{R}{\sigma_\varepsilon}{\id}
    \, \bcon{R,R,GB}{\Psi_{12}}{\id}{\Psi_{12}} \,,
    \qquad \acon{R}{\sigma_{\varepsilon}}{\id}
    =g_R^{-1} A^{(R)}_{\varepsilon,\id} \,,
    \qquad \bcon{R,R,GB}{\Psi_{12}}{\id} {\Psi_{12}}=1 \,,
\end{equation}
where the universal boundary entropy and the bulk-boundary structure constant for $\varepsilon$ in the $\mathcal{M}(6,5)$ BCFT read, respectively \cite{affleck_boundary_1998,caldeira_free_2003,zou_universal_2022}:
\begin{equation}
    g_R=\left(\frac{5-\sqrt{5}}{30}\right)^{\frac{1}{4}} \,,
    \qquad A_{\varepsilon,\id}^{(R)}=\left(\frac{1+\sqrt{5}}{2}\right)^{\frac{3}{2}} \,.
\end{equation}

It may be possible to obtain a simpler ODE by exploiting the null vectors of $\sigma_{\varepsilon}$. According to conformal block counting through the orbifold Verlinde formula \cite{borisov_systematic_1998}, the optimal ODE is expected to be of third order. However, we have not derived this optimal ODE; instead, we work with the generic fourth-order ODE derived earlier.

The fourth order ODE used here for the Potts model is not optimal. Using the optimal third order ODE would remove the degeneracy of the solution corresponding to the exponent $\gamma_3$ and allow to determine directly the physical solution. Alternatively, the physical solution can be determined by demanding that $F_3(1-\zeta)$ can be expressed as a linear combination of $F_k(\zeta)$ for $k \leq 3$.

\section{Conclusion}
\label{sec:conclusion}

In this article, we have outlined a general method for computing R\'enyi entropies in the ground state of a one-dimensional critical system with mixed open boundaries, specifically for an interval starting at one of its ends. Given the significant finite size effects in open chains, we also examined the leading finite-size correction, which results from subleading scaling operators in the expansion of the lattice twist operators. Within the conformal field theory framework, this involves computing certain three-point functions on the upper-half plane in the $\Zbb_N$ cyclic orbifold: two boundary condition changing operators, and one twist operator (bare or excited). 

To achieve this, we derived Fuchsian differential equations satisfied by these correlation functions, by exploiting the null-vectors of the twisted and untwisted representations of its symmetry algebra $\OVir_N$, along with Ward identities obtained from the additional conserved currents of the theory. The Frobenius method then systematically provides a way to find power series solutions to these differential equations. The correlation functions are obtained as particular linear combinations of these conformal blocks, with coefficients given by certain boundary and bulk-boundary structure constants. 

Among the cases we have analysed are the leading and subleading contributions to the one-interval second and third R\'enyi entropies of the Ising model, and the second R\'enyi entropy for the three-state Potts model. We have also derived differential equations for mixed BC twist correlators in the $\Zbb_2$ and $\Zbb_3$ orbifolds of generic BCFTs, as well as an explicit solution for the second R\'enyi entropy.

We analyzed the CFT results, incorporating the first subleading correction, and compared them with numerical data for the critical Ising and three-state Potts spin chains. In both cases, we found remarkable agreement across all mixed boundary condition choices, despite the constraints of limited system sizes. Expanding to larger system sizes could be possible through advanced numerical approaches such as DMRG \cite{PhysRevLett.69.2863,Chepiga_2022} or tensor network methods \cite{Orus_2014}.

An important limitation of our method, first identified in \cite{dupic_entanglement_2018},  is that obtaining a differential equation becomes increasingly difficult as the R\'enyi index $N$ is increased. We have verified, using the fusion rules in \cite{2023JPhA...56T5403E}, that the expected order of the ODEs grows with $N$ for generic minimal models, indicating that more orbifold Ward identities will be needed to obtain the ODEs.

A possible extension of this work would be to generalize the setup for the calculation of R\'enyi entropies of an interval \textit{contained in the bulk}, with mixed BC. However, in this situation, one would have to find a partial differential equation that a four-point function with two twist fields and two BCCOs satisfies. Cardy's doubling trick tells us that such a correlator is given by a linear combination of six-point conformal blocks on the complex plane, so the corresponding differential equation would be partial instead of ordinary.

\clearpage

\appendix
\section*{Appendix}

\section{Hypergeometric differential equation}
\label{app:hypergeom}

In this Appendix, we recall some useful facts about the hypergeometric differential equation. Euler's hypergeometric differential equation is given by
\begin{equation}\label{eq:hypergeometricstandard}
  \eta(\eta-1) f''(\eta)+[(a+b+1) \eta-c] f'(\eta)+a b \,f(\eta)=0 \,,
\end{equation}
and it has the following Riemann scheme:
\begin{equation}
  \begin{tabular}{ccc}
    $0$ & $1$ & $\infty$ \\
    \hline $0$ & $0$ & $a$ \\
    $1-c$ & $c-a-b$ & $b$
  \end{tabular}
\end{equation}
The solutions are constructed using Gauss hypergeometric function
\begin{equation} \label{eq:FGauss}
  \FGauss(a,b;c|\eta) = \sum_{n=0}^\infty \frac{(a)_n (b)_n \eta^n}{(c)_n n!} \,,
  \qquad (q)_n = q(q+1)\dots (q+n-1) \,.
\end{equation}
Following the conventions of \cite{NIST:DLMF}, we give a standard basis of solutions to (\ref{eq:hypergeometricstandard}) around the singular point $\eta=0$:
\begin{equation} \label{eq:solutionseta0}
  \begin{aligned}
    & I_1(\eta)=\FGauss(a, b ; c \mid \eta) \,, \\
    & I_2(\eta)=\eta^{1-c}\FGauss(b-c+1, a-c+1 ; 2-c \mid \eta) \,,
  \end{aligned}
\end{equation}
and around $\eta=1$:
\begin{equation}\label{eq:solutionseta1}
  \begin{aligned}
    & K_1(\eta)=\FGauss(a, b ; a+b-c+1 \mid 1-\eta) \,, \\
    & K_2(\eta)=(1-\eta)^{c-a-b} \FGauss(c-b, c-a ; c-a-b+1 \mid 1-\eta) \,.
  \end{aligned}
\end{equation}
The two bases of solutions are linearly related by 
\begin{equation}
  I_i(\eta)=\sum_{j=1}^{2}P_{ij} K_j(\eta) \,,
\end{equation}
with the fusing matrix $P$
\begin{equation}
  \label{eq_P_matrix}
  P=\left[\begin{array}{cc}
      \frac{\Gamma(c) \Gamma(d)}{\Gamma(c-a) \Gamma(c-b)} & \frac{\Gamma(c) \Gamma(-d)}{\Gamma(a) \Gamma(b)} \\
      \frac{\Gamma(2-c) \Gamma(d)}{\Gamma(1-a) \Gamma(1-b)} & \frac{\Gamma(2-c) \Gamma(-d)}{\Gamma(1-c+a) \Gamma(1-c+b)}
    \end{array}\right] \,,
\end{equation}
and its inverse:
\begin{equation}
  P^{-1}=\left[\begin{array}{cc}
      \frac{\Gamma(1-c) \Gamma(1-d)}{\Gamma(1-c+a) \Gamma(1-c+b)} & \frac{\Gamma(c-1) \Gamma(1-d)}{\Gamma(a) \Gamma(b)} \\
      \frac{\Gamma(1-c) \Gamma(1+d)}{\Gamma(1-a) \Gamma(1-b)} & \frac{\Gamma(c-1) \Gamma(1+d)}{\Gamma(c-a) \Gamma(c-b)}
    \end{array}\right] \,,
\end{equation}
expressed in terms of Euler's Gamma function $\Gamma$,  with $d = c -a -b$. Note that $P^{-1}$ is obtained from $P$ by the change $(a,b,c,d) \to (a,b,1-d,1-c)$.
\medskip

If the parameters of the hypergeometric equation satisfy the relation $c=a-b+1$, then the rescaled function $y(\eta)=\eta^{a/2}f(\eta)$ has Riemann scheme
\begin{equation}
  \begin{tabular}{ccc}
    0 & 1 & $\infty$ \\
    \hline $a/2$ & $0$ & $a/2$ \\
    $b-a/2$ & $1-2b$ & $b-a/2$
  \end{tabular}
\end{equation}
and therefore, it has the same local exponents at $0$ and $\infty$. Correspondingly, in terms of the operator $\theta=\eta\partial_\eta$, the ODE for $y$ reads
\begin{equation}
    \left[(\eta-1) \theta^2  + b(\eta+1) \theta  +\frac{a}{4}(2b-a)(\eta-1) \right] \, y(\eta) = 0 \,,
\end{equation}
and it is invariant under $\eta \mapsto 1/\eta$.
We now introduce the change of variables
\begin{equation} \label{eq:y(eta)=F(zeta)}
    y(\eta) \to F\left[\frac{(1+\sqrt\eta)^2}{4\sqrt\eta} \right] \,,
\end{equation}
which yields the differential equation for the function $F$
\begin{equation} \label{eq:hypergeom.ode.zeta}
  \zeta(\zeta-1) F''(\zeta)+ \frac 12(2b+1)(2\zeta-1) F'(\zeta)+ a(2b-a) \, F(\zeta)=0 \,.
\end{equation}
This is the hypergeometric equation, with modified parameters
\begin{equation}
  \wt{a}=a \,, \quad \wt{b}=2b-a \,, \qquad \wt{c}=b+\frac 12 \,.
\end{equation}
We obtain  the two bases of solutions $(\wt{I}_1,\wt{I}_2)$ and $(\wt{K}_1,\wt{K}_2)$ to \eqref{eq:hypergeom.ode.zeta}, by substituting $(a,b,c) \to (\wt{a},\wt{b},\wt{c})$ in (\ref{eq:solutionseta0}--\ref{eq:solutionseta1}). This yields
\begin{align}
  & \wt{I}_1(\zeta) = \FGauss(a,2b-a;b+1/2 \mid \zeta) \,,
  && \wt{K}_1(\zeta) = \wt{I}_1(1-\zeta) \,, \\
  & \wt{I}_2(\zeta) = \zeta^{1/2-b} \, \FGauss(1/2+a-b, 1/2-a+b ;3/2-b \mid \zeta) \,,
  && \wt{K}_2(\zeta) = \wt{I}_2(1-\zeta) \,.
\end{align}
Let us describe the limiting behaviour of these solutions as $\zeta \to 0$ and $\zeta \to 1$ in the regime $b<1/2$. Clearly one has $\wt{I}_1(0)=1$, whereas $\wt{I}_2(0)=0$. To study the behaviour at $\zeta \to 1$, we use the change of basis
\begin{align} 
  \wt{I}_i(\zeta) = \wt{P}_{i1} \, \wt{I}_1(1-\zeta) + \wt{P}_{i2} \, \wt{I}_2(1-\zeta) \,,
  \qquad i \in \{1,2\} \,,
\end{align}
where $\wt{P}_{ij}$ denotes the matrix element of \eqref{eq_P_matrix} after substituting $(a,b,c) \to (\wt{a},\wt{b},\wt{c})$. On the right-hand side, one has $\wt{I}_1(0)=1$ and $\wt{I}_2(0)=0$. Thus we simply get $\lim_{\zeta \to 1} \wt{I}_i(\zeta) = \wt{P}_{i1}$.

To summarise, in the regime $b<1/2$, the functions $\wt{I}_1,\wt{I}_2$ are continuous in $\zeta$ on the interval $[0,1]$, with limiting values
\begin{align}
  & \wt{I}_1(0)=1 \,,
  && \wt{I}_1(1)=\frac{1}{\pi} \Gamma(b+1/2)\Gamma(3b+1/2) \cos \pi(a-b) \,, \\
  & \wt{I}_2(0)=0 \,,
  && \wt{I}_2(1)=\frac{\Gamma(3/2-b)\Gamma(3b+1/2)}{\Gamma(1-a)\Gamma(1-2b+a)} \,.
\end{align}

\section{Change of variable for Fuchsian differential equations} 
\label{app:change_of_var}

In Appendix \ref{app:hypergeom}, we have shown that, if the hypergeometric ODE in the variable $\eta$ is invariant (after proper rescaling) under $\eta \mapsto 1/\eta$, then the ODE expressed in the variable 
\begin{equation} 
    \zeta(\eta)=\frac{(1+\sqrt\eta)^2}{4\sqrt\eta}
\end{equation}
is also a hypergeometric ODE, with modified parameters. In the present section, we shall extend this result to any Fuchsian ODE which is invariant under $\eta \mapsto 1/\eta$.
\medskip

Consider a generic ODE of Fuchsian type in the variable $\eta$, with singularities at $\eta=0,1$ and $\infty$. That is to say an ODE of the form
\begin{align} \label{eq_generic_ODE}
    \left[ \eta^n (\eta -1)^n  \partial_\eta^n +  \eta^{n-1} (\eta-1)^{n-1}  p_{1}(\eta) \partial_\eta^{n-1} + \cdots + p_n(\eta) \right] y(\eta) =0 \,,
\end{align}
where each $p_j(\eta)$ is a polynomial of degree at most $j$. 
We assume that the ODE is invariant under $\eta \mapsto 1/\eta$.
\medskip

First, we rewrite the ODE in terms of the operator $\theta = \eta \partial_\eta$. Using the identity
 \begin{align}
    \eta^j \partial^j_\eta  = \theta (\theta -1) \cdots (\theta -j+1) \,,
\end{align}
the ODE \eqref{eq_generic_ODE} takes the form 
\begin{align} \label{eq_generic_ODE_theta}
    \left[ (\eta -1)^n  \theta^n +   (\eta -1)^{n-1}  q_1(\eta) \theta^{n-1} + \cdots + q_n(\eta) \right] y(\eta) =0 \,,
\end{align}
where each $q_j(\eta)$ is a polynomial of degree at most $j$. Invariance  of the ODE under $\eta \mapsto 1/\eta$ means that $q_j(\eta)$ is of the form
\begin{align}
    q_j(\eta) = \sum_{k=0}^j q_{j,k} \, \eta^k \,,
    \qquad \text{with} \quad q_{j,j-k} = q_{j,k} \,.
\end{align}
Moreover, for any non-negative integer $m$, we have
\begin{align}
    \eta^{m/2}+\eta^{-m/2} = 2\,T_m(2 \zeta -1) \,,
\end{align}
where $T_m$ is the $m$-th Chebyshev polynomial of the first kind.
We can thus rewrite $q_j(\eta)$ as
\begin{align}
    q_j(\eta) = 
    \eta^{j/2} Q_j(\zeta) \,,
    \qquad \text{where} \quad
    Q_j(\zeta)=\sum_{k=0}^j q_{j,k} \, T_{|j-2k|}(2\zeta-1) \,,
\end{align}
and, by construction, $Q_j(\zeta)$ is a polynomial of degree at most $j$, and it is invariant under $\zeta \to 1-\zeta$.
Hence, we can rewrite the ODE \eqref{eq_generic_ODE} as 
\begin{align} \label{eq_generic_ODE_mixed}
\left[ \left( 4  \sqrt{\zeta(\zeta-1)}\right)^n  \theta^n + \left( 4  \sqrt{\zeta(\zeta-1)}\right)^{n-1}  Q_1(\zeta) \theta^{n-1} + \cdots + Q_n(\zeta) \right] F(\zeta) =0 \,,
\end{align}
where we have applied the change of functions $y(\eta)=F[\zeta(\eta)]$.
Now, we need to express the action of $\theta$ in terms of $\partial_\zeta$.
Using the relation 
\begin{align}
\theta = \frac{\sqrt{\zeta(\zeta-1)}}{2} \partial_{\zeta} \,,
\end{align}
one obtains by induction that, for any integer $j \geq 1$, $\theta^j$ is of the form
\begin{align}
    \theta^j = \sum_{k=1}^j 2^{-j} \beta_{jk}(\zeta) \, \Big(\sqrt{\zeta(\zeta-1)}\Big)^{2k-j} \,\partial_{\zeta}^k \,,
\end{align}
where each $\beta_{jk}(\zeta)$ is a polynomial of degree at most $(j-k)$. The polynomials $\beta_{jk}(\zeta)$ are determined by the recursion relation
 \begin{align}
    \beta_{11}(\zeta)=1 \,, \qquad
    \beta_{j+1,k}(\zeta) = \beta_{j,k-1}(\zeta) + \left[  (2k-j) (\zeta-1/2) + \zeta(\zeta-1) \partial_{\zeta} \right] \beta_{jk}(\zeta) \,,
\end{align}
where by convention $\beta_{jk}(\zeta)=0$ if $k \notin \{1,\dots,j\}$.
Using this recursion, we obtain that $\beta_{jk}(1-\zeta)=(-1)^k \beta_{jk}(\zeta)$.
Finally, we introduce the polynomials
\begin{align}
    P_n(\zeta) = Q_n(\zeta) \,, \quad \text{and} \quad
    P_{k}(\zeta) = \sum_{j=n-k}^n 2^j\beta_{j,n-k}(\zeta) Q_{n-j}(\zeta) 
    \quad \text{for } k=1,\dots,n-1 \,,
\end{align}
and we end up with the ODE in the variable $\zeta$: 
\begin{align} \label{eq_generic_ODE_zeta}
\left[ [\zeta (\zeta-1)]^n  \partial_{\zeta}^n +  [\zeta(\zeta -1)]^{n-1} P_1(\zeta) \partial_{\zeta}^{n-1} + \cdots + P_n(\zeta) \right] F(\zeta) =0 \,,
\end{align}
The ODE \eqref{eq_generic_ODE_zeta} is Fuchsian, since each $P_k(\zeta)$ is a polynomial of degree at most $k$. Moreover, using the property $P_k(1-\zeta) = (-1)^{n-k} P_k(\zeta)$, we find that this ODE is invariant under $\zeta \to 1- \zeta$.

\section{Derivation of the ODE for $\aver{\Psi_{12}\cdot \sigma \cdot \Psi_{12}}$ in the $\mathbb{Z}_3$ orbifold }
\label{app:baretwistthirdrenyi}

We present in this section all the orbifold Ward identities and null-vector conditions we used to derive the third-order differential equation \eqref{eq_ODE_3_leading}.

\subsubsection*{The Ward identities}

We derive linear relations between correlators of the form
\begin{equation}
    \aver{\Psi'_{12}|\sigma^\dag(1)\sigma(\eta)|\Psi''_{12}} \,,
\end{equation}
where $\Psi'_{12}$ and $\Psi''_{12}$ are descendants of $\Psi_{12}$ under $\mathrm{OVir}_N$.
For this, we consider closed-contour integrals of the form
\begin{equation}  \label{eq:Ward-integral}
\oint dz (z-1)^{m_2+1} (z-\eta)^{m_3+1} z^{m_4+1} \, \aver{\Psi'_{12}(\infty) T^{(r)}(z) \sigma^\dag(1) \sigma(\eta)\Psi''_{12}(0)} \,,
\end{equation}
where $m_2 \in \Zbb+r/N$, $m_3 \in \Zbb-r/N$, and $m_4 \in \Zbb$. Applying the Cauchy theorem on the above integral yields a linear relation of the form
\begin{equation} \label{eq:Ward}
\begin{aligned}
   & \sum_{p=0}^\infty a_p(\eta) \aver{(L_{m_1+p}^{(r)}\Psi'_{12})(\infty)  \sigma^\dag(1) \sigma(\eta)\Psi''_{12}(0)}
    = \sum_{p=0}^\infty \Big[ b_p(\eta) \aver{\Psi'_{12}(\infty) (L_{m_2+p}^{(r)}\sigma^\dag)(1) \sigma(\eta)\Psi''_{12}(0)} \\
   & \quad +  c_p(\eta) \aver{\Psi'_{12}(\infty) \sigma^\dag(1) (L_{m_3+p}^{(r)}\sigma)(\eta)\Psi''_{12}(0)}
    + d_p(\eta) \aver{\Psi'_{12}(\infty) \sigma^\dag(1) \sigma(\eta) (L_{m_4+p}^{(r)}\Psi''_{12})(0)} \Big] \,,
\end{aligned}
\end{equation}
where $m_1=-m_2-m_3-m_4-2$, and the functions $a_p,b_p,c_p,d_p$ are the Taylor-series coefficients of simple power functions of $(z,\eta)$, whose exponents are defined in terms of $m_1,\dots,m_4$. For instance:
\begin{equation}
    (1-z)^{m_2+1}(1-\eta z)^{m_3+1} = \sum_{p=0}^\infty a_p(\eta) \, z^p \,,
    \qquad (z-1)^{m_2+1}(z-\eta)^{m_3+1} = \sum_{p=0}^\infty d_p(\eta) \, z^p \,.
\end{equation}
In particular, here we shall set $N=3$ and $m_2,m_3 \geq -1/3$, and thus, due to the null-vector condition $L_{-1/3}^{(1)}\cdot \sigma=0$ and its counterpart for $\sigma^\dag$, the first and second terms in the RHS of \eqref{eq:Ward} vanish.

\begin{description}
\item[Ward 1] Setting $\Psi'_{12}=\Psi''_{12}=L_{-1}^{(1)}\Psi_{12}$, and $(m_1,m_2,m_3,m_4)=(-1,1/3,-1/3,-1)$  and $r=1$ in \eqref{eq:Ward-integral} yields a linear relation of the form
\begin{equation}
\begin{aligned}
    & a_{0|1} \aver{\Psi_{12}|(\rmmode{1}{1})^2\sigma^\dag(1)\sigma(\eta)\rmmode{1}{-1}|\Psi_{12}} + a_{1|1} \aver{\Psi_{12}|\rmmode{1}{1}\rmmode{1}{0}\sigma^\dag(1)\sigma(\eta)\rmmode{1}{-1}|\Psi_{12}} = \\ 
    & \qquad d_{0|1} \aver{\Psi_{12}| \rmmode{1}{1}\sigma^\dag(1)\sigma(\eta)\rmmode{1}{-1}\rmmode{1}{-1}|\Psi_{12}} +d_{1|1}  \aver{\Psi_{12}| \rmmode{1}{1} \sigma^\dag(1)\sigma(\eta)\rmmode{1}{0}\rmmode{1}{-1}|\Psi_{12}} \,.
    \end{aligned}
\end{equation}
\item[Ward 2] Setting $\Psi'_{12}=\Psi_{12}, \Psi''_{12}=(L_{-1}^{(1)})^2\Psi_{12}$, $(m_1,m_2,m_3,m_4)=(-1,1/3,-1/3,-1)$ and $r=1$ in \eqref{eq:Ward-integral} yields a linear relation of the form
\begin{equation}
    \begin{aligned}
    & a_{0|2} \aver{\Psi_{12}|\rmmode{1}{1}\sigma^\dag(1)\sigma(\eta)(\rmmode{1}{-1})^2|\Psi_{12}} = \\
    & d_{0|2}  \aver{\Psi_{12}| \sigma^\dag(1)\sigma(\eta)(\rmmode{1}{-1})^3|\Psi_{12}} 
    +d_{1|2}  \aver{\Psi_{12}|\sigma^\dag(1)\sigma(\eta)\rmmode{1}{0}(\rmmode{1}{-1})^2|\Psi_{12}} \\
    & +d_{2|2}  \aver{\Psi_{12}| \sigma^\dag(1)\sigma(\eta)\rmmode{1}{1}(\rmmode{1}{-1})^2|\Psi_{12}}
    +d_{3|2}  \aver{\Psi_{12}| \sigma^\dag(1)\sigma(\eta)\rmmode{1}{2}(\rmmode{1}{-1})^2|\Psi_{12}} \,.
    \end{aligned}
\end{equation}
\item[Ward 3] Setting $\Psi'_{12}=(L_{-1}^{(1)})^2\Psi_{12}, \Psi''_{12}=\Psi_{12}$, $(m_1,m_2,m_3,m_4)=(-1,1/3,-1/3,-1)$ and $r=1$ in \eqref{eq:Ward-integral} yields a linear relation of the form
\begin{equation}
\begin{aligned}
    & d_{0|3} \aver{\Psi_{12}|(\rmmode{1}{1})^2\sigma^\dag(1)\sigma(\eta)\rmmode{1}{-1}|\Psi_{12}} = \\
    & a_{0|3}  \aver{\Psi_{12}| (\rmmode{1}{1})^3\sigma^\dag(1)\sigma(\eta)|\Psi_{12}}
    +a_{1|3}  \aver{\Psi_{12}| (\rmmode{1}{1})^2\rmmode{1}{0}\sigma^\dag(1)\sigma(\eta)|\Psi_{12}} \\
    & +a_{2|3}  \aver{\Psi_{12}|(\rmmode{1}{1})^2\rmmode{1}{-1} \sigma^\dag(1)\sigma(\eta)|\Psi_{12}} 
    + a_{3|3} \aver{\Psi_{12}|(\rmmode{1}{1})^2\rmmode{1}{-2} \sigma^\dag(1)\sigma(\eta)|\Psi_{12}} \,.
    \end{aligned}
\end{equation}
\item[Ward 4] Setting $\Psi'_{12}=\Psi_{12}, \Psi''_{12}=L_{-1}^{(1)}\Psi_{12}$, $(m_1,m_2,m_3,m_4)=(0,-1/3,1/3,-2)$ and $r=2$ in \eqref{eq:Ward-integral} yields a linear relation of the form
\begin{equation}
    \begin{aligned}
         & d_{0|4}\aver{\Psi_{12}| \sigma^\dag(1)\sigma(\eta) \rmmode{2}{-2}\rmmode{1}{-1} |\Psi_{12}}
         +  d_{1|4}\aver{\Psi_{12}| \sigma^\dag(1)\sigma(\eta) \rmmode{2}{-1}\rmmode{1}{-1} |\Psi_{12}}  \\
         & +d_{2|4}\aver{\Psi_{12}| \sigma^\dag(1)\sigma(\eta) \rmmode{2}{0}\rmmode{1}{-1} |\Psi_{12}} +
         d_{3|4}\aver{\Psi_{12}| \sigma^\dag(1)\sigma(\eta) \rmmode{2}{1}\rmmode{1}{-1} |\Psi_{12}} =0 \,.
    \end{aligned}
\end{equation}
\item[Ward 5] Setting $\Psi'_{12}=L_{-1}^{(1)}\Psi_{12}, \Psi''_{12}=\Psi_{12}$, $(m_1,m_2,m_3,m_4)=(-2,-1/3,1/3,0)$ and $r=2$ in \eqref{eq:Ward-integral} yields a linear relation of the form
\begin{equation}
    \begin{aligned}
        & a_{0|5} \aver{\Psi_{12}| \rmmode{1}{1}\rmmode{2}{2}\sigma^\dag(1)\sigma(\eta) |\Psi_{12}}
        +  a_{1|5} \aver{\Psi_{12}| \rmmode{1}{1}\rmmode{2}{1} \sigma^\dag(1)\sigma(\eta) |\Psi_{12}}  \\ 
         & +a_{2|5} \aver{\Psi_{12}| \rmmode{1}{1}\rmmode{2}{0} \sigma^\dag(1)\sigma(\eta) |\Psi_{12}}
         + a_{3|5} \aver{\Psi_{12}| \rmmode{1}{1}\rmmode{2}{-1} \sigma^\dag(1)\sigma(\eta) |\Psi_{12}} =0 \,.
    \end{aligned}
\end{equation}
\item[Ward 6] Setting $\Psi'_{12}=\Psi_{12}, \Psi''_{12}=L_{-1}^{(1)}\Psi_{12}$, $(m_1,m_2,m_3,m_4)=(-1,-1/3,1/3,-1)$ and $r=2$ in \eqref{eq:Ward-integral} yields a linear relation of the form
\begin{equation}
    \begin{aligned}
   & a_{0|6} \aver{\Psi_{12}|\rmmode{2}{1}\sigma^\dag(1)\sigma(\eta)\rmmode{1}{-1}|\Psi_{12}} 
   = d_{0|6}  \aver{\Psi_{12}| \sigma^\dag(1)\sigma(\eta)\rmmode{2}{-1}\rmmode{1}{-1}|\Psi_{12}} \\
   & +d_{1|6}  \aver{\Psi_{12}| \sigma^\dag(1)\sigma(\eta)\rmmode{2}{0}\rmmode{1} {-1}|\Psi_{12}} 
   +d_{2|6}  \aver{\Psi_{12}| \sigma^\dag(1)\sigma(\eta)\rmmode{2}{1}\rmmode{1}{-1}|\Psi_{12}}
  \,.
  \end{aligned}
\end{equation}
\item[Ward 7] Setting $\Psi'_{12}=\Psi_{12}, \Psi''_{12}=L_{-1}^{(2)}\Psi_{12}$, $(m_1,m_2,m_3,m_4)=(-1,1/3,-1/3,-1)$ and $r=1$ in \eqref{eq:Ward-integral} yields a linear relation of the form
\begin{equation}
\begin{aligned}
   & a_{0|7} \aver{\Psi_{12}|\rmmode{1}{1}\sigma^\dag(1)\sigma(\eta)\rmmode{2}{-1}|\Psi_{12}}
   = d_{0|7}  \aver{\Psi_{12}|\sigma^\dag(1)\sigma(\eta)\rmmode{1}{-1}\rmmode{2}{-1}|\Psi_{12}} \\
   & + d_{1|7} \aver{\Psi_{12}|\sigma^\dag(1)\sigma(\eta)\rmmode{1}{0}\rmmode{2}{-1}|\Psi_{12}}
   + d_{2|7} \aver{\Psi_{12}|\sigma^\dag(1)\sigma(\eta)\rmmode{1}{1}\rmmode{2}{-1}|\Psi_{12}} \,.
    \end{aligned}
\end{equation}
\end{description}

\subsubsection*{The null-vector conditions}

The null-vector condition on $\psi_{12}$ in the mother CFT reads
\begin{equation}
    (g L_{-2} - L_{-1}^2) \cdot \psi_{12} \equiv 0 \,.    
\end{equation}
This yields, in the $\Zbb_N$ orbifold
\begin{equation}
    \left[ g L_{-2}^{(r)} - \frac{1}{N}\sum_{s=0}^{N-1} L_{-1}^{(s)}L_{-1}^{(r-s)} \right] \cdot \Psi_{12} \equiv 0 \,,    
\end{equation}
for any $r \in \Zbb/N$. In the present case, we set $N=3$, and we get
\begin{align}
    & \left[ 3g L_{-2}^{(0)} - L_{-1}^{(0)}L_{-1}^{(0)} - 2L_{-1}^{(1)}L_{-1}^{(2)} \right] \cdot \Psi_{12} \equiv 0 \,, \\
    & \left[ 3g L_{-2}^{(1)} - L_{-1}^{(2)}L_{-1}^{(2)} - 2L_{-1}^{(0)}L_{-1}^{(1)} \right] \cdot \Psi_{12} \equiv 0 \,,\\
    & \left[ 3g L_{-2}^{(2)} - L_{-1}^{(1)}L_{-1}^{(1)} - 2L_{-1}^{(0)}L_{-1}^{(2)} \right] \cdot \Psi_{12} \equiv 0 \,.
\end{align}

\section{Orbifold structure constants}
\label{app:compositeonepointbulkboundary}

In the orbifold BCFT, the operator algebra consists of OPEs of three types. First, there is the operator subalgebra of bulk operators. We shall not directly use the structure constants of this subalgebra in this work, but they have been discussed in \cite{borisov_systematic_1998,dupic_entanglement_2018,headrick_entanglement_2010,ares_crossing-symmetric_2021}. 
The second type of OPE we need to consider in the orbifold BCFT, is the bulk-boundary OPE which encapsulates the singular behaviour of a bulk field as it approaches a conformal boundary. In our calculations, we will only need to work with the OPEs of generic (primary) twist field $\sigma_\phi(z,\zb)$ approaches a (diagonal) conformal boundary $\alpha$:
\begin{equation}
\label{eq_bulk_boundary_OPE}
  \sigma_\phi(x,y) \underset{y \to 0}{\sim}
  \sum_k \acon{\alpha}{\sigma_\phi}{\Psi_k} \, (2y)^{h_{\Psi_k}-2 h_{\sigma_\phi}} \, \left( \Psi_k^{\alpha \alpha}(x)  + \cdots \right) 
\end{equation}
where the sum runs over all the diagonal primary boundary operators $\Psi_k^{(\alpha \alpha)} = \psi_k^{(\alpha \alpha)} \otimes  \cdots \otimes \psi_k^{(\alpha \alpha)}$, and the dots stand for their descendants under the neutral subalgebra \cite{2023JPhA...56T5403E}. We also need to consider the OPEs of orbifold boundary operators. For generic diagonal BCCOs, this takes the form
\begin{equation}
  \Psi_{i}^{(\alpha\beta)}(x_1) \Psi_{j}^{(\beta\gamma)}(x_2)
  \underset{x_1 \to x_2}{\sim}
  \sum_k \bcon{\alpha \beta\gamma}{\Psi_{i}}{\Psi_{j}}{\Psi_k}
  (x_1-x_2)^{h_k-h_{i}-h_{j}} \left( \Psi_k^{(\alpha\gamma)}(x_2) + \cdots \right)
\end{equation}
for $x_1 > x_2$, and where the index $k$ runs over all the primary orbifold BCCOs interpolating between the conformal boundary conditions $\alpha$ and $\gamma$. In the present work we are only concerned with untwisted boundary fields. The corresponding structure constants $\bcon{\alpha \beta\gamma}{\Psi_{i}}{\Psi_{j}}{\Psi_k}$ are straightforward to evaluate as they simply factorize into a linear combination of products of mother BCFT three-point functions. Therefore, we will focus on the bulk-boundary structure constants $\acon{\alpha}{\sigma_\phi}{\Psi_k}$. These structure constants can be computed using factorization and unfolding arguments, along the lines of \cite{2023JPhA...56T5403E},\cite{dupic_imaginary_2019} and \cite{headrick_entanglement_2010}. 

The main result derived in this appendix is that these orbifold structure constants are given in terms of the mother theory data by 
\begin{align}
\label{eq_orbifold_strucuture_constant_master}
\boxed{ \acon{\alpha}{\sigma_{\phi}}{\Psi_{k}} = g_{\alpha}^{1-N} \, N^{-N h_{\psi_k}} \,\left\langle \phi(0) \psi_{k}^{(\alpha\alpha)}(1) \psi_{k}^{(\alpha\alpha)}(\omega) \cdots \psi_{k}^{(\alpha\alpha)}(\omega^{N-1})\right\rangle^{\alpha}_{\mathbb{D}} } \,.
\end{align}
Two particular cases are used in this paper : 
\begin{itemize}
\item when $\Psi_k^{\alpha \alpha} = \id$, this simply becomes
\begin{equation}
\label{orbifold_strucuture_constant_boudnary_id}
    \boxed{  \acon{\alpha}{\sigma_{\phi}}{\id}=g_{\alpha}^{1-N} \, A^{(\alpha)}_{\phi}}
\end{equation}
\item for $N=2$ and $\Phi = \id$ this simplifies to 
\begin{align}
\label{eq:bulk_boundary_phi_k_structure_constant}
 \boxed{\acon{\alpha}{\sigma}{\Psi_{k}} = g_{\alpha}^{-1} \, 2^{-2 h_{\psi_k}} \left\langle  \psi_{k}^{(\alpha\alpha)}(1) \psi_{k}^{(\alpha\alpha)}(-1)\right\rangle^{\alpha}_{\mathbb{D}} = g_{\alpha}^{-1} \, 2^{-4 h_{\psi_k}}}
\end{align}
provided the field $\psi_k$ exists on the boundary $\alpha$. 
\end{itemize} 

The derivation of \eqref{eq_orbifold_strucuture_constant_master} can be done in two steps.  
First, consider the bulk-boundary structure constant $\acon{\alpha}{\sigma}{\id}$, which can be expressed as the following one-point function on the unit disk $\mathbb{D}$:
\begin{equation} \label{eq:onepointdefinitiondisk}
  \acon{\alpha}{\sigma}{\id}
  =\aver{\sigma(0)}_{\mathbb{D}}^\alpha \,,
\end{equation}
This can be interpreted as the following ratio of mother CFT partition functions:
\begin{equation}
  \aver{\sigma(0)}_{\mathbb{D}}^\alpha
  =\frac{\mathcal{Z}^{(\alpha)}_{\mathbb{D}_N}}{\left[\mathcal{Z}^{(\alpha)}_{\mathbb{D}} \right]^N} \,,
\end{equation}
where $\mathbb{D}_N$ denotes the $N$-th covering of the unit disk with a branch point at the origin. The branched covering $\mathbb{D}_N$ is conformally equivalent to the standard unit disk $\mathbb{D}$, but features a conical singularity at the origin. A careful treatment of the corresponding conformal anomaly \cite{sully_bcft_2021}, however, reveals that 
\begin{equation}
\mathcal{Z}^{(\alpha)}_{\mathbb{D}_N} = \mathcal{Z}^{(\alpha)}_{\mathbb{D}}\,,
\end{equation}
leading to the result
\begin{equation}
  \aver{\sigma(0)}_{\mathbb{D}}^\alpha = \left[\mathcal{Z}^{(\alpha)}_{\mathbb{D}} \right]^{1-N} \,.
\end{equation}
Additionally, the partition function $\mathcal{Z}^{(\alpha)}_{\mathbb{D}}$ is equal to the \textit{ground state degeneracy}  $g_\alpha=\aver{0|\alpha}$, which is defined as the overlap between the vacuum state $\ket{0}$ and the boundary state $\ket\alpha$ in the mother BCFT \cite{affleck_boundary_1998}. Consequently, we have
\begin{equation} \label{eq:bulkboundary}
  \acon{\alpha}{\sigma}{\Psi_{\id}}=g_\alpha^{1-N} \,.
\end{equation}
This structure constant is responsible for the well-known universal additive correction $\log g_{\alpha}$ to the boundary entanglement entropy \cite{calabrese_entanglement_2004}. 
Consider now the generic structure constant $\acon{\alpha}{\sigma_{\phi}}{\Psi_{k}} $, which is equal to the following correlation function on the unit disk :
\begin{equation}\label{eq:compositestructure13}
   \acon{\alpha}{\sigma_{\phi}}{\Psi_{k}}   = \left\langle \sigma_{\phi}(0)\Psi_{k}^{(\alpha\alpha)}(1)\right\rangle^{\alpha}_{\mathbb{D}} \,.
\end{equation}
Now the ratio of correlation functions 
\begin{align}
\frac{\left\langle \sigma_{\phi}(0)\Psi_{k}^{(\alpha\alpha)}(1)\right\rangle^{\alpha}_{\mathbb{D}}}{\left\langle \sigma(0)\right\rangle^{\alpha}_{\mathbb{D}}} 
\end{align}
can be interpreted as a correlation function in the mother theory involving $N$ boundary fields $\psi_{k}^{(\alpha \alpha)}$ and one bulk field $\phi$ on the $N$-sheeted branched covering $\mathbb{D}_N$. This covering can be uniformized to the standard unit disk via the holomorphic map $z \to z^{1/N}$, leading to
\begin{align}
\frac{\left\langle \sigma_{\phi}(0)\Psi_{k}^{(\alpha\alpha)}(1)\right\rangle^{\alpha}_{\mathbb{D}}}{\left\langle \sigma(0)\right\rangle^{\alpha}_{\mathbb{D}}} = N^{-N h_{\psi_k}} \left\langle \phi(0) \psi_{k}^{(\alpha\alpha)}(1) \psi_{k}^{(\alpha\alpha)}(\omega) \cdots \psi_{k}^{(\alpha\alpha)}(\omega^{N-1})\right\rangle^{\alpha}_{\mathbb{D}}
\end{align}
Substituting $\left\langle \sigma(0)\right\rangle^{\alpha}_{\mathbb{D}} = g_{\alpha}^{N-1}$ yields the announced result. \\

We note that the structure constant \eqref{eq:bulk_boundary_phi_k_structure_constant}  can also be obtained from the result of \cite{Estienne:2021xzo} for the twist 2-point function on the unit disc with conformal BC $\alpha$. This two-point function is  
\begin{equation}\label{eq:Disc_twist_two_point}
    \aver{\sigma(0)\sigma(x)}_{\mathbb{D}}=g_{\alpha}^{-2}\,2^{-c/3}[|x|^2(1-|x|^2)]^{-2h_\sigma}\mathcal{Z}_{\alpha|\alpha}(\tau)
\end{equation}
where $\mathcal{Z}_{\alpha|\alpha}(\tau)$ is the annulus partition function with boundary conditions $\alpha$ on both boundary components. The annulus has perimeter $1$, and real width $-i \tau/2$, where $\tau$ is given by
\begin{equation}
      \tau(|x|)=i \frac{{ }_2 \mathrm{~F}_1\left(\frac{1}{2}, \frac{1}{2}, 1 ; 1-|x|^2\right)}{{ }_2 \mathrm{~F}_1\left(\frac{1}{2}, \frac{1}{2}, 1 ;|x|^2\right)} \,.
\end{equation}
The limit $|x| \to 1$ corresponds to $\tau \to 0$ and 
\begin{equation}
\tilde{q} = e^{-2i\pi /\tau} \sim \left( \frac{1-|x|^2}{16} \right)^2 \to 0
\end{equation}
Using 
\begin{equation}
\mathcal{Z}_{\alpha|\alpha}(\tau) = \sum_k n_{\alpha \alpha}^k  \chi_k(-1/\tau) 
\end{equation}
and in the absence of degeneracies ($n_{\alpha \alpha}^k \in \{ 0 , 1 \}$), one finds that the conformal block of $\aver{\sigma(0)\sigma(x)}_{\mathbb{D}}$ corresponding to the bulk-to-boundary fusion $  \sigma_\phi(x) \underset{|x| \to 1}{\sim} (1-|x|^2)^{-2h_{\sigma} +  h_{\Psi_k} } \acon{\alpha}{\sigma_\phi}{\Psi_k} \Psi_k $ is proportional to 
\begin{equation}
[|x|^2(1-|x|^2)]^{-2h_\sigma} n_{\alpha \alpha}^k  \chi_k(-1/\tau) 
\end{equation}
and one can read off the corresponding structure constant, namely
\begin{align}
\left( \acon{\alpha}{\sigma}{\Psi_{k}} \right)^2= g_{\alpha}^{-2} \, 2^{-8 h_{\Psi_k}} n_{\alpha \alpha}^k \,.
\end{align}
We thus recover  \eqref{eq:bulk_boundary_phi_k_structure_constant} when $n_{\alpha \alpha}^k =1$.

\section{Alternative derivation of scaling functions in the Ising CFT}
\label{app:fixedmixedIsing}

In this section, we will derive the bare and excited twist contributions to the second and third R\'enyi entropy in the critical Ising chain with fixed mixed BC $\alpha=+,\beta=-$.

In the Ising BCFT, the boundary field that interpolates between the corresponding conformal BC $|\pm\rangle$ is the operator $\psi^{(+-)}_{\varepsilon}$, with conformal dimension $h= 1/2$. The most straightforward approach to this computation is to use equation \eqref{eq:F_{N,phi}unfolded}, as was done in \cite{taddia_entanglement_2013}, to obtain the leading scaling function $\mathcal{F}_{N}^{(+-)}$. Indeed the $2N$-points correlation function in the numerator of \eqref{eq:F_{N,phi}unfolded} can be obtained rather simply using Wick's theorem, bosonization or even Jack polynomials \cite{2001math.....12127F,2007JPhA...40.2509D,2008PhRvL.100x6802B,2009JPhA...42R5209E,Estienne:2010as}. This method would apply equally well to the subleading scaling function $\mathcal{F}_{N,\varepsilon}^{(+-)}$. 

These computations can also be done in the orbifold formalism. In the $\mathbb{Z}_N$ orbifold of this theory, the change in boundary conditions is implemented by the diagonal operator $\Psi^{(+-)}_{\varepsilon}$ defined as in (\ref{eq:diagBCCO}). There the essential observation is that the space of conformal blocks is one-dimensional for the correlation function
\begin{equation}\label{eq:chiralmixedfixeddef}
    \left\langle \Psi^{(- +)}_{\varepsilon}(\infty) \sigma_\phi(z,\bar{z}) \Psi^{(+ -)}_{\varepsilon}(0)\right\rangle
\end{equation}
and is spanned by  
\begin{equation}
  F_{N,\phi}(\eta) = \cblock{\Psi_{\varepsilon}(\infty)}{\ \Psi_{\varepsilon}(0)}{\sigma_{\phi}^{\dag}(1)}{\sigma_{\phi}(\eta)}{\id} 
 \,.
\end{equation} 
This follows from the trivial fusion rule $\Psi_{\varepsilon} \times \Psi_{\varepsilon} \to \id$, coupled to the fact that the fusion channel $\sigma_{\phi} \times \sigma_{\phi}^{\dag} \to \id$ cannot be degenerate (fusion numbers of the form $N_{a \bar{a}}^{\id}$ are always equal to one).  The conformal block in the other fusion channel can be obtained from the fusion rules of the $\mathbb{Z}_N$ orbifold,  found in \cite{borisov_systematic_1998} for $N=2$ and in \cite{2023JPhA...56T5403E} for $N$ prime. 
We will focus on the cases $N = 2$ and $N =3$. For $N=2$ we have 

\begin{equation}
 \cblockv{\Psi_{\varepsilon}(\infty)}{\ \Psi_{\varepsilon}(0)}{\sigma^\dag(1)}{\sigma(\eta)}{\sigma}
 \qquad \text{and} \qquad
 \cblockv{\Psi_{\varepsilon}(\infty)}{\ \Psi_{\varepsilon}(0)}{\sigma_\varepsilon^\dag(1)}{\sigma_\varepsilon(\eta)}{\sigma_\varepsilon}
\end{equation}
while for $N=3$ the blocks are
\begin{equation}
 \cblockv{\Psi_{\varepsilon}(\infty)}{\ \Psi_{\varepsilon}(0)}{\sigma^\dag(1)}{\sigma(\eta)}{\sigma_\varepsilon^{(1)}}
 \qquad \text{and} \qquad
 \cblockv{\Psi_{\varepsilon}(\infty)}{\ \Psi_{\varepsilon}(0)}{\sigma_\varepsilon^\dag(1)}{\sigma_\varepsilon(\eta)}{\sigma^{(2)}}
\end{equation}
where $\sigma_\varepsilon^{(1)} \propto L_{-1/3} \sigma_\varepsilon$ and $\sigma^{(2)} \propto L_{-2/3}\sigma$.
 These fusion rules also imply the leading singular behaviour of the conformal block around the points $\eta\in\{0,1,\infty\}$. The corresponding exponents are given in Table \ref{tab:allexponentsfixedmixed}.
\begin{table}[h]
    \centering
    \begin{tabular}{c|c|c|c}
         & $0$& $1$ & $\infty$  \\
         \hline
        $N=2, \phi=\id$        & $-1$ & $-\frac{1}{16}$ & $-\frac{15}{16}$ \\ 
           \hline
        $N=2,  \phi= \varepsilon$ & $-1$ & $-\frac{9}{16}$  & $-\frac{7}{16}$ \\ 
           \hline
        $N=3,  \phi=\id$ & $-1$ & $-\frac{1}{9}$ & $-\frac{8}{9}$ \\ 
           \hline
        $N=3, \phi= \varepsilon $ & $-1$ & $-\frac{4}{9}$ & $-\frac{5}{9}$  \\ 
    \end{tabular}
    \caption{Singular behaviour of the conformal block of (\ref{eq:chiralmixedfixeddef}) for different $N$ and twist field insertions $\sigma_j(\eta)$.}
    \label{tab:allexponentsfixedmixed}
\end{table}

From the exponents around $\eta\rightarrow 0$ and $\eta\rightarrow 1$ we can determine the generic form of the conformal blocks for the four cases enumerated above to be:
\begin{equation}
   F_{N,\phi}(\eta)=\eta^{-1}(\eta-1)^{-2 h_{\sigma_\phi}}    P_{N,\phi}(\eta) 
\end{equation}
where $P_{N,\phi}(\eta)$ is a degree $2$ polynomial in $\eta$ satisfying
\begin{equation}
P_{N,\phi}(\eta)  = \eta^2  P_{N,\phi}(1/\eta)
\end{equation}
This polynomial can be found by comparing with the OPE 
\begin{equation}\label{eq:OPE}
 (\eta -1)^{2 h_{\sigma_\phi}}   \sigma_\phi(\eta) \sigma^{\dag}_\phi(1)= \id +\frac{2 h_{\sigma_\phi}}{N c} (\eta-1)^2 \, T(1)+\dots
\end{equation}
This implies that 
\begin{equation}
\eta^{-1} P_{N,\phi}(\eta) \sim 1 + \frac{2 h_{\sigma_\phi}}{N c} \bra{\Psi_{12}} T(1) \ket{\Psi_{12}}\, (\eta-1)^2, \qquad (\eta \to 1)
\end{equation}
yielding  
\begin{equation}
    P_{N,\phi}(\eta) = 2 h_{\sigma_\phi} \, (1-\eta)^2 + \eta\,.
\end{equation}
The corresponding scaling functions for $N \in \{2,3\}$ and $\phi \in \{ \id, \varepsilon \}$ follows 
\begin{equation}
\mathcal{F}^{(+-)}_{N,\phi}(\eta) =  \sqrt{2}^{N-1} \left( 1+ 2 h_{\sigma_\phi} \left(\eta^{1/2}-\eta^{-1/2} \right)^2 \right)
\end{equation}
where we used the OPE structure constant of the bulk to boundary fusion $\sigma_{\phi} \to \id$, namely 
\begin{equation}
\acon{\pm}{\sigma_\phi}{\id} = A^{\pm}_{\phi} \, g_{\pm}^{1-N} = \sqrt{2}^{N-1}
\end{equation}
since $A_{\varepsilon}^{\pm}  = A_{\id}^{\pm} =1 $ and $g_{\pm}= 1/\sqrt{2}$.  

More explicitly in terms of $\theta = \pi \ell/L$:
\begin{align}
\mathcal{F}^{(+-)}_{2,\id}(\eta) & = \sqrt{2} \left(1-\frac{\sin ^2 \theta }{4}\right) = \frac{7+ \cos 2 \theta}{4 \sqrt{2}}  \\
\mathcal{F}^{(+-)}_{2,\varepsilon}(\eta) & = \sqrt{2} \left(1-\frac{9 \sin ^2 \theta }{4}\right) = \frac{9 \cos 2 \theta -1}{4\sqrt{2}} \\
\mathcal{F}^{(+-)}_{3,\id}(\eta) & =  2\left( 1 -\frac{4 \sin ^2 \theta }{9} \right) =  \frac{2}{9} (7 + 2 \cos 2 \theta ) \\
\mathcal{F}^{(+-)}_{3,\varepsilon}(\eta) & = 2 \left(1-\frac{16 \sin ^2\theta }{9}\right) = \frac{2}{9} (1 + 8 \cos 2 \theta ) 
\end{align}

\bibliographystyle{unsrt}
\bibliography{references}

\end{document}